\PassOptionsToPackage{square,numbers,comma,sort&compress,super}{natbib} 
\documentclass[aps,prl,twocolumn,a4paper,superscriptaddress]{revtex4-2}

\usepackage[paperwidth=215mm,paperheight=300mm,centering,hmargin=1.6cm,vmargin=2cm]{geometry}
\usepackage{graphicx}
\usepackage{amssymb,amsmath,amsthm,enumitem}

\usepackage{latexsym,stmaryrd}
\usepackage[dvipsnames]{xcolor}
\usepackage[normalem]{ulem}
\usepackage{float}

\usepackage{physics} 
\usepackage{siunitx} 
\usepackage{chemformula} 
\usepackage{textcomp}
\usepackage{gensymb}

\renewcommand{\topfraction}{0.95}
\renewcommand{\bottomfraction}{0.95}
\renewcommand{\textfraction}{0.05}
\renewcommand{\floatpagefraction}{0.95}

\newcommand*{\Scale}[2][4]{\scalebox{#1}{$#2$}}%

\renewcommand{\figurename}{\textbf{Fig.}}

\makeatletter
\renewcommand*{\fnum@figure}{{\normalfont\bfseries \figurename~\thefigure}}
\makeatother

\makeatletter
\def\@hangfrom@section#1#2#3{\@hangfrom{#1#2}{#3}}
\AtBeginDocument{%
  \renewcommand\section{\@startsection{section}{1}{\z@}%
    {-2.5ex plus -1ex minus -.2ex}%
    {0.2ex plus .2ex}%
    {\normalfont\bfseries}}%
  \renewcommand\subsection{\@startsection{subsection}{2}{\z@}%
    {-2.5ex plus -1ex minus -.2ex}%
    {0.2ex plus .2ex}%
    {\normalfont\bfseries}}%
}
\makeatother

\usepackage[utf8]{inputenc}
\usepackage{array}
\usepackage{hyperref}
\usepackage{afterpage}
\usepackage{lipsum}
\makeatletter
\newcommand*{\citenst}[2][]{%
  \begingroup
  \let\NAT@mbox=\mbox
  \let\@cite\NAT@citenum
  \let\NAT@space\NAT@spacechar
  \let\NAT@super@kern\relax
  \renewcommand\NAT@open{[}%
  \renewcommand\NAT@close{]}%
  \citep{#2}%
  \endgroup
}
\makeatother

\makeatletter
\newcommand{\setlabel}[1]{\edef\@currentlabel{#1}\label}
\makeatother

\begin{document}

\title{Broadband silicon photonic phase shifters driven by gradient optical forces}

\author{Guillermo Arregui}
\email{guillermo.arreguibravo@epfl.ch}
\affiliation{Department of Electrical and Photonics Engineering, DTU Electro, Technical University of Denmark, Building 343, DK-2800 Kgs.\ Lyngby, Denmark}
\affiliation{Laboratory of Photonics and Quantum Measurements, Swiss Federal Institute of Technology Lausanne (EPFL), CH-1015 Lausanne, Switzerland}

\author{Sander J{\ae}ger Linde}
\affiliation{Department of Electrical and Photonics Engineering, DTU Electro, Technical University of Denmark, Building 343, DK-2800 Kgs.\ Lyngby, Denmark}

\author{Magnus Vejby Nielsen}
\affiliation{Department of Electrical and Photonics Engineering, DTU Electro, Technical University of Denmark, Building 343, DK-2800 Kgs.\ Lyngby, Denmark}

\author{Bingrui Lu}
\affiliation{Department of Electrical and Photonics Engineering, DTU Electro, Technical University of Denmark, Building 343, DK-2800 Kgs.\ Lyngby, Denmark}

\author{Nikolaj B. Hougs}
\affiliation{Department of Electrical and Photonics Engineering, DTU Electro, Technical University of Denmark, Building 343, DK-2800 Kgs.\ Lyngby, Denmark}

\author{Babak Vosoughi Lahijani}
\affiliation{Department of Electrical and Photonics Engineering, DTU Electro, Technical University of Denmark, Building 343, DK-2800 Kgs.\ Lyngby, Denmark}

\author{S{\o}ren Stobbe}
\affiliation{Department of Electrical and Photonics Engineering, DTU Electro, Technical University of Denmark, Building 343, DK-2800 Kgs.\ Lyngby, Denmark}
\affiliation{NanoPhoton - Center for Nanophotonics, Technical University of Denmark, {\O}rsteds Plads 345A, DK-2800 Kgs.\ Lyngby, Denmark.}

\date{\today}

\small
\begin{abstract}

While initially deployed for optical interconnects, silicon photonics is increasingly being explored as a hardware platform for programmable optical systems, including linear optical processors, neuromorphic photonic networks, quantum photonic circuits and multiplexed sensor arrays. Common to most existing implementations is that light is controlled with electronics and even basic demonstrations wherein light directly controls light remain limited. Here we demonstrate a broadband all-optical silicon photonic phase shifter based on an optomechanically mediated light--light interaction arising from the gradient optical force.  Our device concept relies on slot-mode waveguides suspended by subwavelength gratings, which provide mechanical support while preserving optical confinement. We demonstrate all-optical phase shifting using a guided pump beam co-propagating with the signal beam, with only 60~$\mu$W required to achieve a $\pi$ phase shift in a 178.6 $\mu$m-long device. In addition, we measure the required pump power across a wide parameter space and find quantitative agreement with a lumped force-equilibrium model. Since the actuation relies on an all-optical geometric deformation rather than on material-index tuning, the approach avoids local electrical connections to the active element, carries no Kramers--Kronig absorption penalty, and is naturally compatible with cryogenic quantum photonic platforms.
 \end{abstract}

\maketitle

Silicon photonics has become the dominant platform for large-scale, dense optical integration. The large refractive-index contrast between silicon and silica enables tight optical confinement and compact bends, while CMOS-compatible wafer-scale fabrication processes provide a direct route to high-volume, low-cost manufacturing. Beyond its established role in co-packaged optics and optical interconnects~\cite{reed_silicon_2010}, silicon photonics now underpins complex reconfigurable photonic systems from programmable interferometric meshes~\cite{reck_experimental_1994,bogaerts_programmable_2020}, coherent optical neural networks~\cite{shen_deep_2017}, quantum photonic processors~\cite{harris_quantum_2017}, and fiber-based quantum communication networks~\cite{wehner_quantum_2018}, to application-specific circuits such as tunable filters~\cite{liao_integrated_2014}, optical phased arrays~\cite{sun_large-scale_2013}, and integrated spectrometers~\cite{souza_fourier_2018,yao_integrated_2023}. These are examples of the rapidly growing interest in novel approaches to photonic information processing that go well beyond conventional optical communication systems: Optical signals may carry not only high-bandwidth data, but also, e.g., phase information, quantum states, microwave signals, spectral information, or orbital angular momentum. Common to both existing and nascent photonic information technology is the need for being configured; many quasi-static phase settings define optical transfer functions, implement programmable transformations, and stabilize the circuit operating point.

The key component enabling the reconfigurability is the phase shifter, which must provide $\pi$-scale phase shifts with low loss, low power, compact footprint, and scalable control. In silicon photonics, the dominant mechanisms rely on heating or free-carrier dispersion~\cite{reed_silicon_2010}. Thermo-optic phase shifters are broadband and robust, but typically require milliwatt-level static powers and generate thermal crosstalk~\cite{qiu_energy-efficient_2020}. Carrier-based phase shifters provide high-speed operation and can consume low static power, but require doping and trade efficiency against free-carrier absorption, often leading to millimeter-scale lengths~\cite{baehr-jones_ultralow_2012}. More recently, micro- and nanoelectromechanical (MEMS/NEMS) implementations have achieved sub-nanowatt standby power, insertion loss below 0.2~dB, CMOS-compatible drive voltages, and microsecond switching~\cite{edinger_silicon_2021,baghdadi_dual_2021,kim_programmable_2023}. However, all these approaches require local electrical actuation with metal electrodes and wires, which introduce electrical dissipation, add heat load to cryogenic quantum circuits, and suffer from electrical routing overhead that grows with array size.

All-optical tuning could remove the local electrical interface by using light to set the circuit state. Existing approaches exploit optically induced heating~\cite{yao_ultrahigh_2023}, photocarrier generation and Kerr nonlinearities~\cite{almeida_all-optical_2004,lin_nonlinear_2007}, or optically written phase-change materials~\cite{rios_integrated_2015,delaney_nonvolatile_2021}. These mechanisms generally inherit trade-offs in static or switching power, free-carrier absorption, required optical intensity or resonant enhancement, and heterogeneous material integration. Optical forces provide a distinct route: an on-chip guided control field mechanically reconfigures the waveguide itself, changing the geometry and effective index experienced by a co-propagating signal. In cavity optomechanics~\cite{aspelmeyer_cavity_2014}, radiation pressure has enabled control of mechanical motion down to the quantum level~\cite{chan_laser_2011,wilson_measurement-based_2015,safavi-naeini_squeezed_2013,riedinger_remote_2018} and static reconfiguration of nanophotonic devices~\cite{dorsel_optical_1983,rosenberg_static_2009,wiederhecker_controlling_2009,cai_nanoelectromechanical_2013,ren_experimental_2020}. Still, cavity-enhanced approaches are tied to optical resonances and require wavelength alignment and laser locking~\cite{wiederhecker_broadband_2011}. Travelling-wave gradient-force interactions overcome these limitations by exploiting the exponential dependence of a guided-mode effective index on the separation between adjacent dielectric structures~\cite{povinelli_evanescent-wave_2005}, and have been demonstrated in suspended waveguides coupled to a substrate~\cite{li_harnessing_2008}, another waveguide~\cite{li_tunable_2009,li_broadband_2009}, beams~\cite{fong_high_2010}, and cantilevers~\cite{sauer_nanophotonic_2012}. These approaches have been applied to reconfigurable filters~\cite{deotare_all_2012} and directional couplers~\cite{fong_tunable_2011}, but extending them to a low-power, compact phase shifter requires reconciling four competing demands: nanometric gaps for large forces, mechanical compliance for sizeable phase shifts, low propagation loss, and structural robustness during fabrication. In silicon-on-insulator, this challenge is compounded by the absence of built-in tensile stress, which makes suspended structures particularly vulnerable to collapse during underetching. These challenges may explain why all-optical gradient-force phase shifting was so far not demonstrated in silicon photonics, despite silicon offering advantages in terms of compact footprint, strong gradient forces~\cite{rodrigues_optical_2017}, and scalability.

\begin{figure*}[t!]
\includegraphics[width=\textwidth]{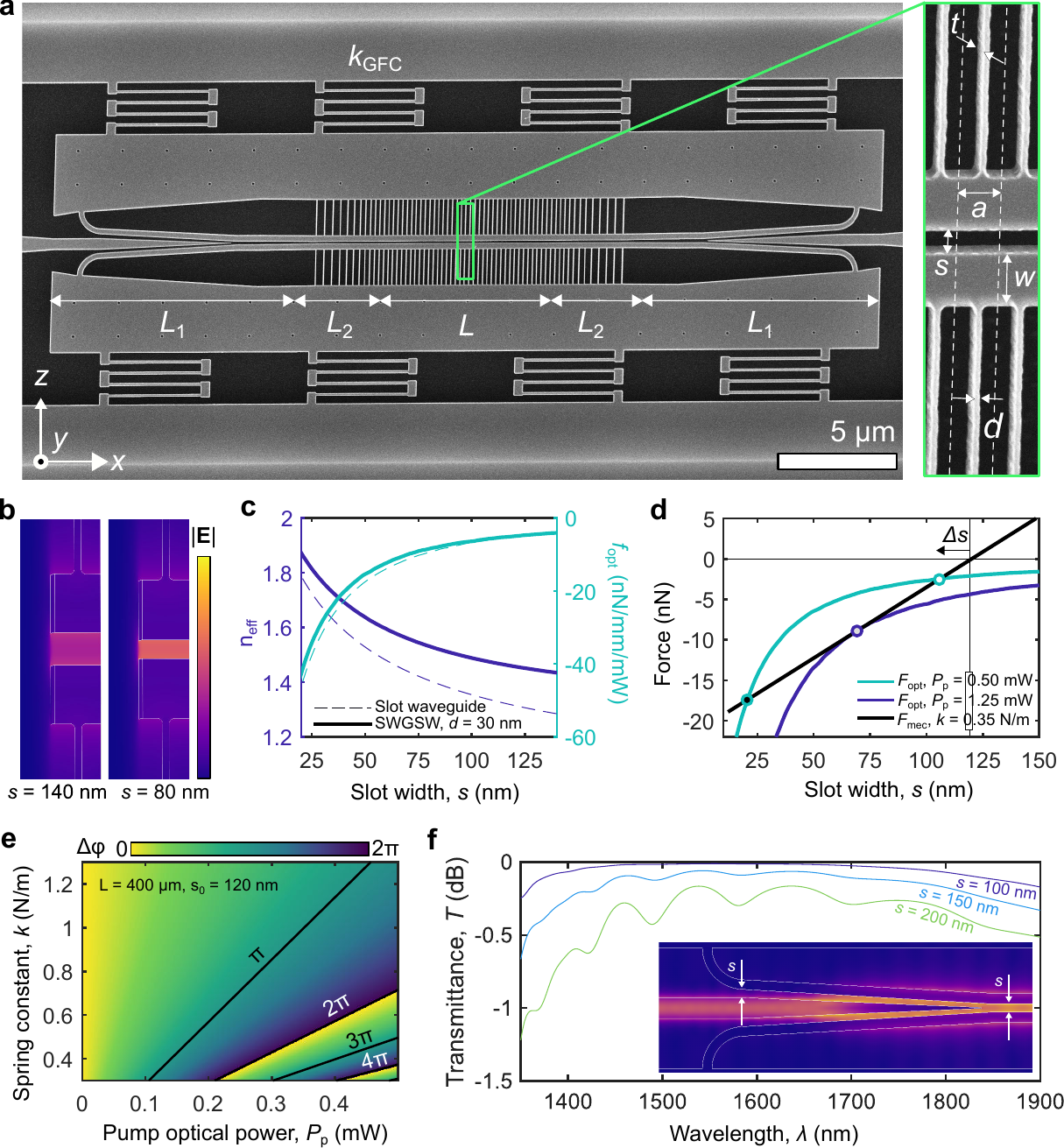}
    \caption{\textbf{Optomechanical phase shifters based on subwavelength-grating slot waveguides (SWGSWs).} \textbf{a}, Scanning electron micrograph (SEM) of a photonic device composed of rectangular-to-slot waveguide couplers of length $L_1$, slot waveguide-to-SWGSW couplers of length $L_2$, and a central SWGSW section of length $L$. The zoom-in shows a tilted view of the SWGSW unit cell and its main geometric parameters. \textbf{b}, Electric field intensity, $\left| \mathbf{E}\right|$, of the fundamental transverse-electric-like guided mode for two different slot widths, $s$. \textbf{c}, Dependence of the effective refractive index, $n_\text{eff}$, and optical force per unit length and power, $f_\text{opt}$, on $s$ for a SWGSW with $d$ = 30 nm and $d$ = 0 nm, where the latter corresponds to a slot waveguide. \textbf{d}, Force equilibrium for a SWGSW ($L$ = 700 $\mu$m, $s_0$ = 120 nm, $k$ = 0.35 N/m) driven at $\lambda$ = 1550 nm and powers $P_{\text{p}}$ of 0.5 mW and 1.25 mW. Intersections between the restoring elastic force, $F_\text{mec}$, and the optical force, $F_\text{opt}$, correspond to either stable (white dot), unstable (black dot), or inflection (gray dot) equilibrium points. \textbf{e}, Achieved phase shift, $\Delta\varphi$, for a SWGSW ($L$ = 400 $\mu$m, $s_0$ = 120 nm) as a function of $k$ and $P_{\text{p}}$. Black solid lines represent isocontours of $\Delta\varphi = n\pi$, $n\in \mathbb{N}$. \textbf{f}, Simulated power transmittance, $T$, of the V-groove rectangular-to-slot adiabatic coupler as a function of wavelength for slot widths $s$ = 150 and 200~nm. Inset: electric field magnitude, $\left|\mathbf{E}\right|$, along the coupler, illustrating adiabatic conversion from the strip mode (left) to the slot mode (right).}
    \label{fig:1}
\end{figure*}

In this work, we meet these competing demands with a phase-shifter architecture built around a suspended subwavelength-grating slot waveguide (SWGSW): the grating cladding doubles as a continuous mechanical support and a low-perturbation optical cladding, eliminating discrete scattering tethers, while a distributed array of guided folded cantilever springs sets the stiffness independently of the device length. Phase shifting is all-optical---a guided pump beam pulls the two halves of the slot together through the attractive travelling-wave gradient force, changing the optical path length experienced by a co-propagating signal at a different wavelength. A compact device with an active length of only 178.6~$\mu$m (a 150-$\mu$m SWGSW plus its adiabatic couplers) achieves a half-wave power of $P_\pi \approx 60~\mu$W with a wavelength-averaged insertion loss of $\sim$1.4~dB. This performance is not accidental: by mapping $P_\pi$ across devices of varying length, slot width, and spring constant, we show that a lumped force-equilibrium model quantitatively captures the entire design space and singles out the initial slot width as the decisive parameter, since narrowing it strengthens the optical force and reduces the propagation loss. Pump-probe measurements further show that the accessible wavelength range is bounded by the grating couplers rather than by the phase shifter, whose purely geometric actuation is intrinsically broadband. These results open a path toward programmable silicon photonic circuits in which broadband phase control is delivered entirely over optical channels without electrical routing to the active elements.

\section{Device concept and modelling}

Figure~\ref{fig:1}\textbf{a} shows a scanning electron micrograph (SEM) of one of our suspended silicon photonic phase shifters. They consist of five different sections in series: two adiabatic couplers connecting rectangular waveguides to slot waveguide at the ends, two intermediate couplers from slot waveguide to subwavelength-grating slot waveguide, and a central subwavelength-grating slot waveguide (SWGSW)~\cite{zhou_fully_2017}. The first two have a fixed geometry (notably their lengths, $L_1$ and $L_2$), while we explore SWGSWs of varying length, $L$, and varying initial slot width, $s_{\text{0}}$. The SWGSW is suspended on each side by a distributed periodic array of $N$ guided folded cantilever springs (GFCSs), each of in-plane stiffness $k_\text{GFC}$. The geometry of the SWGSW unit cell, shown in the zoom-in of Fig.~\ref{fig:1}\textbf{a}, is otherwise fixed. The phase-shifting capabilities of the device rely on the geometrical modification of the cross-section of the SWGSW induced by the optical force exerted by the electromagnetic field propagating in the fundamental transverse-electric-like mode. Figure~\ref{fig:1}\textbf{b} shows the intensity of the electric field, $\left|\mathbf{E}\right|$, of the mode at a wavelength $\lambda$ = 1550 nm for structures with rectangular waveguide width $w$ = 254 nm, periodicity $a$ = 200 nm, tether width $d$ = 30 nm and slot widths $s$ = 140 and 80 nm. For small values of $s$, the field, whose leading polarization is along the $z$ axis, is tightly confined in the air-slot region due to the discontinuity of the electric field perpendicular to the air-silicon interface~\cite{almeida_guiding_2004}. As a consequence, its effective refractive index, $n_{\text{eff}}$, depends biexponentially on $s$, as shown in Supplementary Section S1.1 and illustrated in Fig.~\ref{fig:1}\textbf{c} (left axis), which we utilize to change the optical path length and produce a phase shift, $\Delta\varphi$. The use of a subwavelength grating ($\lambda \gg a$) for the cladding and the small filling fraction employed ($d/a$ = 0.15) ensures that the slot mode of the SWGSW does not deviate much from that sustained by a conventional slot waveguide with a homogeneous air cladding (dashed line in Fig.~\ref{fig:1}\textbf{c}), for which strong attractive gradient optical forces between the two rectangular sections have been predicted~\cite{povinelli_evanescent-wave_2005} and experimentally demonstrated~\cite{li_tunable_2009}. We calculate the transverse, i.e., in the $z$ direction, optical force per unit length and power, $f_{\text{opt}}$, exerted on each side of the SWGSW by direct integration of the Maxwell stress tensor along the silicon-air interfaces of a single unit cell~\cite{rakich_general_2009}. As expected from the strong connection between $\partial n_{\text{eff}}/\partial s$ and $f_{\text{opt}}(s)$~\cite{povinelli_evanescent-wave_2005}, $f_{\text{opt}}$ also displays a non-linear and monotonic behaviour as a function of $s$ (right axis in Fig.~\ref{fig:1}\textbf{c}), and differs only slightly from the case of a conventional slot waveguide. More details on the SWGSW geometry, normal modes, and exerted optical forces can be found in Supplementary Section S1.1.

The optical power, $P_{\text{p}}$, injected into the SWGSW determines the gradient optical force, which deforms the beams and therefore changes $s$. Assuming the waveguide cross section remains constant except for the change in $s$, the optical phase shift on a probe/signal beam at wavelength $\lambda_\text{sig}$ --relative to the undeformed waveguide-- is given by
\begin{equation}
    \Scale[0.92]{\Delta\varphi = \frac{2\pi}{\lambda_\text{sig}}\int_{0}^{L}\left[n_\text{eff}(s_0-u_{z,1}(x)-u_{z,2}(x),\lambda_\text{sig})-n_\text{eff}(s_0,\lambda_\text{sig})\right]dx,}
    \label{eq:phase_shift}
\end{equation}
with $u_{z,i}(x)$ the steady-state displacement profile of each beam induced by a pump beam of power $P_\text{p}$ and wavelength $\lambda_\text{p}$. The exact displacement depends on geometry, e.g., through $s_0$, the mechanical boundary conditions, and how light is coupled into the system. Previous work has explored simple geometries such as doubly-clamped beams or cantilevers~\cite{guo_broadband_2012, ozer_stability_2018, ashour_stability_2019} and applied Euler-Bernoulli beam theory subject to the (non-uniform) pre-calculated load, $f_{\text{opt}}(s)$, to solve for the beam deflection and the expected phase shift. These have resulted in either analytical solutions based on linearization of the force around the initial slot width~\cite{guo_broadband_2012}, which cannot accurately capture their stability condition, or in numerical solutions, which extend the previous to the full non-linear force by using an iterative solver~\cite{ozer_stability_2018}. The latter approach is also used in a finite-element-method (FEM) implementation~\cite{ashour_stability_2019}, which naturally applies to more complex geometries. Nevertheless, the FEM approach makes the exploration of the design space computationally expensive. In addition, all previous schemes are unable to independently change the phase-shifter length, $L$, and the effective spring constant, $k$, which narrows the design space and imposes tight stability constraints due to the cubic relationship between $k$ and $L$ in slender structures~\cite{schmid_fundamentals_2023}. For the devices we explore here (Fig.~\ref{fig:1}\textbf{a}), the design space is widened, and the modeling simplified due to the decoupling of rigidity and device footprint, enabled by the inclusion of subwavelength periodic mechanical anchoring, strip-to-slot waveguide couplers with no mechanical connection, and a distributed array of GFCSs. Neglecting the forces exerted on the coupler regions ($L\gg L_1, L_2$), disregarding bending displacements due to the large rigidity of the platforms on each side of the waveguide, and using the mirror symmetry about the propagation axis, the displacement $u_{z,i}(x)$ of both beams can be considered equal and rigid, $u_z(x) \equiv u_z$, and easily found by the lumped force equilibrium
\begin{equation}
        Nk_\text{GFC}u_z = P_{\text{p}}Lf_\text{opt}(s_0-2u_z).
    \label{eq:force_equilibrium}
\end{equation}
Figure~\ref{fig:1}\textbf{d} represents this force equilibrium for a SWGSW of length $L$ = 700 $\mu$m, initial slot width $s_0$ = 120 nm, a total per-side spring constant $k\equiv Nk_\text{GFC}$ of 0.35 N/m and two different optical drives at $\lambda_{\text{p}}$ = 1550 nm and respective powers $P_{\text{p}}$ of 0.5 mW and 1.25 mW. While the former leads to a stable equilibrium with a slot variation $\Delta s = 2u_z \approx$ 14 nm, the second represents the critical optical power, $P_\text{crit}$, above which a pull-in instability occurs and the two beams irreversibly collapse. The presence of optical propagation losses, $\alpha_\text{o}$, results in a non-uniform distributed load along the structure, which in turn induces rotation and modifies the displacement to $u_z(x) \approx u_{z,0} + \theta_\text{t}x$. The modeled force and torque equilibrium are described in detail in Supplementary Section S1.2. Note that the extended model still disregards vacuum and thermal Casimir forces~\cite{rodrigues_casimir_2016} which may considerably modify the steady-state for very small $s_0$, especially close to the pull-in instability~\cite{babar_self-assembled_2023}.

Under the simple assumptions leading to Eq.(~\ref{eq:force_equilibrium}), the phase shift of Eq.(~\ref{eq:phase_shift}) can be written as
\begin{small}
    \begin{equation}
    \Delta\varphi = \frac{2\pi L}{\lambda_\text{sig}}\left[n_\text{eff}(s_\text{eq},\lambda_\text{sig}) - n_\text{eff}(s_0,\lambda_\text{sig}) \right],
    \label{eq:phase_shift_2}
\end{equation}
\end{small}
where it is important to note that the equilibrium gap, $s_\text{eq} = s_0-2u_z$, depends implicitly on $s_0, k, L, P_\text{p}$, and $\lambda_\text{p}$. The achievable phase shift as a function of pump power $P_\text{p}$ and spring constant $k$ for signal and pump beams both at $\lambda_\text{sig}=\lambda_\text{p}$ = 1550 nm is shown in Fig.~\ref{fig:1}\textbf{e} for a device of length $L$ = 400 $\mu$m and initial slot $s_0$ = 120 nm. The powers $P_{n\pi}$ for which $\Delta\varphi = n\pi$, $n\in \mathbb{N}$, are shown as isocontours and show that for a moderately low spring constant, $k$, of 1 N/m, a power of $P_{\pi} \sim$ 300 $\mu$W suffices to achieve a $\pi$ phase shift, often the maximum phase shift required in applications. Even in the presence of propagation losses on par with those measured experimentally, the half-wave power consumption remains below one milliwatt across a vast parameter space (see Supplementary Section~S1.3).

The preceding analysis defines $P_\text{p}$ at the SWGSW input. In practice, coupling light into the slot mode requires adiabatic conversion from the rectangular bus waveguide via a V-groove coupler of length $L_1$. Figure~\ref{fig:1}\textbf{f} shows the simulated transmittance of this coupler for slot widths $s$ = 150 and 200~nm: both designs maintain losses well below 3~dB over a bandwidth of at least 550~nm. The $s$ = 200~nm coupler, which we use for all phase shifters in this work, incurs $\sim$0.2--0.3~dB per coupler across the C-band; narrowing to $s$ = 150~nm reduces this loss to below 0.1~dB. Extended simulations as a function of geometric parameters are provided in Supplementary Section~S1.7.

\begin{figure}[t!]
\includegraphics[width=\columnwidth]{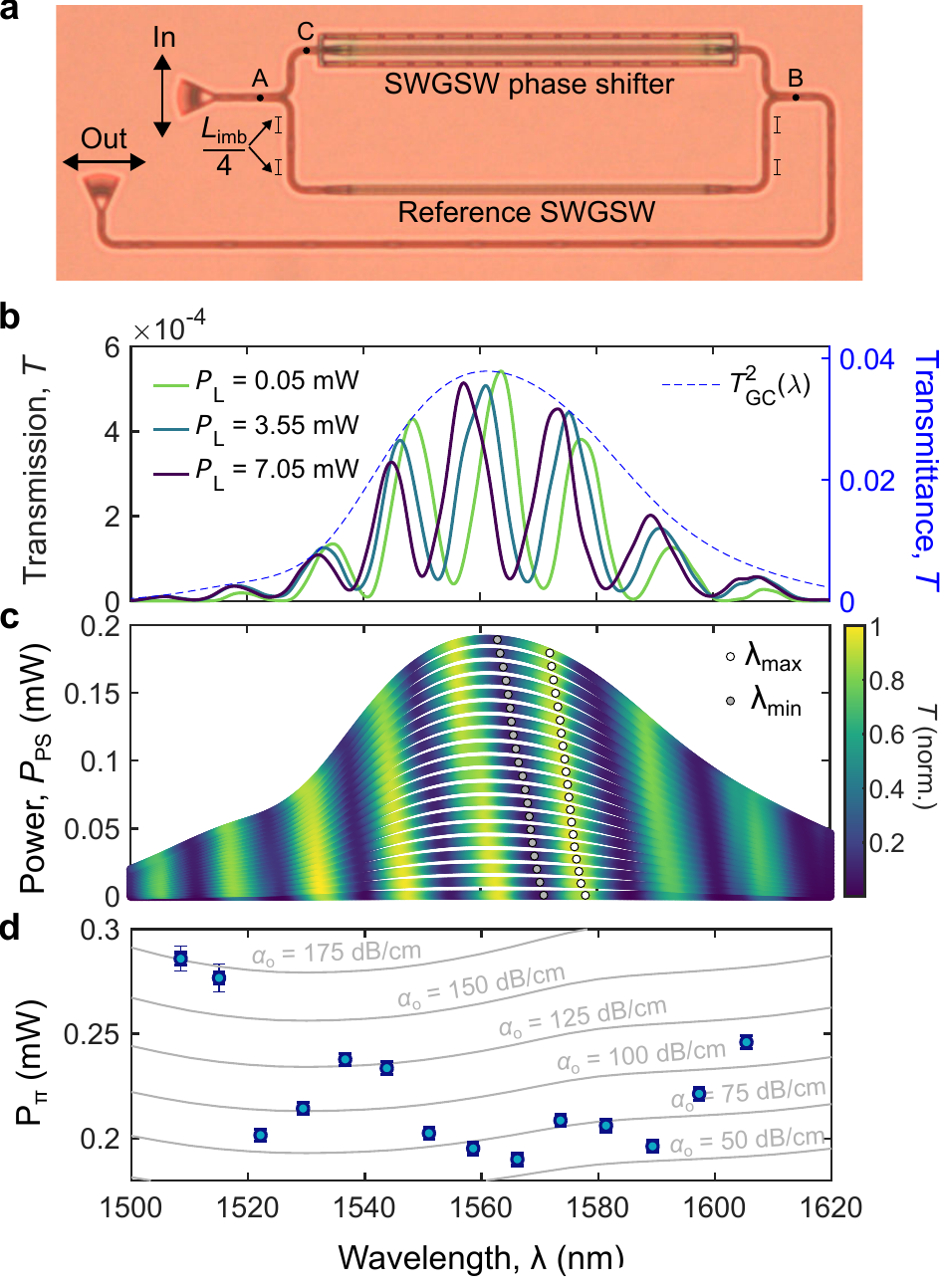}
    \caption{\textbf{Optical characterization of phase shifters using Unbalanced Mach-Zehnder interferometers (UMZIs).} \textbf{a}, Optical microscope image of a UMZI photonic circuit with an active arm (top, SWGSW phase shifter) and a reference arm (bottom, rigidly clamped SWGSW). Cross-polarized Grating Coupler (GC) ports are used as input--output ports. \textbf{b}, Raw end-to-end transmittance through a UMZI including a SWGSW phase shifter ($\{s_0,k,L\}$ = $\{$100 nm, 0.62 N/m, 400 $\mu$m$\}$) at three different laser powers $P_\text{L}$. The square of the GC transmittance, $T^2_\text{GC}$, is shown with a dashed blue line (right axis). \textbf{c}, Scatter colormap of the inferred transmission from point ``A" to ``B" (see panel \textbf{a}) as a function of wavelength and of the inferred optical power, $P_\text{PS}$, at point ``C", i.e., the power at the phase-shifter input. A single pair of transmission extrema are highlighted as $\{\lambda_\text{max},\lambda_\text{min}\}$. \textbf{d}, Inferred $P_\pi$ as a function of wavelength for the device in panels \textbf{b}--\textbf{c}, compared with theoretical predictions for different propagation losses, $\alpha_\text{o}$.}
    \label{fig:2}
\end{figure}

\section{Fabrication and optical characterization}

The spring-suspended SWGSWs are fabricated from silicon-on-insulator (SOI) wafers using electron-beam lithography, anisotropic dry etching, and vapour-phase selective etching (see Supplementary Section S2.1 for details). We characterize the phase-shifting capabilities of the spring-suspended SWGSWs using chip-scale Unbalanced Mach-Zehnder interferometers (UMZIs), shown in Fig.~\ref{fig:2}\textbf{a}, with Y-branch 3-dB power splitters to equally divide and recombine light across the two arms. The upper arm of the UMZI includes the SWGSW phase shifter under study, while the lower reference arm contains the same SWGSW except for the spring suspension system, which makes it so stiff that any optomechanical transduction can be neglected. In addition, the reference arm contains four short suspended ridge waveguides of length $L_\text{imb}/4$ = 10 $\mu$m. Assuming negligible losses in the rectangular waveguides~\cite{hansen_efficient_2023} and slot-width-independent propagation losses within the SWGSW sections, the transmittance of a signal beam across the UMZI (from A to B in Fig.~\ref{fig:2}\textbf{a}), $T_\text{MZI,sig}$, is given by
\begin{equation}
    \Scale[0.88]{T_\text{MZI,sig} = \frac{e^{-\alpha_\text{o}L}}{2}\left[1+\text{cos}(\frac{2\pi n_\text{eff,r}L_\text{imb}}{\lambda_\text{sig}}-\Delta\varphi(k/L,s_0,P_\text{p},\lambda_\text{p},\lambda_\text{sig}))\right],}
    \label{eq:Transmittance_MZI}
\end{equation}
where $n_\text{eff,r}$ is the effective refractive index of the rectangular waveguide and $\Delta\varphi$ is obtained using Eq.(~\ref{eq:phase_shift_2}) (see Supplementary Section~S1.4 for the derivation).

The transmission spectrum of our devices is measured by sweeping the wavelength of an external cavity diode laser (ECDL) and coupling light into and out of the chip using orthogonal free-space grating couplers~\cite{hansen_efficient_2023} addressed via a confocal optical setup~\cite{rosiek_observation_2023}. Figure~\ref{fig:2}\textbf{b} shows the transmission spectrum of a UMZI device including a SWGSW phase shifter ($\{s_0,k,L\}$ = $\{$100 nm, 0.62 N/m, 400 $\mu$m$\}$) for three different values of $P_\text{L}$. Note that these values of $k$ are given assuming Euler-Bernoulli beam theory, which agree within 5$\%$ with the values found by FEM simulations (see Supplementary Section~S1.5). We observe cosine-like fringes with a period of $\sim$ 14 nm, in good agreement with the free-spectral-range $\lambda_\text{sig}^2/(2n_\text{eff,r}L_\text{imb}) = 12.8$ nm found using the simulated $n_\text{eff,r} (\lambda_\text{sig}$ = 1550 nm) = 2.35. More importantly, the fringes exhibit a clear blue-shift as $P_\text{L}$ increases, which agrees with Eq.(~\ref{eq:Transmittance_MZI}) and the fact that $\Delta\varphi$ is always positive. The actual power at the SWGSW phase shifter is strongly wavelength-dependent due to the grating-coupler response, $T_\text{GC}$, the square of which \textemdash obtained from a dedicated experiment in Ref.~\cite{hansen_efficient_2023}\textemdash is shown in Fig.~\ref{fig:2}\textbf{b} and agrees very well with the envelope of the other three curves. Therefore, the observed blue-shift decreases in magnitude as the wavelengths depart from the central wavelength of the coupler ($\lambda \simeq$ 1560 nm). In order to infer the on-chip phase-shifting power requirements as a function of $\lambda_\text{p}$ and compare to theoretical predictions, we calibrate the optical power incident on the input grating coupler (see Supplementary Section S3.2) and use $T_\text{GC}(\lambda)$ to find the optical power $P_\text{PS}$ on the input facet of the coupler between rectangular and slot waveguide (point ``C" in Fig.~\ref{fig:2}\textbf{a}), assuming a perfect 3 dB Y-branch splitter. In addition, we normalize the transmittance data by $T^2_\text{GC}(\lambda)$ to accurately infer the transmission extrema of the UMZI transfer function. This is sufficient, as the other photonic components in the path have a broadband and approximately flat spectral response (see Supplementary Section S4 for a complete experimental characterization of the different components). Figure~\ref{fig:2}\textbf{c} depicts the extracted pump transmittance as a function of $\lambda_\text{p}$ and $P_\text{PS}$. A single pair of transmission minima and maxima are highlighted as $\{\lambda_\text{max},\lambda_\text{min}\}$. The linear evolution of the two extrema with circulating power $P_\text{PS}$ indicates a linear dependence of the phase shift $\Delta\varphi$ with $P_\text{PS}$, which we use to infer the power required to achieve a $\pi$-phase shift, $P_{\pi}$ (see Supplementary Section S3.3 for more details).

Figure~\ref{fig:2}\textbf{d} reports the inferred $P_{\pi}$ as a function of wavelength, along with the theoretical prediction for the SWGSW under test when subject to different levels of propagation losses, $\alpha_\text{o}$. The experimental values agree quantitatively with the model for losses in the range $\alpha_\text{o} \in$ [50--120] dB/cm, which are in turn only slightly larger than the experimentally measured propagation losses for that particular initial slot width (see Supplementary Section S4.1). We attribute the deviation to the 7$\%$ reduction in the experimental transmittance of the rectangular-to-slot waveguide converter relative to the simulated response (see Supplementary Section S4.3), which, along with other device-specific extrinsic scattering effects, might also lead to the observed wavelength dispersion.

\begin{figure}[t!]
\includegraphics[width=\columnwidth]{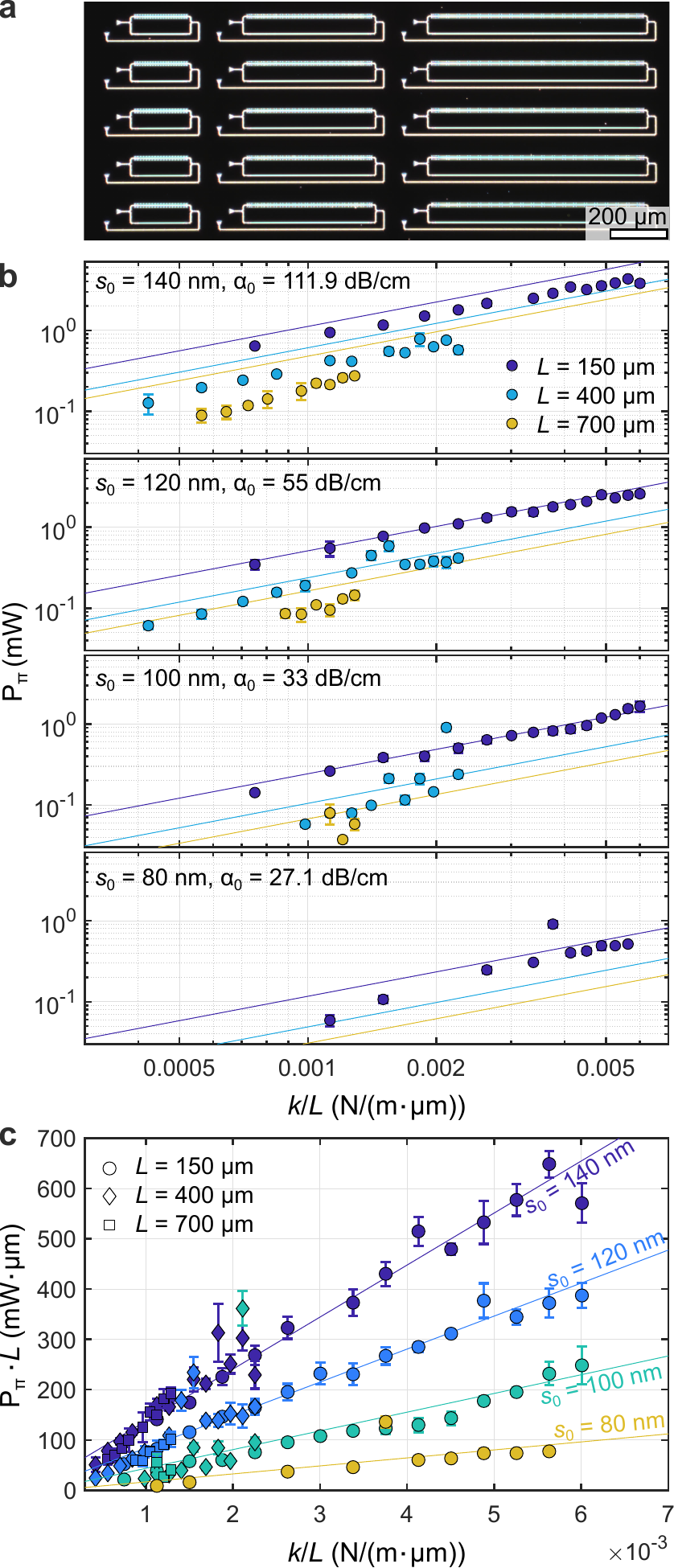}
    \caption{\textbf{Power requirements of optomechanical phase shifters.} \textbf{a}, Dark-field optical microscope image of several devices on a chip. \textbf{b}, Optical power required for achieving a $\pi$ phase shift, $P_\pi$, for various initial slot widths, $s_0$, lengths, $L$, and spring constants, $k$. Theoretical predictions are shown with solid lines, with the same color code as the experimental data. The propagation losses used in the model are indicated. \textbf{c}, Power-length product, $P_\pi\cdot L$, as a function of $k/L$. The solid lines are linear fits to the datasets for different $s_0$.}
    \label{fig:3}
\end{figure}

\section{Mapping of device performance}
To understand the performance limits of the SWGSW optomechanical phase shifter, we fabricate other UMZIs on the same chip. These contain SWGSWs of varying length, $L$ = 150, 400 and 700 $\mu$m, varying initial slot width, $s_0$ = 80, 100, 120 and 140 nm, and varying total spring constant $k \in [0.0563,0.9013]$ N/m. The primary fabrication constraint is in-plane structural collapse of the most compliant devices as they are released during underetching. The origin of these collapses is a combination of stress release of the silicon built-in compressive stress and the strong short-range surface forces, e.g., electrostatic, Casimir, or capillary forces, which are ubiquitous during and after suspension~\cite{babar_self-assembled_2023}. The collapse occurs if the initial gap, $s_0$, is smaller than a critical gap, $s^*_0$, which depends on $k$ and $L$. In contrast, for $s_0 > s^*_0$, the spring-suspended platforms experience a displacement, $\delta s$, and reach a stable equilibrium without collapsing. We first map the available phase space by evaluating which devices collapse (see Supplementary Section S2.3). The resulting phase diagram reveals that the collapse boundary is not a universal function of $k/L$, since shorter devices tend to be more stable than longer ones, even for the same $k/L$. This is most evident for $s_0 = 80$~nm, where only devices with $L = 150~\mu$m could be fabricated successfully at the lowest fabricated spring constants, while longer devices at the same $k/L$ collapsed. We attribute this length dependence to the stochastic nature of surface-force-driven pull-in: the probability of a locally unstable site triggering collapse increases with the total suspended length. Due to charging-induced unwanted collapses upon SEM inspection, the evaluation is done via their optical transmission, which unfortunately does not allow extraction of $\delta s$. Consequently, we consider all uncollapsed geometries as feasible geometries and assume $s_0$ to be that targeted and measured for the lower arm of the UMZI, where the contribution from stress release can be measured (see Supplementary Section S2.2 for a mapping between targeted and fabricated feature sizes).

Figure~\ref{fig:3}\textbf{a} shows a dark-field microscope image of a chip, including multiple fabricated UMZIs, which we address sequentially using the same confocal free-space optical setup and alignment conditions. We then process the wavelength and power-dependent transmittance of all devices, following the protocol described in Fig.~\ref{fig:2}. The inferred on-chip $P_{\pi}$ for all devices that did not collapse is summarized in Fig.~\ref{fig:3}\textbf{b}. The reported values for $P_{\pi}$ are average values over a 40 nm bandwidth around $\lambda$ = 1560 nm, and the error bars correspond to the standard deviation of the extracted values in that range. We also include the theoretical predictions using the propagation losses measured for SWGSWs of the same $s_0$ (see Supplementary Section S4.1), which are also wavelength-averaged over the same range. We note that the dynamic loss due to slot-width dependence is not considered, as all displacements involved are below 10 nm. Figure~\ref{fig:3}\textbf{b} demonstrates good agreement between the functional dependencies of the experimental data and the model, although the model slightly overestimates the required power levels, especially for $s_0$ = 140 nm, and for $L$ = 700 $\mu$m. We attribute the former disagreement to a possible overestimation of the propagation losses for $s_0$ = 140 nm by the independent measurements, and the latter to the expected reduction of $s_0$ by the aforementioned $\delta s$, an effect that is more prominent for longer devices. Interestingly, for a fixed value of $k/L$, the product $P_\pi\cdot L$ has no apparent length dependence and depends solely on the initial slot width, which is evident from Fig.~\ref{fig:3}\textbf{c}. This indicates that the displacements involved are small enough for a linearization to hold, with $k/L$ setting the equilibrium position and the total acquired phase scaling as $\Delta\varphi \simeq LP_p$. The initial slot width, $s_0$, thus emerges as the primary design knob for device efficiency: A smaller $s_0$ simultaneously reduces $P_\pi\cdot L$, through the stronger optical force, and reduces the insertion loss, since the measured propagation losses grow approximately quadratically with $s_0$ (see Supplementary Section~S4.1). This doubly-favorable scaling makes the smallest fabricable $s_0$ the natural design target, with the collapse boundary setting the practical limit.

Among all characterized geometries, the phase shifter with the best overall balance of $P_\pi$ and insertion loss is $\{s_0,k,L\}$ = $\{$80~nm, 0.169~N/m, 150~$\mu$m$\}$ (178.6~$\mu$m total active length including the input and output couplers), yielding $P_\pi$ = 59~$\pm$~9~$\mu$W and an estimated wavelength-averaged total insertion loss of $\sim$1.4~dB (1.0~dB from the two V-groove strip-to-slot mode converters ($\sim$0.5~dB each) and $\alpha_\text{o}L \approx 0.4$~dB from propagation; see Supplementary Sections~S4.1--S4.3). A direct cutback measurement on complete devices ($s_0$ = 100~nm, $L$ = 400~$\mu$m) yields a total insertion loss of $\sim$2~dB, validating this component-wise budget (see Supplementary Section~S4.4). For $s_0$ = 100~nm and $L = 700~\mu$m, $P_\pi$ reaches as low as 37.4~$\pm$~3.2~$\mu$W albeit at a rather prohibitive estimated insertion loss of $\sim$3.4~dB.

\section{Control of a signal beam}
\begin{figure*}[t!]
\includegraphics[width=0.9\textwidth]{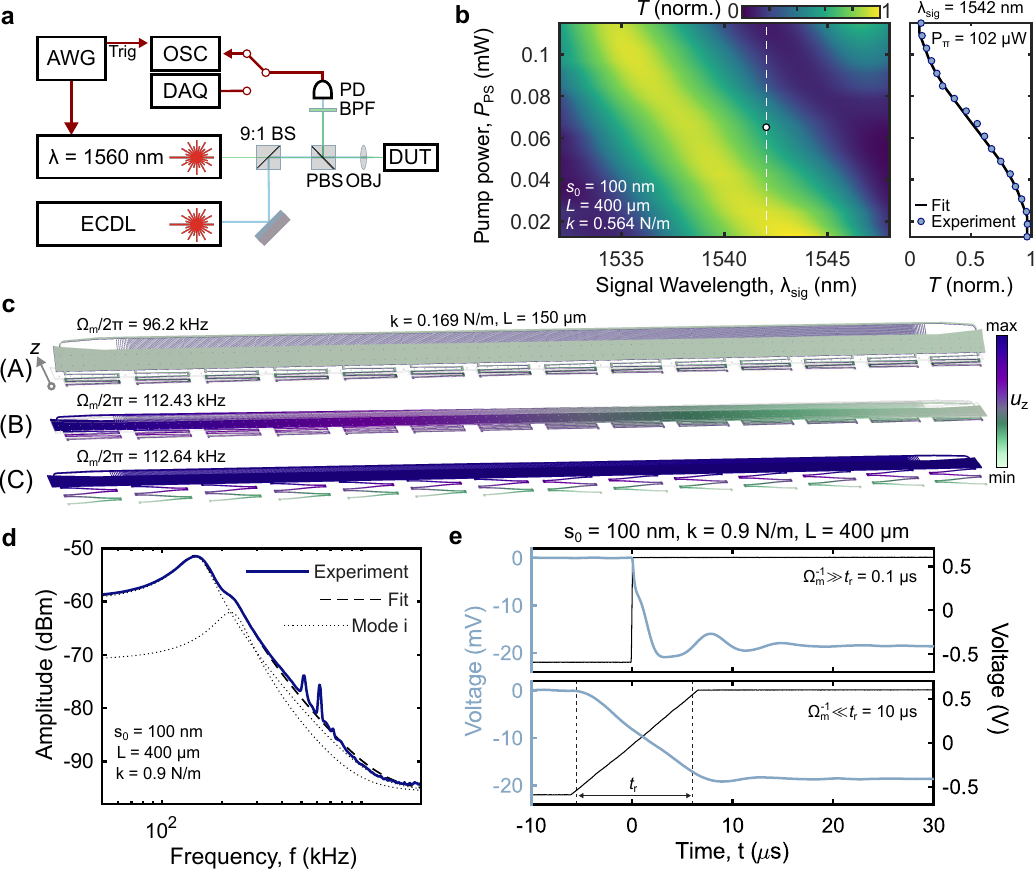}
    \caption{\textbf{Control of a signal beam by an optical pump.} \textbf{a}, Schematic of the optical setup employed for pump/probe measurements. \textbf{b}, Normalized transmittance of the probe beam across a UMZI including a SWGSW phase shifter ($\{s_0,k,L\}$ = $\{$100 nm, 0.564 N/m, 400 $\mu$m$\}$) as a function of probe wavelength and the on-chip pump power, $P_\text{PS}$. \textbf{c}, Simulated mechanical mode shapes (deformation) and in-plane displacement $u_z$ (colormap) for one suspended SWGSW platform ($\{s_0,k,L\}$ = $\{$100~nm, 0.169~N/m, 150~$\mu$m$\}$) obtained by FEM. \textbf{d}, Optomechanical response spectrum of a SWGSW acquired by sinusoidally modulating the pump power and recording the resulting probe intensity modulation. The dashed line is a fit to two Lorentzian lineshapes, whose individual contributions are represented with dotted lines.  \textbf{e}, Transient response of the probe signal under trapezoidal pump pulses with two ramp durations $t_r$. Top: $t_r \ll \Omega_\text{m}^{-1}$, Bottom: $t_r \gg \Omega_\text{m}^{-1}$. The solid black line represents the modulation signal sent to the pump laser and the solid blue line the photodetected probe signal.}
    \label{fig:4}
\end{figure*}

For simplicity, the results presented so far evaluate the phase shift undergone by the same laser light field used for optomechanical actuation. However, a more realistic scenario for applications of the proposed phase shifters would be controlling the phase of a signal beam with a pump beam at another wavelength within the broadband system response. In this section, we demonstrate that a (strong) pump beam coupled into the transverse-electric-like fundamental slot mode of the SWGSW can control the phase of a (weak) probe beam coupled into the same optical mode at another wavelength. We use the same ECDL as in the previous section ($\lambda_\text{p}$ = 1560 nm) for optomechanical actuation and an additional ECDL for the probe beam. Both beams are co-propagating and coupled into and out of the UMZI through the same two grating couplers. The out-coupled pump beam is filtered before detection using a bandpass filter (BPF) centered around the probe wavelength ($\lambda$ = 1540 nm, FWHM = 12 nm, 33.1 dB suppression), as shown in Fig.~\ref{fig:4}\textbf{a}. We estimate the circulating power immediately before the SWGSW phase shifter following the calibration procedure described earlier and acquire the transmission spectra of the probe beam for varying pump powers. The color map of Fig.~\ref{fig:4}\textbf{b} represents the normalized transmittance for a UMZI including a SWGSW ($\{s_0,k,L\}$ = $\{$100 nm, 0.564 N/m, 400 $\mu$m$\}$), after correcting for the response of the grating couplers, the BPF, and the detection of the residual pump transmittance. Similar to the observation in Fig.~\ref{fig:2}\textbf{b}, we observe a blueshift of the UMZI fringes as $P_\text{PS}$ varies, which evidences the phase shift undergone by the probe beam. We fit the response at each wavelength using Eq.(~\ref{eq:Transmittance_MZI}) (see the right panel of Fig.~\ref{fig:4}\textbf{b} for an example) and find $P_\pi$ in $\lambda \in [1530-1550]$ nm to be $P_\pi$ = 100 $\pm$ 8 $\mu$W, which is in good agreement with the value found directly for the pump in Fig.~\ref{fig:3}, $P_\pi$ = 96 $\pm$ 4 $\mu$W. A similar level of agreement is found for all devices characterized in the pump-probe configuration.

\section{Phase-shifting dynamics}
In addition to the static response of the SWGSW phase shifters, we employ the setup shown in Fig.~\ref{fig:4}\textbf{a} to determine their dynamics. We set the probe wavelength and pump power so that the probe beam is on the phase-quadrature of the UMZI response (i.e. the white dot in Fig.~\ref{fig:4}\textbf{b}) and modulate the pump laser power using an arbitrary waveform generator (AWG). The applied modulation amplitude is low enough to ensure the optical signal is proportional to the displacement, enabling optical detection of the mechanical motion. In photonic MEMS and NEMS devices, the operational dynamics are fundamentally governed by the mechanical resonance frequencies and the coupling of the mechanical modes to the optical field~\cite{errando-herranz_mems_2020, edinger_silicon_2021, grottke_optoelectromechanical_2021}. Reaching this mechanics-limited regime requires efficient charge delivery to the actuating electrodes. However, placing metal contacts near the optical mode may introduce parasitic absorption, while routing them further away incurs parasitic RC loads that can suppress the usable bandwidth below the mechanical resonance~\cite{errando-herranz_mems_2020}. By contrast, the all-optically actuated phase shifter demonstrated here eliminates electrical actuation altogether, so the temporal response is governed purely by the mechanical susceptibility of the two suspended platforms. The mode shape of three simulated mechanical modes is shown in Fig.~\ref{fig:4}\textbf{c}, for one side of a suspended SWGSW phase shifter ($\{s_0,k,L\}$ = $\{$100 nm, 0.169 N/m, 150 $\mu$m$\}$). The lowest-frequency mode (A) is a slot-width-preserving out-of-plane flexural mode, whereas mode (B) is the antisymmetric in-plane mode, whose equal opposite displacements of the two platforms cancel to a net zero change in slot width. As a result, both modes are optomechanically dark. The symmetric in-plane mode (C), which rigidly narrows the slot, is the only one that imprints a net phase shift on the output field; we refer to it as the fundamental mode. The eigenfrequencies of the three modes are shown in Fig.~\ref{fig:4}\textbf{c}, and the complete mechanical decoupling of the rigid SWGSW platform from its surroundings leads to quantitative agreement \textemdash within less than 5$\%$ error\textemdash with a mass-on-a-spring model based on the design value for $k$ and the system effective mass, $m_\text{eff}$, e.g., $\Omega_\text{m} \equiv \sqrt{k/m_\text{eff}}$ = 2$\pi\times$111.59 kHz for the case represented in Fig.~\ref{fig:4}\textbf{c}. The same level of agreement is found across all devices (Supplementary Section S1.5). Figure~\ref{fig:4}\textbf{d} shows the optomechanical response spectrum of a different device ($\{s_0,k,L\}$ = $\{$100 nm, 0.9 N/m, 400 $\mu$m$\}$), acquired by sinusoidally modulating the pump power and recording the resulting probe intensity modulation. The dominant resonance ($\Omega_\text{m}/2\pi \approx 152$~kHz, $Q_\text{m} \approx 3$) is identified as the fundamental symmetric in-plane mode. The weaker resonance at $\Omega_\text{m}/2\pi \approx 233$~kHz is attributed to the second symmetric in-plane mode (FEM prediction: $245.2$~kHz), whose higher-order displacement profile along the waveguide length partially cancels the integrated slot-width change, which reduces the effective optomechanical coupling and yields a proportionally smaller spectral amplitude. The two peaks therefore arise from structurally distinct mechanical modes rather than pump-induced splitting of the fundamental resonance~\cite{eichenfield_picogram-_2009}, as discussed in Supplementary Section~S1.6. The measured value of $\Omega_\text{m}$ falls $\sim$10$\%$ below the predicted value, consistent with an $\sim$18$\%$ reduction in spring constant attributed to the finite dimensional accuracy in the SEM-extracted cantilever widths and a reduced silicon Young's modulus ($E \approx$ 139 GPa), in agreement with previous measurements on GFCs of comparable dimensions~\cite{weis_electrostatic_2026}. The remaining spectral features, with anomalously narrow textwidths, are attributed to noise pickup in the detection electronics.

The achievable switching time of the phase shifter is probed by applying trapezoidal pump pulses with ramp durations, $t_r$, varying relative to $\Omega_\text{m}^{-1} \approx 1.0~\mu$s, as shown in Fig.~\ref{fig:4}\textbf{e}. For $t_r \ll \Omega_\text{m}^{-1}$, the abrupt change in the optical force coherently excites the mechanical resonance(s), and the probe signal shows damped oscillations. The resonance frequency and ringdown time, $\tau_m$, agree with the response in Fig.~\ref{fig:4}\textbf{d}, i.e., $\tau_\text{m} \approx 2Q_\text{m}/\Omega_\text{m} \approx 5.5~\mu$s (top panel). For $t_r \gg \Omega_\text{m}^{-1}$, the platforms follow the optical force adiabatically, settling without ringing (bottom panel). Because $Q_\text{m} \approx 3$ at ambient pressure, $\tau_\text{m}$ is only a few microseconds irrespective of the ramp shape: Fast and adiabatic switching are effectively equivalent in the present conditions, with the platform always settling within $\sim\!\tau_\text{m}$~\cite{papon_nanomechanical_2019}. In low-loss environments --vacuum or cryogenic operation, where $Q_\text{m} > 10^3$ is typical for structures of comparable dimensions~\cite{weis_electrostatic_2026}-- $\tau_\text{m}$ extends to milliseconds; in this case, adiabatic switching becomes essential to suppress ringing and achieve deterministic phase reconfiguration of the proposed optomechanical phase shifters.

\section{Conclusions and outlook}
We have demonstrated all-optical phase shifters in a silicon-on-insulator platform based on suspended subwavelength grating slot waveguides, in which gradient optical forces deform the slot width and impart a phase shift on a co-propagating signal beam. Characterization across a wide parameter space ($L$ = 150--700~$\mu$m, $s_0$ = 80--140~nm, $k$ = 0.0563--0.9013 N/m) confirms the quantitative agreement with a lumped force-equilibrium model that accounts for finite propagation losses. The best-performing device is 178.6~$\mu$m-long and achieves $\pi$-phase modulation with a moderate pump optical power of $P_\pi \approx 60~\mu$W, while incurring an insertion loss on the probe of $\sim$1.4~dB. The travelling-wave geometry and the wideband waveguide couplers we employ place no fundamental optical bandwidth restriction on the pump and probe wavelength, which in practice is limited only by the input--output grating couplers. In addition, the phase switching settling time is shown to be limited by the mechanical frequency of the fundamental in-plane resonance, $\Omega_\text{m}^{-1}$.

Three improvement axes emerge from the identified device scaling. The current insertion loss of $\sim$1.4~dB exceeds that of state-of-the-art electrostatic NEMS phase shifters, which have reported values $<$0.3~dB~\cite{edinger_silicon_2021,baghdadi_dual_2021}, primarily because of the strip-to-slot mode converters. We have demonstrated experimentally that the same coupler design with a slot width of $s_\text{t}$ = 150 nm reduces the coupler contribution to the insertion loss by $\sim$0.5~dB (see Supplementary Section~S4.3), and simulations indicate that for $s_\text{t}$ = 100 nm, it can go below 0.05~dB (see Supplementary Section~S1.7). Beyond the couplers, the propagation loss depends strongly on both the slot width, $s_0$, and the tether width, $d$: Measurements on suspended circuits with $d = 20$~nm and $s_0$ down to 40~nm show propagation losses approaching $\sim$10~dB/cm (Supplementary Section~S4.5), which would reduce the propagation contribution to below 0.2~dB for $L = 150~\mu$m. The combination of optimized couplers and reduced $s_0$ and $d$ projects a potential total insertion loss below 0.3~dB, but whether thin-tether, narrow-slot geometries can survive the suspension release step with adequate yield remains an open question. Since the stiffness, $k$, is set by the springs, reducing platform mass by using perforated or truss-like geometries can therefore decrease the switching time without incurring a $P_\pi$ penalty. Simple lumped-element modelling indicates that a truss platform with $\sim$15\% fill fraction is possible and can increase $\Omega_\text{m}$ by a factor of 2--3, reducing the minimum switching time accordingly.

Beyond these performance metrics, the all-optical actuation enables capabilities that are difficult or impossible to replicate with electrically driven devices. First, the complete removal of metal electrodes and electrical routing eliminates parasitic optical absorption, RC bandwidth constraints, and electromagnetic susceptibility. Second, because the pump is delivered over the same optical fiber as the signal, a single fiber carrying multiple pump wavelengths can independently address an array of phase shifters via on-chip wavelength demultiplexing, scaling the number of controllable degrees of freedom with no additional routing overhead. Third, the actuation principle is purely geometric and does not rely on any material-specific electro-optic or thermo-optic effects. Thus, the same design transfers to silicon nitride~\cite{fong_tunable_2011}, GaAs~\cite{qvotrup_integration_2025}, or diamond~\cite{khanaliloo_single-crystal_2015} platforms without re-optimizing the drive mechanism. Finally, the device is naturally suited to cryogenic quantum photonic circuits~\cite{qvotrup_integration_2025,beutel_cryo-compatible_2022}, where optical fibers already serve as the signal channel and metallic wires would conduct heat between temperature stages and substantially increase the heat load at the coldest stages. In particular, the proposed architecture is directly applicable to erbium-based C-band quantum networks in silicon~\cite{gritsch_optical_2025}, and, since its operating range and performance is limited by the grating-coupler transmission window rather than by the device itself, it naturally covers the wavelengths of other silicon luminescent centers~\cite{deabreu_waveguide-integrated_2023,saggio_cavity-enhanced_2024,hollenbach_wafer-scale_2022}. As phase control underlies directional couplers~\cite{fong_tunable_2011}, optical switches, and beam splitters in programmable photonic circuits, our work constitutes an important step toward all-optical computing based on gradient forces.

\section{Acknowledgments}
We gratefully acknowledge financial support from the European Research Council (Grant No.\ 101045396 -- SPOTLIGHT), the Danish National Research Foundation (Grant No.\ DNRF147 -- NanoPhoton), the Innovation Fund Denmark (Grant No.\ 4356-00007B -- EQUAL), the European Union's Horizon research and innovation programme (Grant No.\ 101098961 -- NEUROPIC), and the Villum Foundation (Grant No.\ 84095 -- HYPER). G.~A. acknowledges the European Union’s Horizon 2021 research and innovation programme under Marie Skłodowska-Curie Action (Grant no. 101067606 -- TOPEX). G.~A.\ acknowledges Amirali Arabmoheghi for valuable discussions.

\bibliographystyle{naturemag-etalnoitalics}
\makeatletter
\def\urlprefix{\@ifnextchar\url\@swallow@biburl{}}
\def\@swallow@biburl\url#1.{}  
\makeatother
\bibliography{OM_Phase_shifters}

\clearpage
\onecolumngrid  

\setcounter{figure}{0}
\setcounter{equation}{0}
\setcounter{table}{0}
\setcounter{section}{0}
\renewcommand{\thefigure}{S\arabic{figure}}
\renewcommand{\theequation}{S\arabic{equation}}
\renewcommand{\thetable}{S\arabic{table}}
\renewcommand{\figurename}{\textbf{Figure}}
\renewcommand{\bibname}{References}

\setcounter{topnumber}{2}
\setcounter{bottomnumber}{2}
\setcounter{totalnumber}{5}
\renewcommand{\topfraction}{.99}
\renewcommand{\bottomfraction}{.99}
\renewcommand{\textfraction}{0.01}
\renewcommand{\floatpagefraction}{0.99}

\begin{center}
{\Large\bfseries Supplementary Information}\\[1ex]
{\large Broadband silicon photonic phase shifters driven by gradient optical forces}
\end{center}
\bigskip
\suppressfloats[t]  

\section{S1. Modelling of subwavelength grating slot waveguide optomechanical phase shifters} \setlabel{S1}{sec:modelling}

\subsection{S1.1. Effective refractive index and optical force}
\setlabel{S1.1}{subsec:neff_and_opt}

\begin{figure*}[t]
    \includegraphics[width=0.8\textwidth]{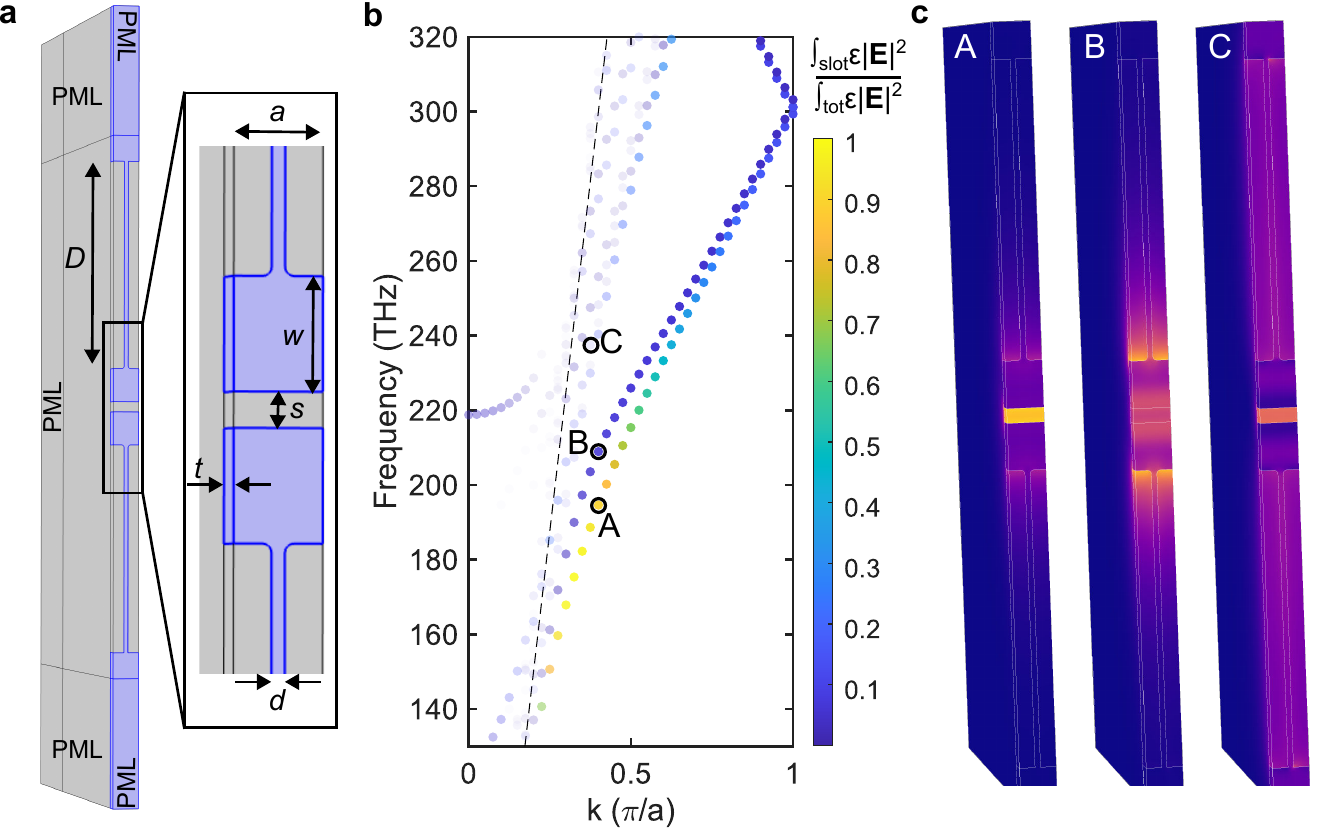}
    \caption{\textbf{Photonic band structure of the subwavelength grating slot waveguide (SWGSW).} \textbf{a}, Geometry of the SWGSW unit cell used for the finite-element analysis. \textbf{b}, Band diagram of the SWGSW for $a$ = 200 nm, $s$ = 80 nm, $w$ = 254 nm, $d$ = 30 nm, $D$ = 3 $\mu$m and $t$ = 220 nm. The modes whose electric field intensities $\left|\mathbf{E}\right|$ are shown in \textbf{c} are highlighted as A/B/C.}
    \label{fig:SWGSW_bands}
\end{figure*}

We calculate the effective refractive index, $n_{\text{eff}}$, of the fundamental slot-guided mode of the subwavelength grating (SWG) slot waveguide, which we abbreviate from now as SWGSW, using a commercial finite-element-method (FEM) eigensolver (COMSOL Multiphysics). A single unit cell is solved by imposing Floquet boundary conditions,
\begin{equation}
\label{eq:fbc}
    \mathbf{E}(\mathbf{r}+a\mathbf{\hat{e}}_x) = \mathbf{E}(\mathbf{r})e^{i\beta_\text{SWGSW}a},
\end{equation}
for all points $\mathbf{r} \in \Gamma_{B,1}$, with $\Gamma_{B,1}$ the domain boundary at the leftmost $yz$ plane (Fig.~\ref{fig:SWGSW_bands}\textbf{a}) and $\mathbf{\hat{e}}_x$ the vector connecting points in $\Gamma_{B,1}$ and $\Gamma_{B,2}$. The distance $a$ denotes the periodicity of the SWGSW. The eigenproblem is run for a fine grid of Bloch wavevectors $\beta_\text{SWGSW} \in [0, \pi/a]$, with which we obtain a band diagram as shown in Fig.~\ref{fig:SWGSW_bands}\textbf{b}. Figure~\ref{fig:SWGSW_bands}\textbf{b} displays only modes with an effective quality factor $Q_\text{eff} > 1$, which suppresses the large number of perfectly-matched-layer (PML) and radiation modes that inevitably arise in open-boundary simulations. The transparency of each dot encodes the fraction of electromagnetic energy confined to the two silicon beams forming the waveguide and the color the fraction of energy in the slot region, providing a visual indicator of the slot-like character of each band. The electric field profiles of the three highlighted modes are shown in Fig.~\ref{fig:SWGSW_bands}\textbf{c}: mode A is the fundamental slot-guided mode used for operation of the phase shifters; mode B is also waveguide-confined, corresponding to the antibonding supermode of the two parallel silicon beams; modes C and those at higher frequencies are also guided below the air light cone but their spatial field extends laterally over the SWG cladding regions rather than remaining tightly confined to the slot.

\begin{figure*}[t!]
\includegraphics[width=0.8\textwidth]{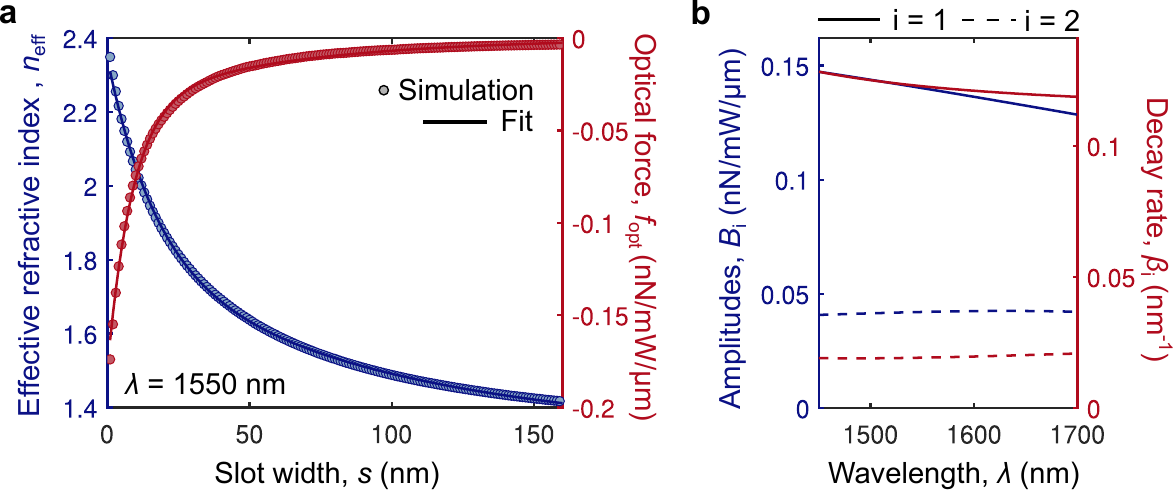}
    \caption{\textbf{Effective refractive index and optical forces in subwavelength-grating slot waveguides (SWGSW).} \textbf{a}, (Left y-axis) The numerically obtained effective index, $n_{\text{eff}}(s)$, and (right y-axis) optical force per unit power and length, $f_\text{opt}(s)$, for a wavelength $\lambda$ = 1550 nm are shown with solid dots. Fits to bi-exponential functions (Eqs.~\ref{eq:optical_force_dobule_exponential} and~\ref{eq:neff_double_exponential}) are represented with solid lines. \textbf{b}, Extracted amplitudes, $B_\text{i}$, and decay rates, $\beta_\text{i}$, from fitting the simulated $f_\text{opt}(s)$ with Eq.(~\ref{eq:optical_force_dobule_exponential}), as a function of wavelength.}
    \label{fig:EffectiveIndex_And_Force} 
\end{figure*}

As the FEM simulation is run for fixed wavevectors, we find the effective refractive index, $n_{\text{eff}}$, for the wavelength of interest by linear interpolation. We repeat this evaluation for varying slot width $s$, keeping the rest of the geometry unchanged. The extracted $n_{\text{eff}}(s)$ is shown with solid blue dots in Fig.~\ref{fig:EffectiveIndex_And_Force}\textbf{a} for a wavelength $\lambda$ = 1550 nm. Interestingly, $n_{\text{eff}}(s)$ is well fitted with a biased double exponential function,
\begin{equation}
\label{eq:neff_double_exponential}
   n_{\text{eff}}(s) \approx A_1e^{-\alpha_1s} + A_2e^{-\alpha_2s} + n_{\text{eff},\infty},
\end{equation}
as evidenced by the solid blue line in Fig.~\ref{fig:EffectiveIndex_And_Force}\textbf{a}. In the transverse direction, $z$ in Fig. 1\textbf{a} in the main text, the optical force exerted by the light field on the slot, $f_\text{opt}(s)$, is calculated numerically via
\begin{align}
    T_{ij} &= \varepsilon\varepsilon_0(E_i E_j - \frac{1}{2}\delta_{ij}\Vec{E}^2) + \frac{1}{\mu\mu_0}(B_i B_j - \frac{1}{2}\delta_{ij}\Vec{B}^2),\\
    &= 
    \begin{bmatrix}
        -\frac{1}{2}\varepsilon E_z^2 - \frac{1}{2\mu}B_y^2 & 0 & 0\\
        0 & -\frac{1}{2}\varepsilon E_z^2 + \frac{1}{2\mu}B_y^2 & 0\\
        0 & 0 & \frac{1}{2}\varepsilon E_z^2 - \frac{1}{2\mu}B_y^2
    \end{bmatrix},\\
    f_\text{opt} &= \oint_\mathcal{S}\overset{\leftrightarrow}{T}\cdot d\textbf{a} - \epsilon_0\mu_0 \frac{d}{dt}\int_\mathcal{V}\textbf{S}d\tau,\\
    &= \oint\left[\frac{1}{2}\varepsilon E_z^2 - \frac{1}{2\mu}B_y^2\right]\cdot d\mathbf{a},\qquad \mathbf{\hat{n}}=\mathbf{\hat{z}},
\end{align}
in the case of linearly polarized light propagating along $\mathbf{\hat{x}}$ with an E-field, $E_z$, pointing in the z-direction and B-field, $B_y$, pointing along the y-direction. The electromagnetic momentum term $\epsilon_0\mu_0\,d\int_\mathcal{V}\mathbf{S}\,d\tau/dt$ in the first line vanishes upon time-averaging for CW illumination, reducing $f_\text{opt}$ to the surface integral of the Maxwell stress tensor. This force is equally well described by a double exponential, which is unsurprising given that $f_\text{opt}(s)$ and $n_{\text{eff}}$ are related via
\begin{equation}
\label{eq:CMT}
    f_\text{opt}(s) = \frac{1}{c}\frac{\partial n_\text{eff}}{\partial s},
\end{equation}
for parallel dielectric waveguides~\cite{povinelli_evanescent-wave_2005} and our SWGSWs deviate only slightly from this geometry (see Fig.~1c in the main text). This is evidenced by the solid red dots and line in Fig.~\ref{fig:EffectiveIndex_And_Force}\textbf{a}. From now on, the optical force is written as
\begin{equation}
\label{eq:optical_force_dobule_exponential}
       f_\text{opt}(s) \approx B_1e^{-\beta_1s} + B_2e^{-\beta_2s}.
\end{equation}
As the devices we explore here are broadband, we depict in Fig.~\ref{fig:EffectiveIndex_And_Force}\textbf{b} the fitted values for $B_1$, $\beta_1$, $B_2$ and $\beta_2$, as a function of wavelength over the range $\lambda \in [1450, 1650]$. The parameters vary smoothly and over a limited range and so do $A_1$, $\alpha_1$, $A_2$ and $\alpha_2$.

\subsection{S1.2. Static displacement}
\setlabel{S1.2}{subsec:force_equilibrium}

In the main text, the steady-state mechanical displacement of the SWGSW is calculated from a lumped model that assumes a point particle subject to a non-linear optical force and a linear restoring force in the opposite direction. However, the presence of optical propagation losses (see Section~\ref{subsec:propagation_losses} for details on their experimental evaluation) leads to a power that decays along the propagation direction $x$ as
\begin{equation}
\label{eq:power_decay}
    P(x) = P_\text{p}e^{-\alpha_\text{o}x}.
\end{equation}
This leads to an inhomogeneous distributed load, $f_\text{o}(x)$, on each half of the SWGSW,
\begin{equation}
\label{eq:powerdcay}
    f_\text{o}(x) = P_\text{p}e^{-\alpha_\text{o}x}f_\text{opt}(s(x)),
\end{equation} 
which prevents the point particle approximation as the two suspended parts can rotate. 

In our structures, we choose to maximally distribute the restoring force and to use guided folded cantilever springs with 4 folded segments in series. For a target total per-side spring constant $k$ in a structure of length $L$, this leads to $N$ equal springs per side, with $N$ given by
\begin{equation}
    \label{eq:N_springs}
    \begin{aligned}
        N = \arg\min_{n} \left| L - nL_\text{spring} \right| \quad \mathrm{s.t.} \quad & nL_\text{spring} \leq L, \\
       & L_\text{spring} = \left[ \frac{Etw_s^3}{k}\frac{N}{4} \right]^{1/3} + 2\left[ W_\text{c}+d_\text{spring} \right],
    \end{aligned}
\end{equation}
where $L_\text{spring}$ is the length taken up by each individual spring, including the spacing between springs, $2d_\text{spring}$, $E$ is the Young's modulus, $w_s$ and $t$ are the width and thickness of the cantilever cross-section, and $W_\text{c}$ is the width of each truss connecting individual cantilevers. In essence, the value of $N$ is the largest integer rendering the inequality true in Eq.(~\ref{eq:N_springs}). For modelling purposes, we assume such discrete distribution to be equivalent to a continuous distributed spring of spring constant per unit length given by $k/L$.

Given the large width of the suspended platform attached to the SWGSW and the dense distribution of springs just mentioned, the two regions can be treated as rigid bodies in comparison to the spring constant of the cantilevers. The waveguide displacement to first order therefore takes the form,
\begin{equation}
    u_z(x) \equiv u_{z,0} - \theta_t x,
\end{equation}
where $\theta_t$ is the angle taken by the displaced structure in the small displacement approximation, which holds for all structures we explore here. Consequently, the slot width entering the force in Eq.(~\ref{eq:powerdcay}) reads
\begin{equation}
\label{eq:slot}
    s(x) \equiv s_0 -2u_{z,0} + 2\theta_t x,
\end{equation}
where, as in the main text, we assume symmetry between the two structures.

\begin{figure*}[t!]
\includegraphics[width=0.9\textwidth]{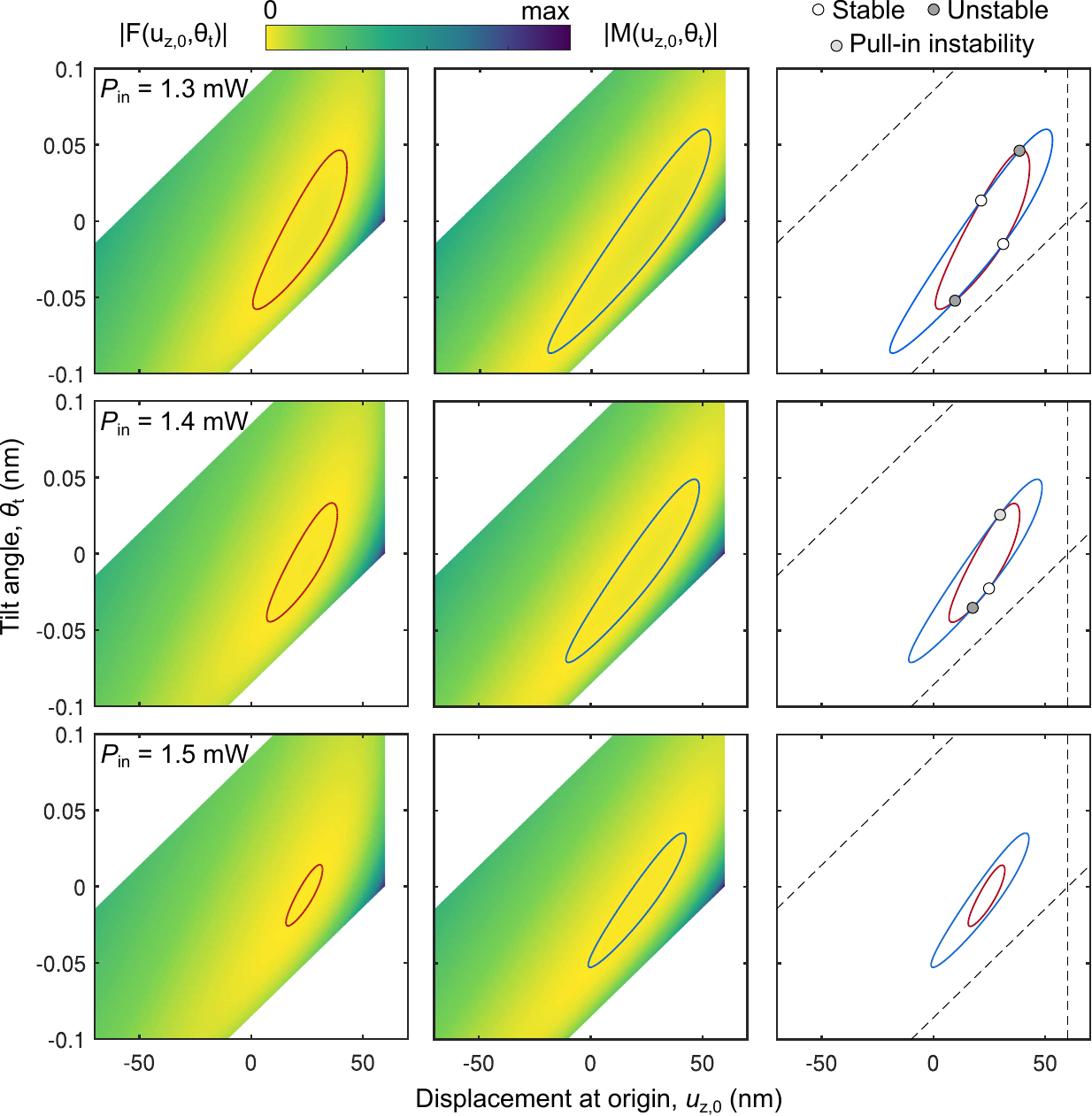}
    \caption{\textbf{Static displacement of a spring-suspended subwavelength grating slot waveguide (SWGSW).} The colormaps represent the sum of the forces, $F(u_{z,0},\theta_t)$, and of the torques, $M(u_{z,0},\theta_t)$, acting on each side of a SWGSW as a function of $u_{z,0}$ and $\theta_t$, with the displacement profile given by $u_z(x) \equiv u_{z,0} - \theta_t x$. These are represented for three different input powers, $P_\text{p}$. The closed red and blue lines represent the contours where the functions are equal to zero and their intersections indicate points of either stable equilibrium, unstable equilibrium or, for a specific power, the pull-in instability.}
    \label{fig:StaticDisplacement} 
\end{figure*}

With the two forces described above and under the ansatz of Eq.(~\ref{eq:slot}), the force and torque equilibrium read:
\begin{subequations}
\begin{equation}
    \int_0^L f_\text{o}(x)dx - \int_0^L \left(\frac{k}{L}\right)u_z(x)dx = 0,
\end{equation}
\begin{equation}
    \int_0^L xf_\text{o}(x)dx - \int_0^L x\left(\frac{k}{L}\right)u_z(x)dx = 0,
\end{equation}
\end{subequations}
where the springs are considered to be infinitely stiff in rotation. Using all the expressions given above, this leads to a system of two equations for the pair $\{u_{z,0},\theta_t\}$,

\begin{subequations}
\label{eq:force_and_torque}
    \begin{equation}
        \Scale[0.75]{
        F(u_{z,0},\theta_t) = \left\{\begin{aligned}
        &\sum_{i = 1,2} \left[\frac{B_iP_\text{p}e^{-\beta_i(s_0-2u_{z,0})}}{2\beta_i\theta_t+\alpha_\text{o}}\left(1-e^{-(2\beta_i\theta_t+\alpha_\text{o})L}\right)\right] - ku_{z,0} + \frac{k\theta_tL}{2}, & \quad 2\beta_i\theta_t \neq -\alpha_\text{o} \\
        &\sum_{i = 1,2} \left[B_iP_\text{p}e^{-\beta_i(s_0-2u_{z,0})}L\right] - ku_{z,0} + \frac{k\theta_tL}{2}, & \quad 2\beta_i\theta_t = -\alpha_\text{o}
        \end{aligned}\right\}
        = 0,}
    \end{equation}
    \begin{equation}
    \Scale[0.75]{
        M(u_{z,0},\theta_t) = \left\{\begin{aligned}
        &\sum_{i = 1,2} \left[\frac{B_iP_\text{p}e^{-\beta_i(s_0-2u_{z,0})}}{(2\beta_i\theta_t+\alpha_\text{o})^2}\left[1-e^{-(2\beta_i\theta_t+\alpha_\text{o})L}[(2\beta_i\theta_t+\alpha_\text{o})L+1]\right]\right] - \frac{ku_{z,0}L}{2} + \frac{k\theta_tL^2}{3}, & \quad 2\beta_i\theta_t \neq -\alpha_\text{o} \\
        &\sum_{i = 1,2} \left[B_iP_\text{p}e^{-\beta_i(s_0-2u_{z,0})}\frac{L^2}{2}\right] - \frac{ku_{z,0}L}{2} + \frac{k\theta_tL^2}{3}, & \quad 2\beta_i\theta_t = -\alpha_\text{o}
        \end{aligned}\right\}
        = 0.}
    \end{equation}
\end{subequations}
The colormaps in Fig.~\ref{fig:StaticDisplacement} represent the functions $F(u_{z,0},\theta_t)$ and $M(u_{z,0},\theta_t)$ for a structure with geometric parameters $\{s_0,k,L,\alpha_\text{o}\}$ = $\{$120 nm, 0.35 N/m, 700 $\mu$m, 20 dB/cm$\}$, and at three different input powers for a laser drive at $\lambda$ = 1550 nm. The regions depicted in white represent areas where the two structures are in contact, i.e., $s$ = 0 in Eq.(~\ref{eq:slot}), which go far beyond the validity of this model and, in any case, will lead to irreversible in-plane collapse. We represent also the contour lines for $F(u_{z,0},\theta_t) = 0$ and $M(u_{z,0},\theta_t)=0$. The two equilibrium conditions are satisfied simultaneously at the points where the contour lines intersect, which are represented in the rightmost column of Fig.~\ref{fig:StaticDisplacement}. For $P_\text{p}$ = 1.2 mW, we find two stable and two unstable equilibrium displacements. When $\alpha_\text{o} = 0$, the two equilibrium positions with $\theta_t > 0$ coincide with those shown in Fig.~1d of the main text. The one with a smaller displacement is stable, while the other is unstable, as shown via a stability analysis. The two equilibrium positions with negative $\theta_t$ correspond to deformations for which the reduced optical force due to optical losses is compensated by a narrower slot width. These equilibria are physically inaccessible when $P_\text{p}$ is ramped from zero; under that protocol the device always settles into the $\theta_t > 0$ branch.

\subsection{S1.3. Phase shift}
\setlabel{S1.3}{subsec:phase_shift}

\begin{figure*}[t!]
\includegraphics[width=0.9\textwidth]{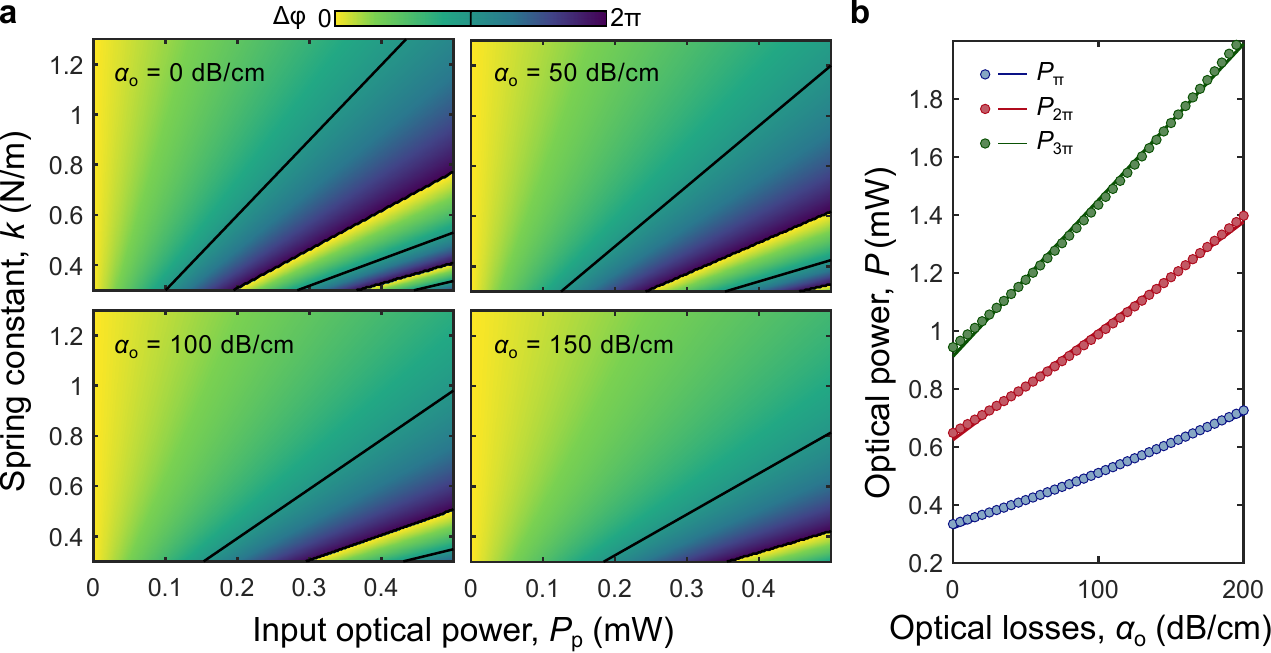}
    \caption{\textbf{Phase shift of a signal beam in an optically-pumped spring-suspended subwavelength grating slot waveguide (SWGSW).} \textbf{a}, Phase shift as a function of pump optical power and spring constant for a SWGSW phase shifter with $\{s_0,L\}$ = $\{$120 nm, 400 $\mu$m$\}$. Each colormap is evaluated for the indicated propagation losses and $\lambda_\text{p} = \lambda_\text{sig} = 1550$ nm. \textbf{b}, Power required for achieving $\Delta\varphi = n\pi$ phase shifts as a function of propagation losses. Solid lines are obtained from first-order polynomial fits.}
    \label{fig:phase_shift_maps} 
\end{figure*}

For a signal optical wavelength $\lambda_\text{sig}$, we evaluate the phase shift $\Delta\varphi$ as
\begin{equation}
 \label{eq:phase_shift_integral}
    \Delta\varphi = \frac{2\pi}{\lambda_\text{sig}}\int_0^L\left[n_\text{eff}(s_0 -2u_{z,0} + 2\theta_t x,\lambda_\text{sig}) - n_\text{eff}(s_0,\lambda_\text{sig}) \right]dx,
\end{equation}
where, for simplicity, we assume that the signal to be phase-shifted does not produce any displacement of the structure. This corresponds to the case where the optical power carried by the signal, $P_\text{sig}$, is much lower than that carried by the pump wave that induces displacement, $P_\text{p}$. Alternatively, the term $n_\text{eff}(s_0,\lambda_\text{sig})$ in Eq.(~\ref{eq:phase_shift_integral}) may be replaced by the appropriate \textit{initial} effective refractive index considering the displacement induced by the signal itself. With the expression for $n_{\text{eff}}(s)$ introduced in Eq.(~\ref{eq:neff_double_exponential}), the phase shift simplifies to

\begin{equation}
\label{eq:phase_shift_simplified}
    \Delta\varphi = \frac{2\pi}{\lambda_\text{sig}} \sum_{i = 1,2} \left[\frac{e^{2\alpha_\text{i,sig}u_{z,0}}}{2\alpha_\text{i,sig}\theta_t}\left[ 1-e^{-2\alpha_\text{i,sig}\theta_t L}\right] - L\right]A_\text{i,sig}e^{-\alpha_\text{i,sig}s_0},
\end{equation}
with $A_\text{i,sig}$ and $\alpha_\text{i,sig}$ the parameters fitted at the wavelength $\lambda_\text{sig}$ and $u_{z,0}$ and $\theta_t$ the steady state deformation induced by the pump beam (see Section~\ref{subsec:force_equilibrium}). The colormaps in Fig.~\ref{fig:phase_shift_maps}\textbf{a} show the phase shift achieved in a SWGSW with parameters $\{s_0,L\}$ = $\{$120 nm, 400 $\mu$m$\}$, varying spring constant, $k$, and varying the optical power of the pump, $P_\text{p}$. Four different values for the propagation losses of the pump beam are represented and, in all panels, we choose $\lambda_\text{p} = \lambda_\text{sig}$ = 1550 nm. The case for $\alpha_\text{o}$ = 0 dB/cm coincides with Fig. 1e in the main text, with the small difference that the panel here is based on the double exponential fit for the effective refractive index and the optical force (Eqs.~\ref{eq:neff_double_exponential} and~\ref{eq:optical_force_dobule_exponential}) instead of their numerical counterparts. Fig.~\ref{fig:phase_shift_maps}\textbf{b} depicts the variation in the powers required for achieving $\Delta\varphi = n\pi$, $P_{n\pi}$ for $n$ = $\{1, 2, 3\}$ as a function of the propagation losses for a fixed spring constant of $k$ = 1 N/m. We observe their dependence to be well approximated by a first-order polynomial, especially for $P_\pi$. This is why the different theoretical curves in Fig.~2\textbf{d} in the main text appear linearly offset relative to each other.

\subsection{S1.4. Unbalanced Mach-Zehnder interferometer with a phase-shifting arm}
\setlabel{S1.4}{subsec:MZI_probing}

\begin{figure*}[t!]
\includegraphics[width=0.8\textwidth]{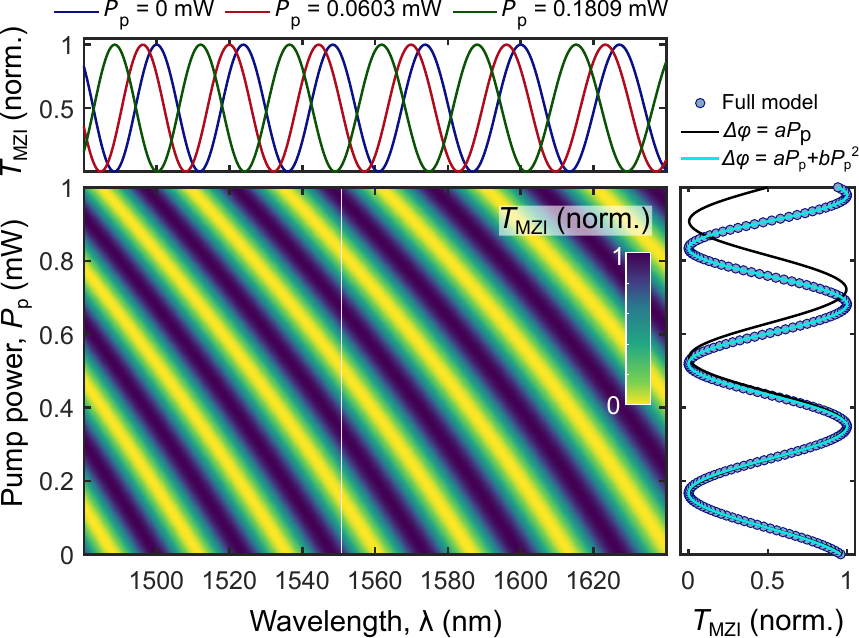}
    \caption{\textbf{Transfer function of an unbalanced Mach-Zehnder interferometer (UMZI) with an optically-pumped spring-suspended subwavelength grating slot waveguide (SWGSW) in one arm.} \textbf{a}, Calculated colormap of the normalized UMZI transmittance $T_\text{MZI,sig}$ as a function of signal wavelength $\lambda$ and incident pump power $P_\text{p}$, evaluated using Eq.(~\ref{eq:MZI_pumped}). The extrema of the transmittance blue shift with increasing pump power. The blue and green solid lines in the top panel indicate the pump powers required to achieve a $\pi$ phase shift. \textbf{b}, Vertical cut of the colormap at $\lambda = 1550$ nm showing the extracted phase shift $\Delta\varphi$ as a function of pump power. The data is fitted with both linear and quadratic functions for the phase shift, illustrating the onset of optomechanical non-linearity at high input powers.}
    \label{fig:MZI_transmittance} 
\end{figure*}

We employ chip-scale unbalanced Mach-Zehnder interferometers (UMZI) to characterize the phase-shifting capability of the spring-suspended SWGSWs. Y-branch 50/50 power splitters divide the input signal equally and recombine the two arms at the output. The top arm of the UMZI includes the SWGSW phase shifter under study, while the bottom arm contains the same SWGSW except that it is not suspended on springs, which makes it rigid, and four short suspended ridge waveguides of length $L_\text{imb}/4$ = 10~$\mu$m, suspended as part of the same release step as the rest of the circuit. We assume the power splitters to have negligible insertion losses --above 98.3\% transmittance over 200 nm wavelength range according to frequency-domain FEM simulations-- and to be perfectly balanced due to our high-precision nanofabrication~\cite{hansen_inverse_2024}. In the absence of deformation in the phase shifting arm, i.e., for infinitesimally small input power of the pump beam, and neglecting the propagation losses in the additional suspended rectangular waveguide sections -- shown to be below 7 dB/cm~\cite{hansen_inverse_2024} -- the total electric field of the signal beam at the output of the UMZI is given by
\begin{equation}
        E_\text{out,sig} = \frac{E_\text{in,sig}}{2}e^{-\frac{\alpha_\text{o}}{2}L}e^{-i\omega t}e^{i\beta_\text{SWGSW}L}(1+e^{i\beta_\text{r}L_\text{imb}}),
\end{equation}
where $\alpha_\text{o}/2$ is the field-amplitude attenuation coefficient, $\omega$ is the electromagnetic wave angular frequency, and $\beta_\text{r}$ is the propagation constant of the fundamental TE-like mode of the rectangular waveguide sections at $\omega$. Note that the pump beam is also transmitted, so the electric field here refers to the field at the signal wavelength. Therefore, the transmitted power at $\lambda_\text{sig}$ reads
\begin{equation}
        P_\text{out,sig} = \frac{P_\text{in,sig}}{2}e^{-\alpha_\text{o}L}\left[1+\text{cos}(\beta_\text{r}L_\text{imb})\right],
\end{equation}
which is wavelength-dependent via the propagation constant $\beta_\text{r}$. For larger pump input power, the transmittance is given by
\begin{equation}
    T_\text{MZI,sig} = \frac{e^{-\alpha_\text{o}L}}{2}\left[1+\text{cos}(\beta_\text{r}L_\text{imb}-\Delta\varphi(L,k,s_0,\alpha_\text{o},P_\text{p},\lambda_\text{p},\lambda_\text{sig}))\right],
    \label{eq:MZI_pumped}
\end{equation}
where $\Delta\varphi$ can be obtained using Eq.(~\ref{eq:phase_shift_simplified}) after solving for the force equilibrium. We note that we have disregarded the dependence of $\alpha_\text{o}$ on the slot width in Eq.(~\ref{eq:MZI_pumped}). This is justified because the slot width changes during normal operation are small relative to the equilibrium value, so the variation in $\alpha_\text{o}$ over the operating range is negligible compared to the other terms in the transmittance. Correctly accounting for this dependence would, however, be necessary to predict the pull-in instability accurately, since loss-induced asymmetry in the force profile becomes significant near collapse. The colormap in Fig.~\ref{fig:MZI_transmittance} represents Eq.(~\ref{eq:MZI_pumped}), re-normalized to the maximum transmittance, as a function of the pump power incident on the phase-shifting arm. We observe that the wavelengths at which $T_\text{MZI,sig}$ attains its extrema blue shift as we increase the pump power. Above a particular pump power, the non-linearity of the optical force with slot width (Eq.(~\ref{eq:optical_force_dobule_exponential})), leads to a non-linear dependence of the phase shift with power, $\Delta\varphi(P_\text{p})$ = $aP_\text{p}$ + $bP_\text{p}^2$ + $\mathcal{O}(P^3)$. This is shown clearly in the right panel of Fig.~\ref{fig:MZI_transmittance}, which represents a vertical cut of the colormap at $\lambda$ = 1550 nm with corresponding fits assuming either a linear or a quadratic relation for $\Delta\varphi(P_\text{p})$. We note that such non-linearity is, for most devices explored in this work, only relevant for input powers much larger than those required to achieve a $\pi$ phase shift (see blue and green solid lines in the top panel of Fig.~\ref{fig:MZI_transmittance}).

Due to the wavelength dependence of a number of quantities, e.g., $\beta_{\text{r}}$ or $\alpha_\text{o}$, in Eq.(~\ref{eq:MZI_pumped}), the ideal procedure to extract the phase shift as a function of $P_\text{p}$ is to fit the UMZI transmittance at each signal wavelength $\lambda_\text{sig}$ as is done in Fig.~\ref{fig:MZI_transmittance}. However, for limited changes of $\Delta\varphi$ due to practical pump power limitations, an alternative extraction procedure based on tracking the wavelength shift of the transmission extrema is preferred. The UMZI extrema occur at wavelengths satisfying $\beta_\text{r}(\lambda_\text{sig})L_\text{imb} - \Delta\varphi = n\pi$ for integer $n$. When this condition is met, both at $\lambda_n^{(0)}$ in the absence of pump and at $\lambda_n(P_\text{p})$ under pump power $P_\text{p}$, then substituting $\beta_\text{r} = 2\pi n_\text{eff,r}/\lambda$, yields
\begin{equation}
    \Delta\varphi(P_\text{p}) = 2\pi n_\text{eff,r} L_\text{imb} \left(\frac{1}{\lambda_n(P_\text{p})} - \frac{1}{\lambda_n^{(0)}}\right) \approx -\frac{2\pi n_\text{eff,r} L_\text{imb}}{\bigl(\lambda_n^{(0)}\bigr)^2}\,\Delta\lambda_n,
    \label{eq:phase_from_shift}
\end{equation}
where $\Delta\lambda_n = \lambda_n(P_\text{p}) - \lambda_n^{(0)}$ and the approximation holds for $|\Delta\lambda_n| \ll \lambda_n^{(0)}$. The induced phase shift is therefore linearly proportional to the fringe displacement. Since consecutive extrema $n$ and $n+1$ differ by $\pi$ in accumulated phase, a $\pi$ phase shift converts the $n$-th maximum into the adjacent $(n+1)$-th minimum, so that $P_\pi$ satisfies
\begin{equation}
    \lambda_n^{\text{max}}(P_\pi) = \lambda_{n+1}^{\text{min},(0)}.
    \label{eq:Ppi}
\end{equation}
The extraction of $P_\pi$ from experimental data is described in Section~\ref{subsec:phase_shift_extraction}.

\subsection{S1.5. FEM validation of spring constants and mechanical resonance frequencies}
\setlabel{S1.5}{subsec:freq_comparison}

\begin{figure}[t!]
    \includegraphics[width=0.9\columnwidth]{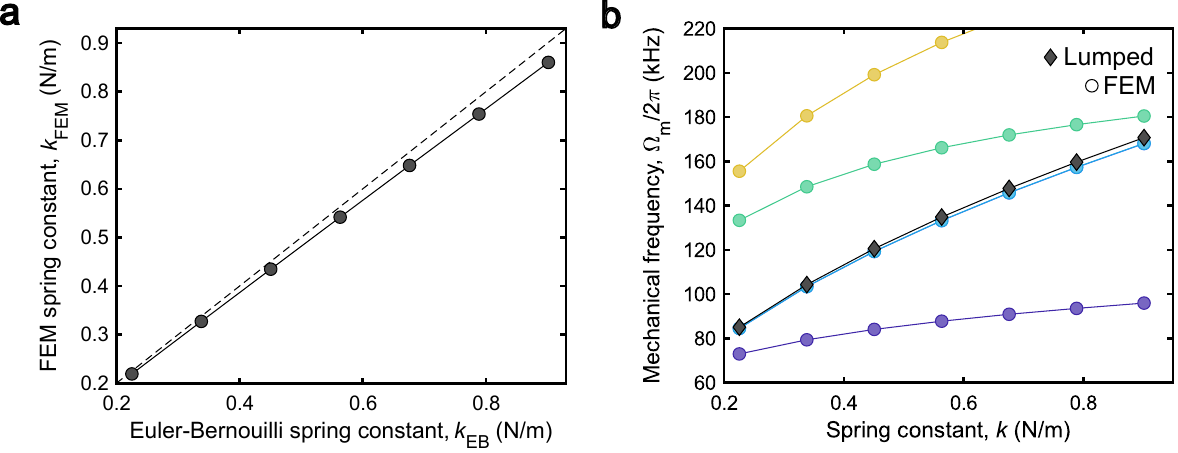}
    \caption{\textbf{Comparison of mechanical response obtained from FEM simulation and lumped-element model for the SWGSW phase shifter.} \textbf{a}, FEM-extracted spring constant, $k_\text{FEM}$, versus the Euler--Bernoulli prediction $k_\text{EB}$ for arrays of $N$ parallel guided folded cantilever springs. The dashed line is the identity. \textbf{b}, Five lowest FEM mechanical eigenfrequencies (circles, connected by lines as guides to the eye) as a function of $k_\text{EB}$. Black diamonds show the lumped-element prediction $\Omega_\text{m} = \sqrt{k/m_\text{eff}}$. Both panels are given for a suspended platform with $L = 400~\mu$m.}
    \label{fig:FEM_modes}
\end{figure}

We validate the lumped-element predictions for the spring constant and mechanical resonance frequencies against FEM simulations for suspended platforms with $L = 400~\mu$m and $s_0 = 100$~nm, varying the guided folded cantilever (GFC) arm length, $L_s$, to span spring constants from approximately 0.2 to 0.9~N/m. Figure~\ref{fig:FEM_modes}\textbf{a} compares the spring constant extracted from a FEM static analysis --- in which a uniform lateral load is applied to the slot sidewall and the resulting displacement infers $k_\text{FEM}$ --- against the Euler--Bernoulli prediction for $N$ parallel GFC springs,
\begin{equation}
    k_\text{EB} = N k_\text{GFC}, \qquad k_\text{GFC} = \frac{1}{N_\text{b}}\frac{12 EI}{L_s^3},
    \label{eq:kEB_SI}
\end{equation}
where $N_\text{b} = 4$ is the number of cantilever beams per GFC, $I = tw_s^3/12$ is the second moment of area of the GFC arm cross-section (width $w_s = 0.1~\mu$m, height $t = 0.22~\mu$m), and $E = 169$~GPa is the Young's modulus of silicon in the $\langle 110 \rangle$ direction. The two estimates agree to within $\sim5$\% across the full spring-constant range, confirming that the GFC array dominates the restoring force and that coupling between individual springs is negligible. We note that for the stiffer springs, the agreement degrades because the stiffness of the connecting trusses starts becoming relevant, as already observed in our previous work~\cite{weis_electrostatic_2026}.

To calculate the mechanical resonance frequency of the fundamental in-plane mode of each suspended platform, we calculate the effective mass as follows:
\begin{equation}
    m_\text{eff} = m_\text{eff,platform} + m_\text{eff,waveguide} + m_\text{eff,springs},
    \label{eq:meff_SI}
\end{equation}
where $m_\text{eff,platform} = \rho t A_\text{platform}$ is the effective mass of the rigid silicon platform (including the trapezoidal end sections and subtracting the periodic square perforations of side $L_h$ on a grid of pitch $\Delta x \times \Delta y$); $m_\text{eff,waveguide} = \rho t (w L_\text{platform} + d L_t N_t)$ accounts for the effective mass of the waveguide beams (width $w$, length $L_\text{platform}$) and the $N_t = L/a$ periodic tethers (width $d$, length $L_t$, pitch $a$); and
\begin{equation}
    m_\text{eff,springs} = \rho h N \left[ N_\text{b}\,\frac{13}{35}\,w_s L_s + (N_\text{b}-2)W_\text{c}L_\text{c} + 2W_a L_a \right],
\end{equation}
is the spring contribution, where the factor $13/35$ is the Rayleigh effective-mass fraction for a uniform cantilever vibrating in its first flexural mode~\cite{tang_laterally_1989}, and the remaining terms account for the folded-beam connectors (width $W_\text{c}$, length $L_\text{c}$) and anchors (width $W_a$, length $L_a$). Figure~\ref{fig:FEM_modes}\textbf{b} shows the five lowest FEM eigenfrequencies as a function of the (Euler-Bernoulli) spring constant together with the lumped-element prediction $\Omega_\text{m} = \sqrt{k/m_\text{eff}}$. The lumped model tracks the fundamental symmetric in-plane mode --- the mode that narrows the slot and imprints a net phase shift, identified in Fig.~4\textbf{c} of the main text --- quantitatively across the full range, with deviations below $\sim5$\%, which we attribute to the deviation of $k$ shown in Fig.~\ref{fig:FEM_modes}\textbf{a}. The remaining FEM branches correspond to higher-order structural modes (out-of-plane, rocking, and torsional) that do not contribute substantially to the optomechanical phase-shifting operation.

\subsection{S1.6. Mechanical normal modes and optical spring effect}
\setlabel{S1.6}{subsec:normal_modes}

The SWGSW phase shifter consists of two suspended silicon platforms (the left and right beams of the slot), each with total in-plane spring constant, $k$, and effective mass, $m_\text{eff}$, giving a bare mechanical resonance frequency, $f_\text{m} = (1/2\pi)\sqrt{k/m_\text{eff}}$, set by the guided folded cantilever springs (see previous section). At pump power, $P_\text{p}$, the equilibrium position of each platform is described by the displacement profile, $u_z(x) = u_{z,0} - \theta_t x$, established in Section~\ref{subsec:force_equilibrium}, where $u_{z,0}$ is the rigid-body component and $\theta_t$ is the tilt angle arising from the non-uniform optical load due to propagation losses. The equilibrium slot width at position, $x$, is accordingly (Eq.(~\ref{eq:slot})):
\begin{equation}
    s_\text{eq}(x) = s_0 - 2u_{z,0} + 2\theta_t x,
    \label{eq:seq_normal_modes}
\end{equation}
where $\{u_{z,0}, \theta_t\}$ solve the force and torque equilibrium of Section~\ref{subsec:force_equilibrium}. We denote by $\delta u_1$ and $\delta u_2$ the small time-dependent perturbations of each platform about this equilibrium, and treat the device as a lumped system with $s_\text{eq}$ evaluated at the midpoint $x = L/2$. The two platforms are not mechanically connected across the slot; their only coupling is mediated by the optical gradient force, which depends on $s_\text{eq}$ and is therefore sensitive to the position of both platforms. Linearizing the equations of motion around the static equilibrium, the perturbations $\delta u_1$ and $\delta u_2$ obey:
\begin{align}
    m_\text{eff}\,\delta\ddot{u}_1 &= (k_\text{opt} - k)\,\delta u_1 + k_\text{opt}\,\delta u_2, \label{eq:eom1_SI} \\
    m_\text{eff}\,\delta\ddot{u}_2 &= k_\text{opt}\,\delta u_1 + (k_\text{opt} - k)\,\delta u_2, \label{eq:eom2_SI}
\end{align}
where
\begin{equation}
    k_\text{opt} \equiv -\frac{\partial f_\text{opt}}{\partial s}\bigg|_{s_\text{eq}} L P_\text{p} \geq 0,
    \label{eq:kopt_SI}
\end{equation}
is the optical spring stiffness per platform. Its non-negativity follows from $\partial f_\text{opt}/\partial s \leq 0$ (the attractive force increases as the slot narrows). Equations~(\ref{eq:eom1_SI})--(\ref{eq:eom2_SI}) decouple in the symmetric and antisymmetric coordinates $q_+ = \delta u_1 + \delta u_2$ and $q_- = \delta u_1 - \delta u_2$:
\begin{align}
    m_\text{eff}\,\ddot{q}_+ &= (2k_\text{opt} - k)\,q_+, \label{eq:sym_SI} \\
    m_\text{eff}\,\ddot{q}_- &= -k\,q_-. \label{eq:anti_SI}
\end{align}
leading to the two normal-mode frequencies,
\begin{align}
    f_+ &= f_\text{m}\sqrt{1 - \frac{2k_\text{opt}}{k}}, \label{eq:fsym_SI} \\
    f_- &= f_\text{m}. \label{eq:fanti_SI}
\end{align}
The symmetric mode ($f_+$) is softened by the optical field and appears below $f_\text{m}$, while the antisymmetric mode ($f_-$) sits exactly at the bare mechanical frequency. Note also that Eq.~(\ref{eq:fsym_SI}) implies $f_+ \to 0$ as $2k_\text{opt} \to k$, which coincides with the pull-in instability condition: the symmetric mode goes soft as the device approaches collapse.

The optomechanical coupling of each mode to the optical field is determined by how it modifies the slot width. For the symmetric mode ($q_+$), both platforms move in phase: $\Delta s = -q_+$, giving a non-zero variation of the effective refractive index $\partial n_\text{eff}/\partial q_+ \neq 0$. This mode is therefore \textit{bright}: it modulates the optical phase and appears with a large signal in the transduced mechanical spectrum. For the antisymmetric mode ($q_-$), the two platform displacements cancel: $\Delta s = 0$ to first order, so $\partial n_\text{eff}/\partial q_- = 0$. This mode is \textit{dark}: it produces no optomechanical transduction signal in the perfectly symmetric case. Therefore, in the limit of perfect symmetry ($k_1 = k_2 = \bar{k}$), the antisymmetric mode is completely invisible in the optomechanical response. This ideal picture is modified by the fabrication imperfections unavoidably present in real devices. If fabrication disorder leads to slightly unequal spring constants $k_1 = \bar{k} + \Delta k/2$ and $k_2 = \bar{k} - \Delta k/2$, the symmetric and antisymmetric coordinates are no longer exact eigenmodes. Solving the coupled eigenvalue problem then yields
\begin{equation}
    f_\pm^2 = \frac{1}{(2\pi)^2}\left[\frac{\bar{k}-k_\text{opt}}{m_\text{eff}} \mp \frac{1}{m_\text{eff}}\sqrt{k_\text{opt}^2 + \left(\frac{\Delta k}{2}\right)^2}\right].
    \label{eq:disordered_freqs_SI}
\end{equation}
Several consequences follow directly from this expression. From the first term, the mid-point of the two mode frequencies, $\sqrt{(\bar{k}-k_\text{opt})/m_\text{eff}}/(2\pi)$, lies below $f_\text{m}$ regardless of disorder and decreases with pump power. From the second term, since $\sqrt{k_\text{opt}^2 + (\Delta k/2)^2} > k_\text{opt}$ for any $\Delta k \neq 0$, the upper mode $f_-$ always exceeds the bare frequency $f_\text{m}$: disorder raises the dark mode $f_-$ \textit{above} the bare mechanical frequency while pushing the bright mode $f_+$ further below its symmetric-case value. In the limit $\Delta k \to 0$ Eq.~(\ref{eq:disordered_freqs_SI}) reduces to Eqs.~(\ref{eq:fsym_SI})--(\ref{eq:fanti_SI}). Importantly, disorder also mixes the two eigenvectors: the true eigenstates are rotated from the symmetric/antisymmetric basis by a mixing angle $\theta$ satisfying
\begin{equation}
    \tan(2\theta) = \frac{\Delta k}{2k_\text{opt}},
    \label{eq:mixing_angle_SI}
\end{equation}
so the formerly dark mode acquires a finite transduction amplitude proportional to $\sin\theta$, making it weakly visible in the optomechanical spectrum with a signal amplitude reduced by a factor $\tan\theta = \Delta k / (2k_\text{opt} + \sqrt{4k_\text{opt}^2+\Delta k^2})$ relative to the bright mode. The dark mode visibility therefore increases at lower pump powers (as $k_\text{opt} \to 0$) and decreases at higher pump powers (as $k_\text{opt}$ grows). In addition, in a traveling-wave optomechanical system such as the SWGSW (no optical resonance cavity), the gradient force is conservative and carries no retardation. Consequently, there is no optical damping or amplification, and the quality factor of both modes is governed solely by intrinsic mechanical dissipation and equals $Q_\text{m,i}$ for each mode independently of $P_\text{p}$.

\begin{figure*}[t!]
    \includegraphics[width=\textwidth]{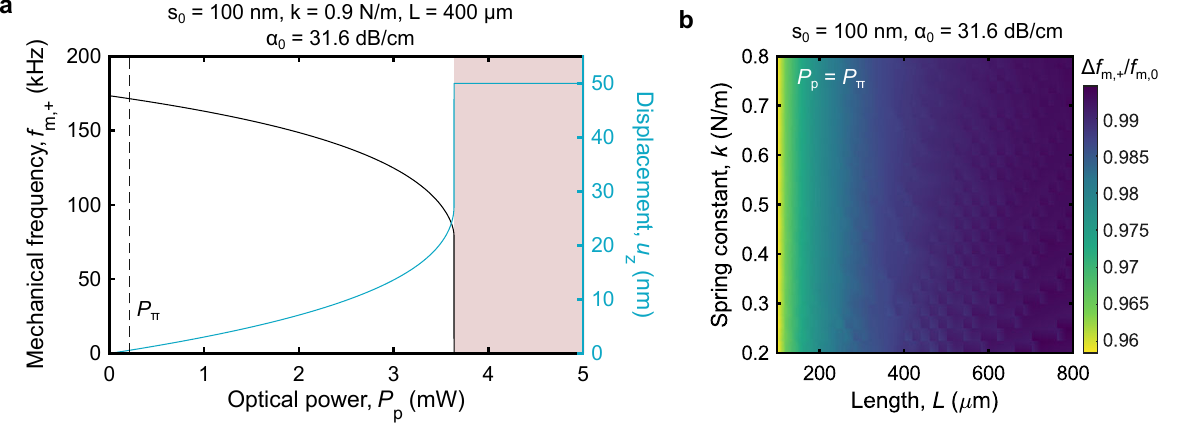}
    \caption{\textbf{Predicted spring frequency shift of the symmetric mode due to optical forces.} \textbf{a}, Symmetric mode frequency $f_+$ as a function of pump power for a device with $\{s_0, k, L\} = \{100~\text{nm},\, 0.9~\text{N/m},\, 400~\mu\text{m}\}$. The vertical dashed line marks $P_p = P_\pi$ and the shaded area marks the region where the device would have collapsed. \textbf{b}, Relative frequency shift $\Delta f / f_\text{m}$ of the symmetric mode evaluated at $P_\text{p} = P_\pi$, shown as a function of $k$ and $s_0$.}
    \label{fig:opticalSpring}
\end{figure*}

Figure~\ref{fig:opticalSpring}\textbf{a} shows the predicted symmetric mode frequency $f_+$, normalized to the bare frequency $f_\text{m}$, as a function of pump power for a representative device ($\{s_0, k, L\} = \{100~\text{nm},\, 0.9~\text{N/m},\, 400~\mu\text{m}\}$), evaluated self-consistently by solving the force equilibrium of Section~\ref{subsec:force_equilibrium} at each $P_\text{p}$ and computing $k_\text{opt}$ from the analytical derivative of the bi-exponential fit of Section~\ref{subsec:neff_and_opt}. The softening is monotonic and the vertical dashed line marks the half-wave power $P_\pi$ for that device. In the linearized regime (small displacements), the relative shift at $P_\text{p} = P_\pi$ reduces to:
\begin{equation}
    \frac{\Delta f}{f_\text{m}} \equiv 1 - \frac{f_+(P_\pi)}{f_\text{m}} \approx \frac{k_\text{opt}(P_\pi)}{k} \approx \frac{u_\pi}{b(s_0)},
    \label{eq:Deltaf_linearized}
\end{equation}
where $u_\pi$ is the per-platform displacement at $P_\pi$ and $b(s_0) = f_\text{opt}(s_0)\,/\,|\partial f_\text{opt}/\partial s|_{s_0}$ is the local decay length of the optical force, evaluated from the bi-exponential fit at the nominal slot width. Because $u_\pi \propto P_\pi/k \propto 1/L$ (for fixed $s_0$), the ratio $u_\pi / b(s_0)$ depends primarily on $s_0$ and is nearly independent of $k$ and $L$ in the linear regime. Figure~\ref{fig:opticalSpring}\textbf{b} shows $\Delta f/f_\text{m}$ at $P_\text{p} = P_\pi$ as a function of $k$ and $L$ for all device geometries characterized in this work. In all cases the predicted shift is below 5$\%$, which is far smaller than the frequency gap between the two spectral features observed at $\approx 152$~kHz and $\approx 233$~kHz in Fig.~4\textbf{d} of the main text. This quantitative comparison confirms that these two peaks cannot be assigned to the optically-split $f_+$ and $f_-$ mode pair.

\subsection{S1.7. Simulation of additional photonic components}
\setlabel{S1.7}{subsec:additional_devices}

The modelling of the SWGSW provided above disregards a few important aspects that need to be considered when comparing experimental data to the model. Notably, coupling into the slot mode of interest is achieved via V-groove mechanically disjoint rectangular-to-slot waveguide couplers, which were initially proposed and experimentally demonstrated in Ref.~\cite{wang_ultracompact_2009} for silicon waveguides embedded in silicon oxide. This coupler adiabatically transforms the mode profile of the rectangular waveguide into that of a waveguide with two slots, which end up merging in the center to form the target slot waveguide. These couplers inevitably have a finite insertion loss, but, more importantly, the presence of the two slots in the coupler region and the associated field enhancement~\cite{almeida_guiding_2004} leads to additional optical forces. In addition to this circuit element, we have also designed a coupler that transforms the slot waveguide mode into the slot SWGSW waveguide mode. This subsection provides simulation details of these photonic components, whose independent experimental characterization is described in Section~\ref{sec:losses}.

\subsubsection{Rectangular-to-slot waveguide coupler}

\begin{figure*}[t!]
\includegraphics[width=0.64\textwidth]{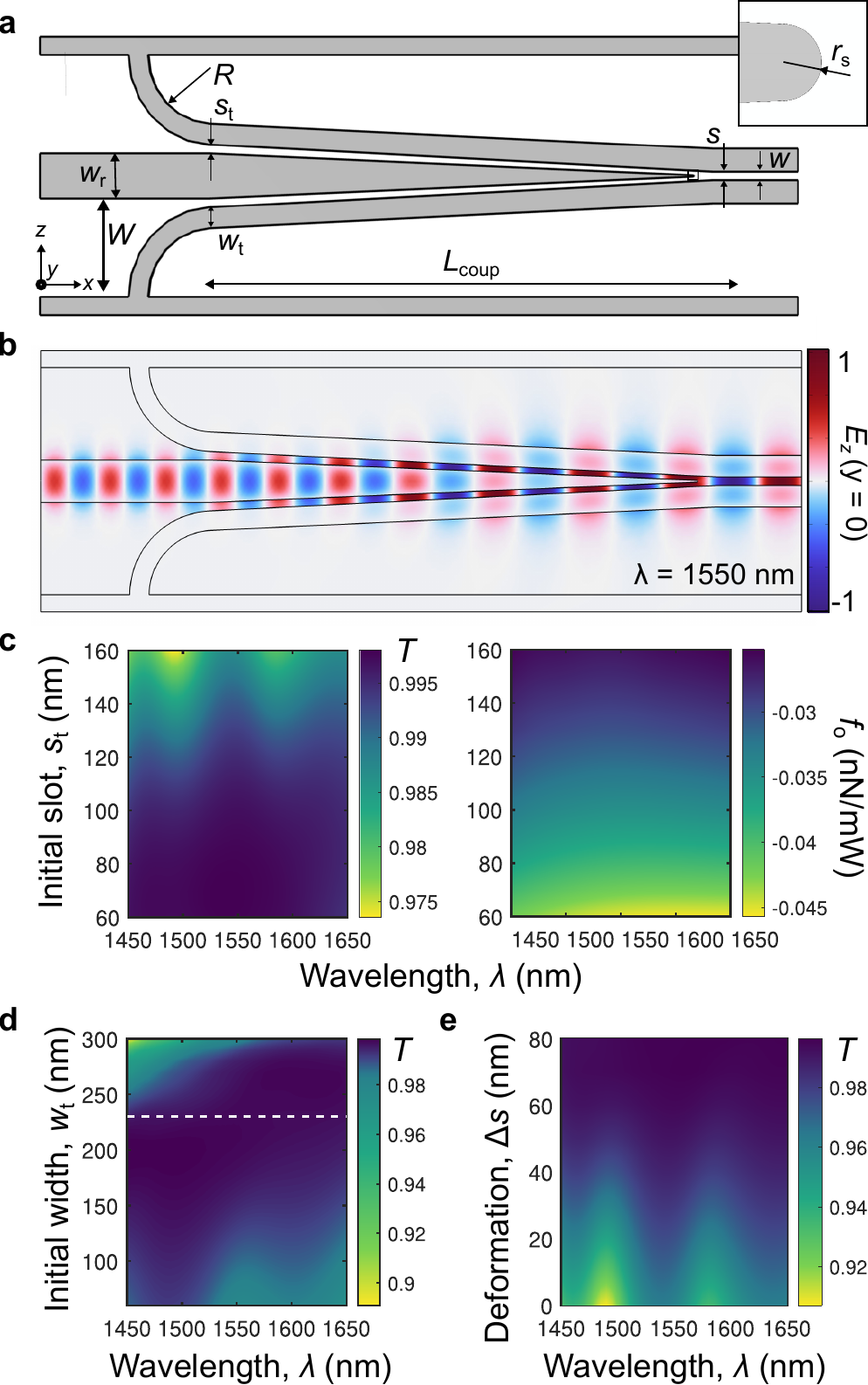}
    \caption{\textbf{Strip-to-slot suspended waveguide coupler.} \textbf{a}, Geometry and parametrization of the coupler. Parameters $\{w_\text{r}, W, R, L_\text{coup}, r_\text{s}, w \}$ are fixed to $\{$500, 1000, 1000, 6000, 10, 254$\}$ nm. \textbf{b}, Electric field, $E_z$, at $y$ = 0 for an excitation from the left into the fundamental TE-like mode. \textbf{c}, Wavelength-dependent (left) transmittance and (right) optical force as a function of the initial slot, $s_\text{t}$. The remaining parameters are fixed to $\{w_\text{t}, s\}$ = $\{$230, 120$\}$ nm. \textbf{d}, Transmittance as a function of $w_\text{t}$, for the case $\{s_\text{t}, s\}$ = $\{$100, 100$\}$ nm. \textbf{e}, Same for parameters $\{w_\text{t}, s_\text{t}, s \}$ = $\{$230, 200, 200$\}$ nm and as a function of a rigid displacement, $\Delta s$, of each side towards the axis.}
    \label{fig:Vgroovecoupler} 
\end{figure*}

The geometry and all relevant geometric parameters of the V-groove mode converters we employ are detailed in Fig.~\ref{fig:Vgroovecoupler}\textbf{a}. The working principle of the employed coupler is that of an adiabatic mode converter which transforms the mode profile of the rectangular waveguide into that of a waveguide with two slots, which end up merging in the center to form the target slot waveguide. The taper is achieved by modifying the geometric features in a linear fashion. The mode conversion is represented in Fig.~\ref{fig:Vgroovecoupler}\textbf{b}, which depicts the $z$-component of the electric field in the mid-plane of the silicon device layer for a wavelength $\lambda$ = 1550 nm. As expected from the presence of the two slots and the fact that they need a width of at least of $s/2$ close to the slot waveguide, the field intensity is enhanced in both slots, which leads to additional optical forces. These forces are exemplified in Fig.~\ref{fig:Vgroovecoupler}\textbf{c} (colormap on the right), which shows the integrated force on one of the two beams of the coupler as the initial slot width,$s_\text{t}$ is swept. This is represented for a coupler with parameters $\{w_\text{r}, W, R, w_\text{t}, L_\text{coup}, r_\text{s}, s, w \}$ = $\{$500 nm, 1 $\mu$m, 1 $\mu$m, 230 nm, 6 $\mu$m, 10 nm, 120 nm, 254 nm$\}$. As expected, increasing $s_\text{t}$ reduces the field strength at the sidewalls and therefore lowers the attractive force per unit power. In principle, the presence of such additional force is not undesired and it may contribute to lower the power requirements to operate the devices. In practice, the simplicity of the modelling described in previous sections is partially lost when the coupler contributes significantly to the optical forces, especially for short devices. Therefore, we restrict the design space to couplers with an initial slot width $s_\text{t}$ = $s$, a compromise chosen in order to not sacrifice too much in transmittance, which degrades as $s_\text{t}$ increases (left colormap of Fig.~\ref{fig:Vgroovecoupler}\textbf{c}). 
An important difference between this geometry and that of Ref.~\cite{wang_ultracompact_2009} is the extension of the two silicon sections leading to the slot waveguide beyond the point where the coupling region \textit{starts}. This is done by using bends of radius $R$ and anchoring the end of the bends to the surrounding frame. A similar---though not symmetric---approach was employed in Ref.~\cite{grottke_optoelectromechanical_2021}. Such anchoring allows for enhanced mechanical stability compared to a geometry based on two slender triangular cantilevers, which we saw led systematically to in-plane stiction for small initial slot widths, $s$. Interestingly, an anchored geometry also leads to a better nominal transmission. Specifically, we show in Fig.~\ref{fig:Vgroovecoupler}\textbf{d} the transmittance as the initial width $w_\text{t}$ is swept, which shows the presence of an optima in the range $w_\text{t} \in 200 - 240$ nm and we choose $w_\text{t}$ = 230 nm for all the employed couplers. Figure~\ref{fig:Vgroovecoupler}\textbf{e} illustrates the expected variation in coupler transmission due to rigid deformation as the optical pump power increases, for a structure with $s$ = 200 nm. Consistent with previous pictures, when the deformation reduces the slot widths, the transmission increases.

Finally, we note that the design includes a sharp tip at the slot-waveguide edge of the coupler, as any widening of that region leads to a fast decrease in transmission, as it was also shown in Ref.~\cite{wang_ultracompact_2009}. All reported simulations have a rounding of $r_\text{s}$ = 10 nm at the edge of the coupler, as shown on the inset of Fig.~\ref{fig:Vgroovecoupler}\textbf{a}, which corresponds to the minimum solid radius of curvature of the nanofabrication process we employ~\cite{babar_self-assembled_2023}.

\subsubsection{Slot waveguide to subwavelength grating slot waveguide coupler}

\begin{figure*}[t!]
\includegraphics[width=0.76\textwidth]{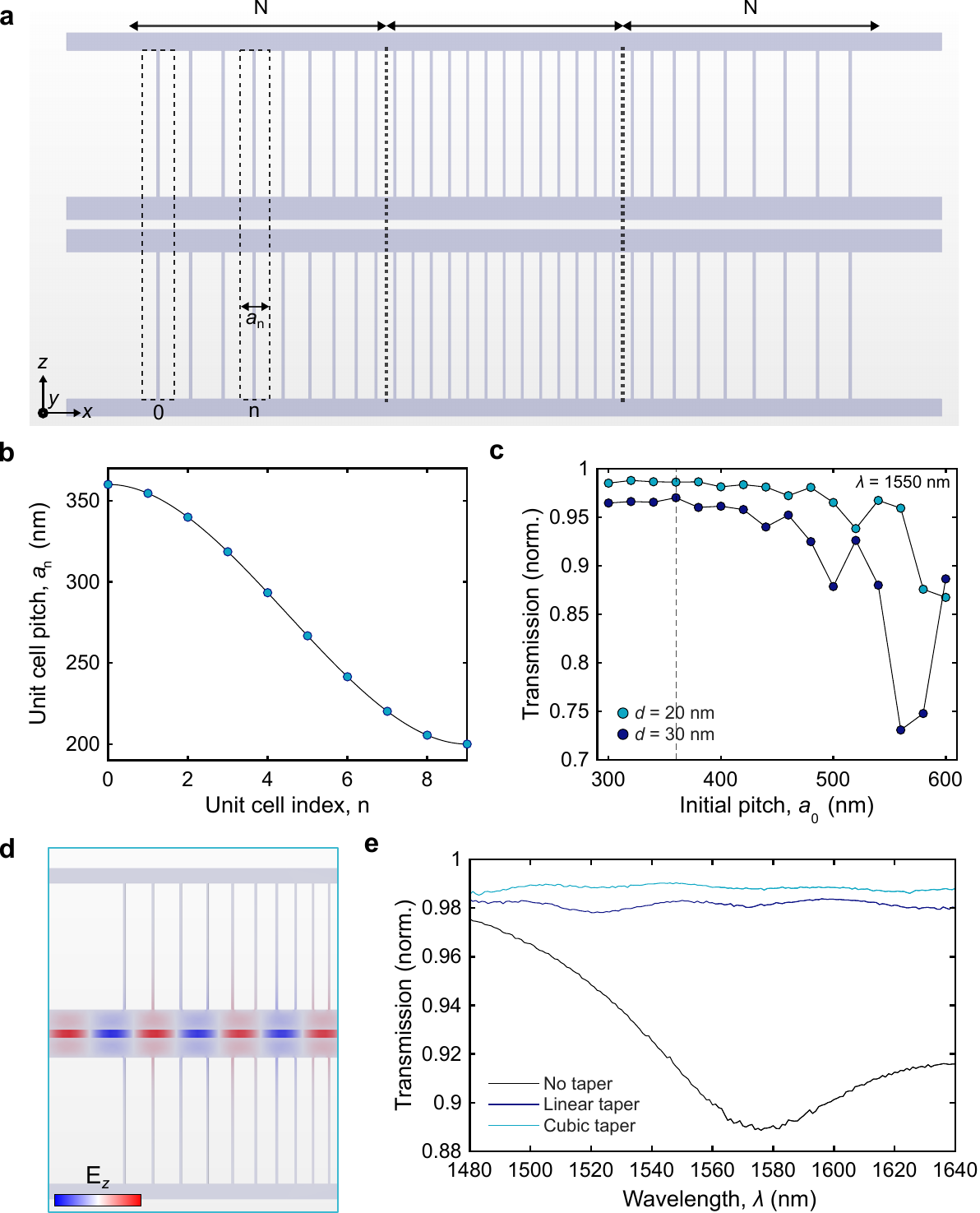}
    \caption{\textbf{Slot waveguide to subwavelength grating slot waveguide (SWGSW) coupler.} \textbf{a}, Schematic of the input and output couplers illustrating the transition between the uniform slot waveguide and the SWGSW by tapering the periodicity $a$ of the unit cell. \textbf{b}, Layout of a specific cubic taper realization with an initial pitch $a_0 = 360$ nm. \textbf{c}, Calculated transmittance of the coupler pair at $\lambda = 1550$ nm as a function of the initial pitch $a_0$, showing performance degradation for larger initial pitches and beam widths. \textbf{d}, Electric field, $E_z$, at $y$ = 0 for an excitation from the left into the fundamental slot mode. \textbf{e}, Wavelength-dependent transmittance spanning $\lambda \in [1450, 1650]$ nm for the optimal $a_0 = 360$ nm design, comparing the performance of a cubic taper, a linear taper, and a direct transition.}
    \label{fig:SWtoSWGSW} 
\end{figure*}

In the previous subsection, we have discussed how to efficiently couple from a rectangular strip waveguide into a slot waveguide. In this subsection, we discuss a simple strategy to couple light into the SWGSW mode from the slot waveguide. Because the mode profile and effective refractive index of the SWGSW slot-guided mode already resembles closely that of the slot waveguide (see Fig. 1\textbf{c} in the main text), such a coupler can be implemented with minimal modification of the geometry. We have explored two different design strategies: A design in which the tether length is tapered --and thus transformed from a doubly clamped beam to a cantilever-- and a design where the periodicity of the unit cell, $a$, is tapered. Despite convincing numerical results for both cases, the former approach leads to long cantilevers that easily deform and collapse, so the latter approach has been selected and is shown in Fig.~\ref{fig:SWtoSWGSW}\textbf{a}. We compare linear and cubic tapering functions to the case without tapering, i.e., a sharp step-like transition. The discrete cubic tapering function gives the period as
\begin{equation}
    a_n = a\left[1-\left(1-\frac{a_{0}}{a}\right)\left(2\left(\frac{n}{N}\right)^3 - 3\left(\frac{n}{N}\right)^2 + 1\right)\right], \quad n \in [0,N],
    \label{taper_func}
\end{equation}
with $a$ = 200 nm the pitch of the SWGSW and $a_0$ the pitch for the initial unit cell in the taper. A specific realization for $a_0$ = 360 nm of such tapering is shown in Fig.~\ref{fig:SWtoSWGSW}\textbf{b}.
We select $N = 9$ and solve for the coupler transmittance for models with varying $a_0 \in [300, \ 600] \ \mathrm{nm}$. Note that in the frequency-domain simulations we include both the input and output coupler (as in the schematic of Fig.~\ref{fig:SWtoSWGSW}\textbf{a}) and assume negligible reflectance, such that the single-coupler transmittance is estimated via $T_\text{coup} = \sqrt{T_\text{tot}}$. Figure~\ref{fig:SWtoSWGSW}\textbf{c} shows the transmittance at $\lambda$ = 1550 nm as a function of $a_0$: When $a_0$ increases, the transmittance is reduced non-monotonically. We attribute this to the fact that the most isolated beams start to behave as single tethers rather than as part of the SWGSW taper, i.e., both the subwavelength nature and the adiabaticity are simultaneously lost. Likewise, introducing a wider beam for the tethers ($d$ = 30 nm) degrades the performance. We therefore select $a_0$ = 360 nm, whose simulated electric field, $E_z$, is shown in Fig.~\ref{fig:SWtoSWGSW}\textbf{d}, illustrating the adiabatic transfer of the fundamental slot mode from the slot waveguide into the SWGSW. This design leads to a transmittance of approximately 99$\%$ in the range $\lambda\in$ [1450, 1650] nm, as evidenced by Fig.~\ref{fig:SWtoSWGSW}\textbf{e}.  The cubic taper shows approximately 1$\%$ improvement compared to the linear taper, but both significantly outperform the case without tapering. All simulations presented in this section are carried out for a tether width $d$ = 20\,nm, but in the fabricated structures, we adopt $d$ = 30\,nm to provide additional margin against fabrication variability; all SWGSW phase shifters and associated photonic circuits in this work are fabricated with 30-nm wide tethers.

\section{S2. Sample fabrication and structural characterization.}
\setlabel{S2}{sec:fabrication}

\subsection{S2.1. Fabrication process flow}
\setlabel{S2.1}{subsec:process}

The SWGSW devices are fabricated from a commercial silicon-on-insulator (SOI) substrate (SOITEC) including a 220 nm silicon device layer and a 2 $\mu$m buried silicon oxide layer. The first fabrication step is the deposition of 10 nm of poly-crystalline chromium and 10 nm of poly-crystalline silicon on top of the silicon device layer, which is subsequently used as a two-layer hard mask. The patterns are exposed in a manually spin-coated 50 nm-thick chemically semi-amplified resist (CSAR) using a 100 keV 100 MHz JEOL9500FSZ electron-beam writer. The transfer of the patterns into the silicon device layer is achieved by a series of dry etching steps~\cite{hoang_nguyen_cr_2021}, all of them performed in the same processing chamber: 1) the poly-silicon layer is etched using the resist mask, 2) the resist is stripped, 3) the chromium layer is etched using the poly-silicon mask, 4) the remaining poly-silicon and the silicon device layer are etched using a low-power switched reactive-ion etch, and 5) the chromium layer is etched. The process ends with a selective isotropic etching of the buried silicon oxide layer to suspend the devices. This is performed using an anhydrous hydrofluoric-acid (99.995\%) vapour phase etcher (SPTS Primaxx uEtch), using ethanol as a catalyst. We employ a low process pressure of 131 Torr, and a slow etching recipe (etch rate of approximately 14 nm/min) to minimize the probability of water condensation in the process chamber leading to in-plane capillary-force stiction of long suspended devices. The fabrication process is equivalent to that employed for studying surface-force-assisted in-plane collapses in our previous work~\cite{babar_self-assembled_2023} and is described in detail in Refs.~\cite{albrechtsen_nanometer-scale_2022, nguyen_core_2020, hoang_nguyen_cr_2021}. The critical dimensions achieved by this process and their systematic deviation from the lithographic mask are quantified in Section~\ref{subsec:SEM_sizes}.

\begin{figure}[t!]
    \centering
    \includegraphics[width=\textwidth]{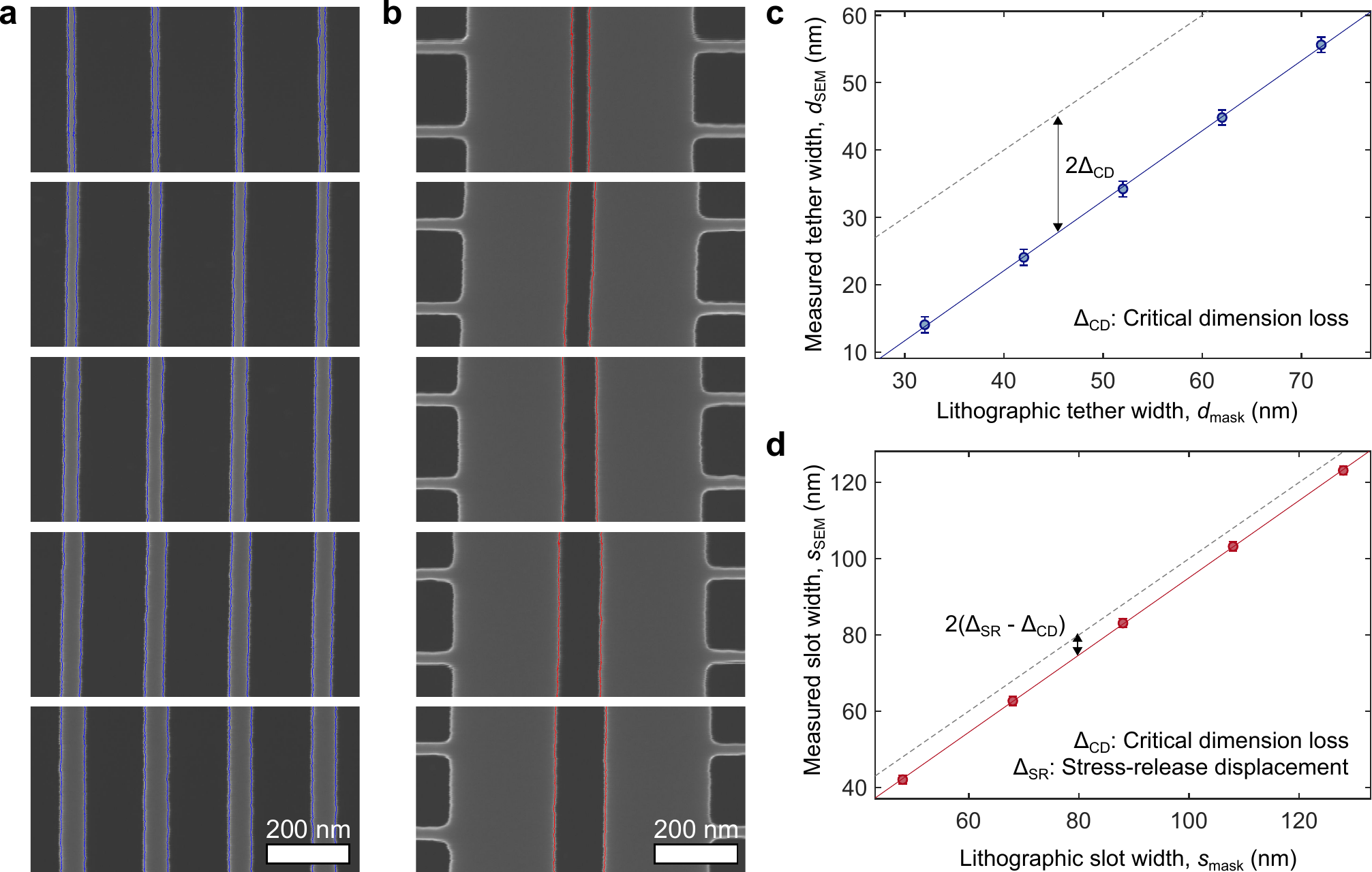}
    \caption{\textbf{Scanning electron microscopy (SEM) analysis of critical dimensions.} \textbf{a,} SEM images of SWGSW tethers of different width $d$. \textbf{b,} SEM images of SWGSWs with different slot width, $s_0$. Shear in the images in \textbf{a} and \textbf{b} are due to horizontal drift as the SEM raster-scans, and does not impact the reported widths. \textbf{c,} Mapping of lithographic dimension onto fabricated dimension for $d$. \textbf{d} Mapping of lithographic dimension onto fabricated dimension for $s_0$. $\Delta_\text{CD}$ represents the critical dimension loss due to lithography and etching and $\Delta_\text{SR}$ the additional change on $s_0$ due to built-in stress release.}
    \label{fig:Fab1}
\end{figure}

\subsection{S2.2. Scanning electron microscopy characterization of fabricated devices}
\setlabel{S2.2}{subsec:SEM_sizes}

Prior to fabrication of the suspended SWGSW phase shifters, we fabricate a series of SWGSWs with the objective of exploring the minimum feature sizes achievable within such geometry, i.e. the resolution of our nanofabrication process for SWGSWs. Notably, the slot width, $s$, and the tether width, $d$, are important parameters in the behaviour of the phase shifters due to their respective role in terms of optical forces and propagation losses. For efficient devices, both parameters require values between 10 and 100 nm. After fabrication, we do scanning electron microscopy (SEM) characterization of the suspended devices and systematic image analysis to extract the critical dimensions from the fabricated geometries. Figure~\ref{fig:Fab1}\textbf{a} shows representative high-magnification (250x) SEM images of the tethers of SWGSWs with tether width varying in steps of 10 nm. We include (blue contours) the edges found by our automated image analysis routine, where the edge is identified as the pixel of maximum brightness at the air-silicon interfaces. Figure~\ref{fig:Fab1}\textbf{b} depicts the same analysis for the slot width, based on SWGSWs with slot width varying in steps of 20 nm, but a fixed tether width. Comparing the beam widths in the lithographic mask, i.e., $d_\text{mask}$ = [32,42,52,62,72] nm, and the SEM-extracted beam widths, i.e., $d_\text{SEM}$ = [$14.1 \pm 1.2, 24.1 \pm 1.2, 34.2 \pm 1.2, 44.8 \pm 1.1, 55.6 \pm 1.1$] nm, we find that the critical dimension (CD) loss --the growth of hole features and subsequent shrinkage of material features due to fabrication effects, e.g., etching-- is approximately size-independent, as confirmed by the slope (1.01) of the linear fit in Fig.~\ref{fig:Fab1}\textbf{c}. We extract a CD loss of $\Delta_\text{CD}$ = 9.7 nm, which represents the displacement of all lithographic edges towards the solid areas. This value is also found when evaluating the deviation of the width of the two rectangular waveguides forming the SWGSW, and is larger than our expectation from a previous fabrication process, which was 6 nm per edge. Interestingly, the same analysis for the slot widths (Fig.~\ref{fig:Fab1}\textbf{d}) shows that the inferred feature growth is different. We attribute this to the static displacement induced by the release of the built-in silicon device layer stress, $\Delta_\text{SR}$ = 12.9 nm, when the structures are suspended. We also confirm this by comparison to SEM images acquired before the final release step. Consequently, the deviation between perfect pattern transfer and the fabricated slot widths (distance between black dashed and red lines in Fig.~\ref{fig:Fab1}\textbf{d}) is given by $2|\Delta_\text{CD}-\Delta_\text{SR}|$ = 6.3 nm.
The exact values of $\Delta_\text{CD}$ and $\Delta_\text{SR}$ depend on the fabrication batch and the location on the chip (due to loading effects), but we estimate the variations of these quantities to be below 3 nm across different fabrication batches. Therefore, we employ the values reported here to correct the lithographic mask in order to obtain the desired fabricated geometry. All figures following this section and showing results as a function of slot width, $s$, use values based on such corrections and thus the reported values of $s$ should be read as targeted values, typically achieved to within 2--3~nm.

\subsection{S2.3. Mapping the feasible parameter space for subwavelength-grating slot waveguide phase shifters}
\setlabel{S2.3}{subsec:collapsedUncollapsed}

\begin{figure}[t!]
    \centering
    \includegraphics[width=0.8\textwidth]{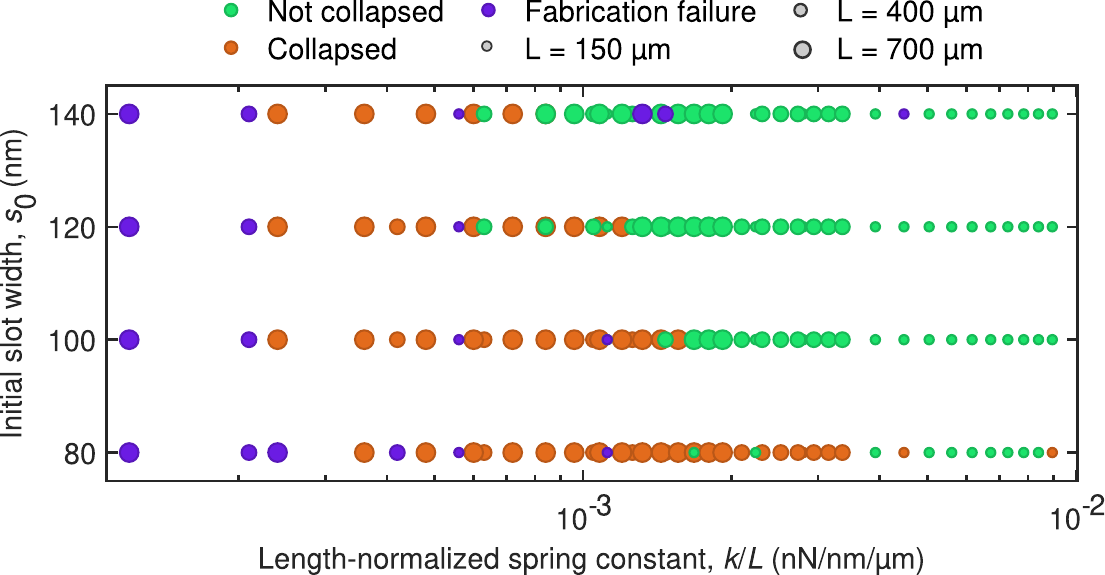}
    \caption{\textbf{Classification of fabricated SWGSW phase shifters.} Ternary classification of fabricated devices on a single chip as a function of the length-normalized spring constant, $k$/$L$, and the targeted fabricated slot width $s$. The three categories are defined as: the slot is not closed after release (green), the slot is closed after release due to an in-plane collapse (orange) and the device suffered from fabrication issues (blue). For completion, the length is indicated via the size of the solid dots.}
    \label{fig:Fab2}
\end{figure}

An important aspect to consider is the yield of our fabrication process, because efficient all-optical optomechanical phase shifting requires low spring constant values in the in-plane actuation direction ($k <$ 1 N/m) and narrow slots ($s <$ 150 nm). As the structures are released during HF vapour isotropic etching, the linear elastic restoring force of the guided folded cantilever springs balances the (generally) non-linear surface forces that act between the two slot facets (namely Casimir-van-der-Waals, capillary, electrostatic forces, etc.). This equilibrium persists until a critical slot width is reached, below which the two sides collapse in-plane via a pull-in instability. We have recently studied such in-plane collapses in the same suspended silicon photonics platform, with the goal of understanding the phase-space for self-assembly of nanostructures~\cite{babar_self-assembled_2023}. Here we replicate such an experiment using the fabricated spring-suspended SWGSWs, whose lengths are far beyond the widths of the platforms explored in Ref.~\cite{babar_self-assembled_2023}. We classify the fabricated devices as being: a) \textit{collapsed}, i.e, the gap between the two sides of the SWGSW closes, b) \textit{not collapsed}, i.e., the slot of the SWGSW is open, or c) a \textit{fabrication failure}, i.e., the device collapsed out-of-plane onto the substrate or had other fabrication-related issues. As in our previous work, we observe a clear phase-space separation in the parameter space defined by the initial slot width, $s_0$, and the length-normalized spring constant $k/L$ between the regions where devices collapse in-plane (orange dots in Fig.~\ref{fig:Fab2}) and the region where devices do not collapse in-plane (green dots in Fig.~\ref{fig:Fab2}). Except for a few devices for the smaller slot width ($s_0$ = 80 nm) and lower $k/L$, where we report in-plane collapses, the phase separation is clear and provides quantitative design rules for our SWGSW phase shifters. 

\begin{figure}[t!]
    \centering
    \includegraphics[width=0.8\textwidth]{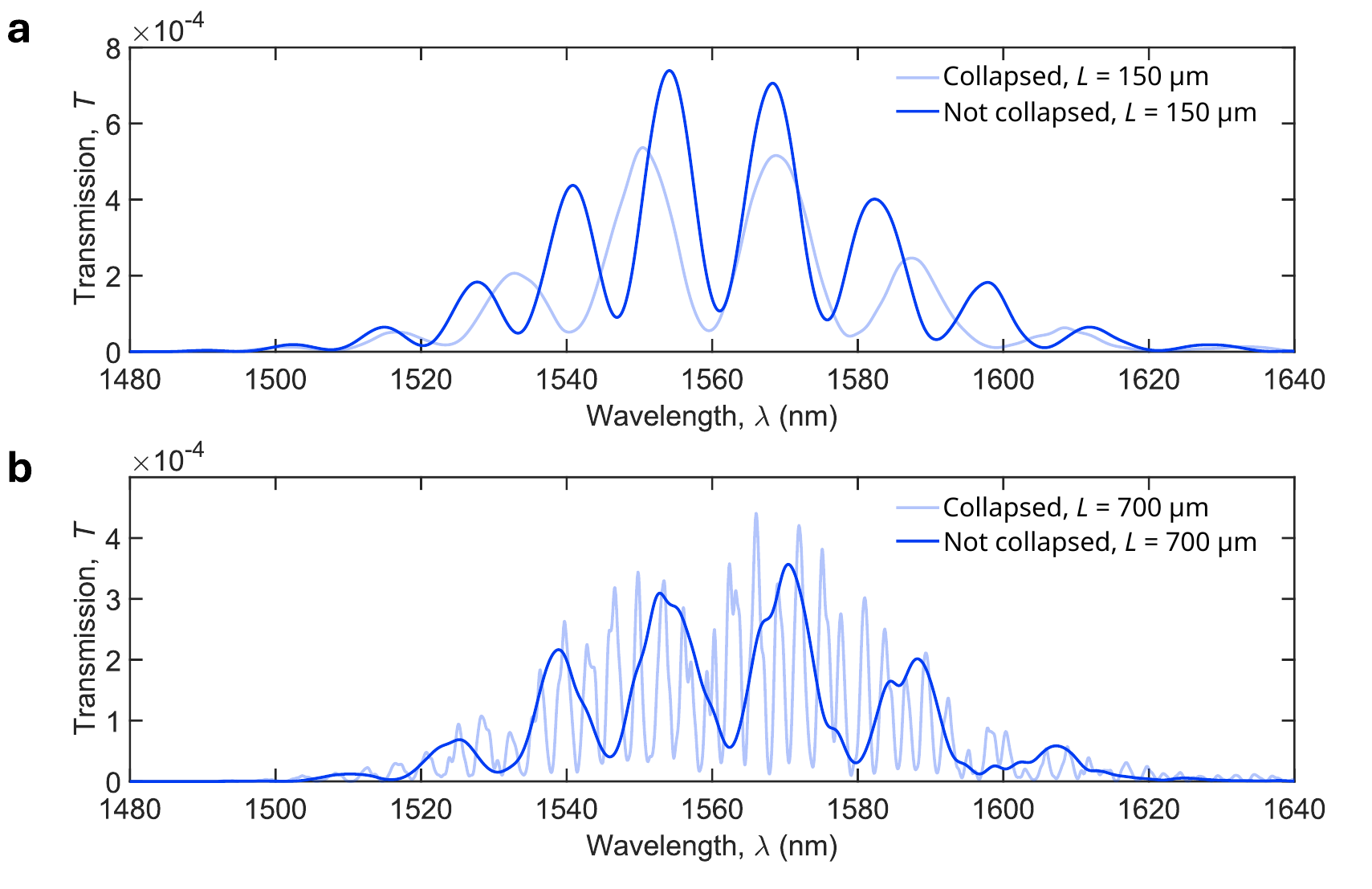}
    \caption{\textbf{Characteristic transmission traces of collapsed and un-collapsed devices}. \textbf{a}, Normalized transmission of two SWGSWs of length $L$ = 150 $\mu$m, respectively un-collapsed (dark blue line) and collapsed (light blue line). The comparison illustrates the clear transmission drop of the envelope. \textbf{b}, Similar comparison for two SWGSWs of length $L$ = 700 $\mu$m, highlighting the significant increase in fringe count that results from the large optical path imbalance.}
    \label{fig:collapsecomparison}
\end{figure}

We note that, compared to our work in Ref.~\cite{babar_self-assembled_2023}, the ternary classification in Fig.~\ref{fig:Fab2} is not achieved through inspection of SEM images --as we observe our devices to collapse upon electron beam incidence and charging-- but by inspection of the optical transmission through the UMZIs and its dependence with power. We analyze three features: 1) the number of spectral fringes, 2) the maximum transmittance and 3) the tuning of the fringes with input power. Devices labelled \textit{not collapsed} exhibit 7-10 fringes within the grating coupler bandwidth, can be tuned via the input optical power and have their maximum transmittance all within a 5-20\% range from the maximum transmittance observed, when comparing devices of identical length and slot width. Devices labelled \textit{collapsed} have a much larger number of fringes for long devices, as the imbalance increases when the SWGSW collapses and becomes a (self-assembled) rectangular waveguide, and a transmittance between 0\% and 50\% from the maximum transmittance observed. For short devices we observe minimal change in fringe count but a drop in transmission of at least 20\%. Devices labelled \textit{fabrication failure} have neither of the latter spectral features and a transmittance lower than 20\% of the maximum transmittance observed. Representative spectra are shown in Fig.~\ref{fig:collapsecomparison}.

\section{S3. Data analysis.}
\setlabel{S3}{sec:dataanalysis}

In this section we detail a number of data processing aspects that are relevant to understand the extracted quantities based on transmission measurements.

\begin{figure}[t!]
    \centering
    \includegraphics[width=0.8\textwidth]{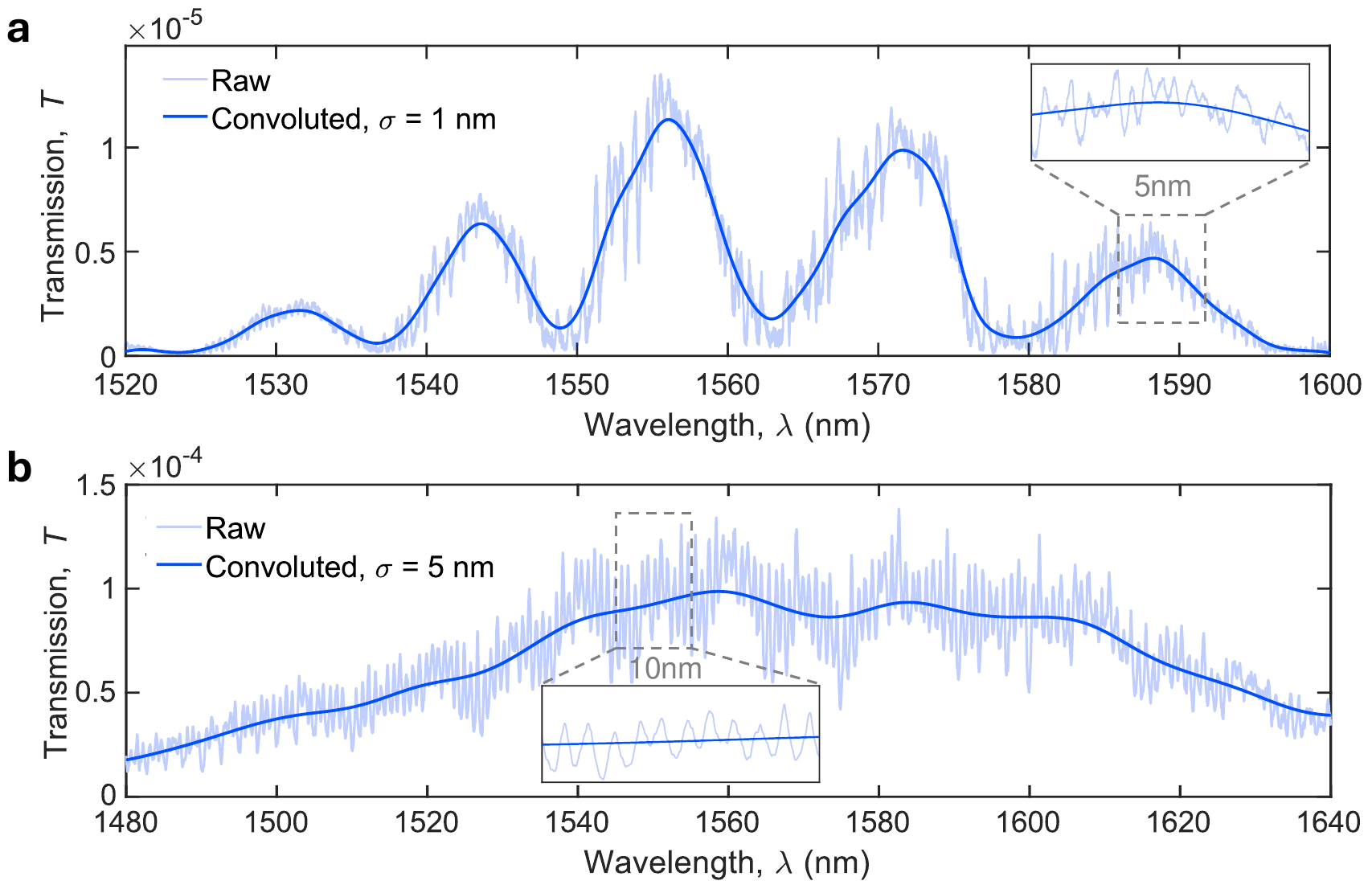}
    \caption{\textbf{Spectral smoothing of raw transmission spectra.} \textbf{a,} Raw transmission (light blue) and convolution with a Gaussian of $\sigma=1$ nm (dark blue) for an unbalanced Mach-Zehnder interferometer (UMZI) of parameters $\{s_0,L,k, P_\text{PS}\}$ are fixed to $\{100\:\text{nm}, 400\:\mu\text{m}, 0.92\:\text{N/m}, 0.21\:\text{mW}\}$. $\sigma=1$ nm. \textbf{b,} Propagation loss smoothing, parameters $\{s_0,L\}$ are fixed to $\{100\:\text{nm}, 480\:\mu\text{m}\}$. $\sigma=5$ nm.}
    \label{fig:smoothing}
\end{figure}

\subsection{S3.1. Spectral smoothing}
\setlabel{S3.1}{subsec:smoothing}

We employ free-space grating couplers to couple in and out of the photonic circuits under test. Two different designs are included: a high-efficiency \textit{narrowband} shape-optimized apodized grating coupler for the circuits including the phase shifters~\cite{hansen_efficient_2023}, and a low-efficiency broadband one-ring circular coupler~\cite{faraon_dipole_2008} for the circuits used for measuring the different loss mechanisms (see Section~\ref{sec:losses}). In both cases, the finite in-plane reflections at the couplers lead to narrow spectral fringes --much narrower than any other spectral feature we observe-- that vary from circuit to circuit as the total optical path length changes~\cite{hansen_efficient_2023}. To discriminate the effect of such fringes from the spectral features of interest, the raw transmitted power spectra, $P_\text{T}(\lambda)$, are smoothed via a linear convolution with a normalized Gaussian filter,
\begin{subequations}
    \begin{equation}
        T_\text{S}(\lambda) = \int^{\infty}_{\infty} T(\lambda')g(\lambda-\lambda')d\lambda',
    \end{equation}
    \begin{equation}
        g(\lambda) = \frac{1}{\sigma\sqrt{2\pi}}e^{ -\frac{\lambda^2}{2\sigma^2}},
    \end{equation}
\end{subequations}
with $\sigma$ the standard deviation of the Gaussian. Figure~\ref{fig:smoothing} shows two characteristic transmission curves of the devices we explore. On top, the transmission across one of the UMZIs, for which we fix $\sigma$ = 1 nm. On the bottom, the transmission across one of the 'O'-shaped photonic circuits used for the propagation loss extraction (see Fig.~\ref{fig:SWGSW_Loss}), for which we fix $\sigma$ = 5 nm. The values of $\sigma$ are chosen to allow the fringes to be smoothed out without considerably changing the result of interest, as determined by analysis of the result in question as a function of $\sigma$. The reason for the different values of $\sigma$ employed is linked to (1) the type of grating coupler employed, (2) the total length of the circuits, and (3) the presence of certain resonances in the propagation loss circuits induced by periodic features on the photonic circuit.

\subsection{S3.2. Power calibration}
\setlabel{S3.2}{subsec:power_calibration}

\begin{figure}[t!]
    \centering
    \includegraphics[width=0.8\textwidth]{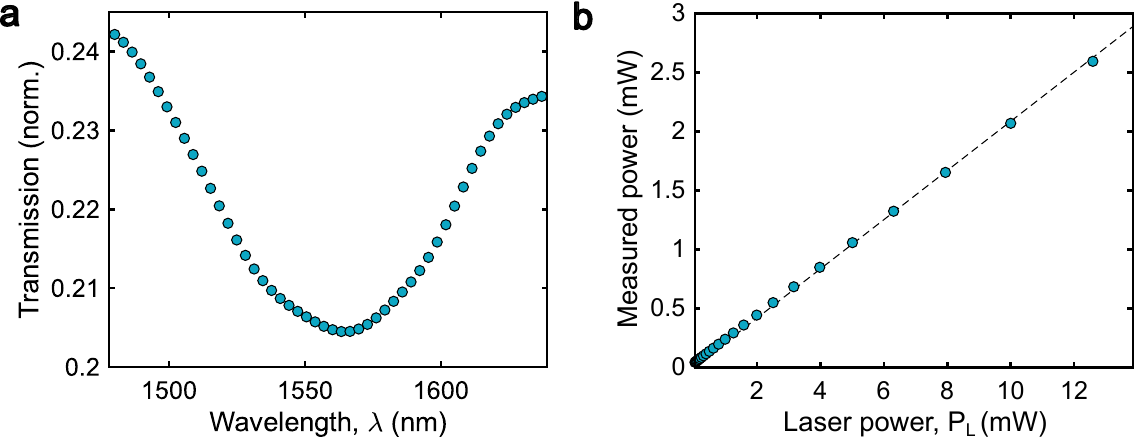}
    \caption{\textbf{Optical setup transmission characterization.} \textbf{a,} Measured transmission spectrum of the full optical setup. \textbf{b,} Measured power incident on the chip at $\lambda = 1.56\ \mu\text{m}$ as a function of laser output power.}
    \label{fig:power_calibration}
\end{figure}

To evaluate the true efficiency of the optomechanical phase shifter and be able to compare the measured phase shift to the model, we require knowledge of the power $P_\text{PS}$ at the input of the SWGSW phase shifter. This quantity differs from the laser output power $P_\text{L}$ due to a series of transmission factors accumulated along the optical path from the source to the device. Tracing this path, the pump power at the SWGSW input is
\begin{equation}
    P_\text{PS}(\lambda) = \frac{1}{2}\,T_\text{setup}(\lambda)\,T_\text{GC}(\lambda)\,P_\text{L},
    \label{eq:power_calibration}
\end{equation}
where $T_\text{setup}(\lambda)$ is the wavelength-dependent transmission of the optical setup (fiber and free-space optics), $T_\text{GC}(\lambda)$ is the on-chip grating coupler response, and the factor of $1/2$ accounts for the 50/50 Y-branch splitter, assumed to be perfectly balanced. Propagation losses in the short rectangular and slot waveguide sections between these elements are neglected, as they are small compared to the coupler insertion losses. Figure~\ref{fig:power_calibration}\textbf{a} shows the measured transmission spectrum of the full setup, recorded with a calibrated powermeter at the focal plane of the microscope objective. The wavelength dependence is weak, considerably less pronounced than that of the grating coupler (Fig.~2\textbf{b} in the main text). Figure~\ref{fig:power_calibration}\textbf{b} shows the measured power at $\lambda = 1.56\ \mu\text{m}$ as a function of laser power, confirming the linearity of the setup. We note that $P_\text{PS}$ as defined in Eq.(~\ref{eq:power_calibration}) is referenced to a plane upstream of the rectangular-to-slot and slot-to-SWGSW couplers, which are an integral part of the device. It therefore differs from the power $P_\text{p}$ at the SWGSW input used in the model of Section~\ref{subsec:force_equilibrium}, the two being related by the simulated coupler transmittance; the model values are accordingly rescaled when experimental and theoretical efficiencies are compared (Section~\ref{subsec:additional_devices}).

\subsection{S3.3. Extraction of $P_\pi$}
\setlabel{S3.3}{subsec:phase_shift_extraction}

\begin{figure}[t!]
    \centering
    \includegraphics[width=0.85\textwidth]{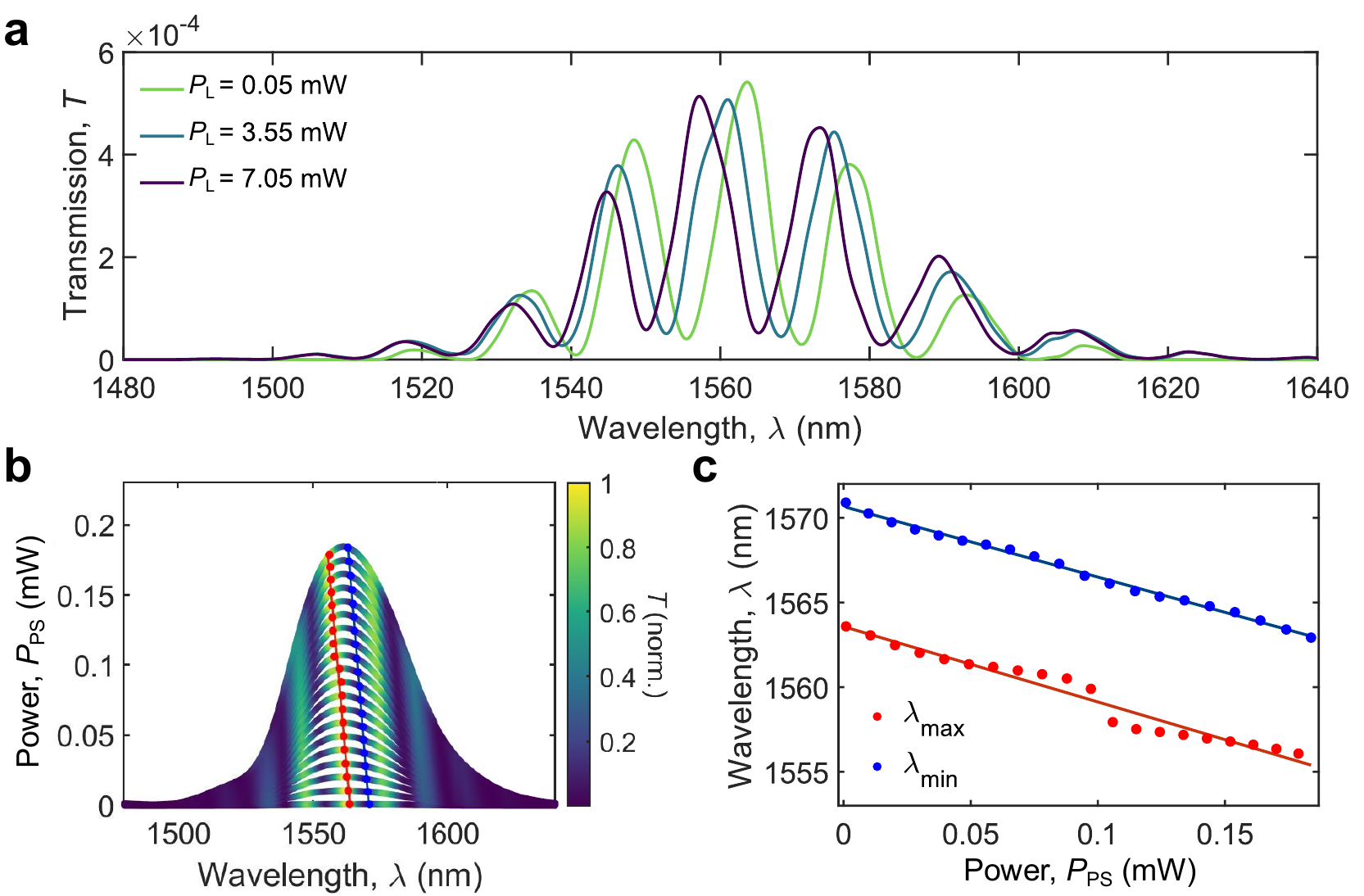}
    \caption{\textbf{Extraction of $P_\pi$ for an unbalanced Mach-Zehnder interferometer (UMZI) including an optomechanical phase shifter.} \textbf{a,} Smoothed UMZI transmission spectra for three different laser powers $P_\text{L}$ for a structure with $\{s_0,L,k\} = \{100\:\text{nm}, 400\:\mu\text{m}, 0.92\:\text{N/m}\}$. \textbf{b,} Wavelength positions of the tracked local maxima (red) and minima (blue) as a function of the power reaching the phase shifter $P_\text{PS}$, extracted from spectra of the type shown in \textbf{a}. \textbf{c,} Linear fits $\Delta\lambda = A_\text{max/min}\,P_\text{PS}$ to the maxima and minima shifts, from which $P_\pi$ is extracted via Eq.(~\ref{eq:Ppi_final}).}
    \label{fig:phaseextraction}
\end{figure}

The smoothed UMZI transmission spectra obtained as described in Section~\ref{subsec:smoothing} (Fig.~\ref{fig:phaseextraction}\textbf{a}) are first corrected for the wavelength-dependent grating coupler response $T_\text{GC}(\lambda)$, which enters twice in the detected signal (input and output couplers). Dividing by $T_\text{GC}^2(\lambda)$ removes the spectral envelope imposed by the grating and allows for an improved retrieval of the UMZI response maxima and minima. For each laser power, the on-chip pump power $P_\text{PS}(\lambda)$ at the SWGSW phase shifter input is then determined from the laser output power $P_\text{L}$ following the calibration of Section~\ref{subsec:power_calibration}.

The wavelengths of all local maxima and minima and the powers at which they occur are extracted from the corrected spectra as shown in Fig.~\ref{fig:phaseextraction}\textbf{b}. From that, the wavelength shift associated with the $n$-th fringe, $\Delta\lambda_n = \lambda_n(P_\text{PS}) - \lambda_n^{(0)}$, is computed for all fringes relative to the lowest-power measurement. These are plotted as a function of $P_\text{PS}$ in Fig.~\ref{fig:phaseextraction}\textbf{c}, where the linear scaling with power expected from the small-displacement regime (Section~\ref{subsec:MZI_probing}) is clearly visible. Fitting the maxima and minima shifts separately yields
\begin{equation}
    \Delta\lambda_\text{max} = A_\text{max}\,P_\text{PS}, \qquad \Delta\lambda_\text{min} = A_\text{min}\,P_\text{PS},
    \label{eq:linear_fits}
\end{equation}
as shown in Fig.~\ref{fig:phaseextraction}\textbf{c}. Since both extrema experience the same phase shift, the two slopes $A_\text{max}$ and $A_\text{min}$ should be equal by symmetry and serve as a mutual consistency check. Applying the condition of Eq.(~\ref{eq:Ppi}) and averaging the two slopes for robustness gives
\begin{equation}
    P_\pi = \frac{2\left(\lambda_{n+1}^{\text{min},(0)} - \lambda_n^{\text{max},(0)}\right)}{A_\text{max} + A_\text{min}} \approx \frac{\text{FSR}}{A_\text{max} + A_\text{min}},
    \label{eq:Ppi_final}
\end{equation}
where $\lambda_n^{\text{max},(0)}$ and $\lambda_{n+1}^{\text{min},(0)}$ are the zero-pump positions of the $n$-th maximum and adjacent $(n+1)$-th minimum, and the approximation uses $\lambda_{n+1}^{\text{min},(0)} - \lambda_n^{\text{max},(0)} \approx \text{FSR}/2$.

\section{S4. Measurement of optical losses.}
\setlabel{S4}{sec:losses}

A key figure of merit for assessing the scalability of the proposed phase shifters is the insertion loss (IL) per device. For the architecture employed here, this has three contributions: the rectangular-to-slot waveguide coupler (and its reciprocal), the slot-to-SWGSW coupler (and its reciprocal), and propagation losses in the SWGSW section. All three depend---to different degrees---on the slot width and may therefore evolve as the structure deforms under pump illumination. To explore this dependence systematically, we characterize each loss pathway independently using dedicated test devices on the same chip as the UMZIs. To extract the loss of interest, we employ the cutback method, i.e., we measure the transmission across photonic circuits including a varying number (or length) of the lossy component under test and ensure that all other lossy elements (in number and length) are kept fixed. 

For most photonic components, the ensemble-averaged logarithm of the transmittance follows
 \begin{equation}
    \overline{\log(T)} = -\alpha(\lambda)M + \overline{\log(T_0(\lambda))},
    \label{eq:cutback}
 \end{equation}
where $\alpha(\lambda)$ is the wavelength-dependent loss we seek to obtain, $\overline{\log(T_o(\lambda))}$ encompasses a constant IL loss independent of the component under test and $M$ may either be the length of the waveguide, $L$, for which the propagation losses are analyzed, $M \equiv L \in \mathbb{R}_{\geq 0}$, or the number of photonic components $N$ placed in series whose losses are analyzed, $M \equiv N \in \mathbb{N}_0$. In practice, the ensemble averaging in Eq.(~\ref{eq:cutback}), which is particularly relevant when evaluating extrinsic propagation losses due to the stochastic nature of fabrication imperfection such as sidewall roughness, is ensured by performing transmission measurements on sets of $S$ nominally identical circuits for each value of $M$.

\subsection{S4.1. Propagation losses of subwavelength grating slot waveguides}
\setlabel{S4.1}{subsec:propagation_losses}

\begin{figure*}[t!]
    \includegraphics[width=\textwidth]{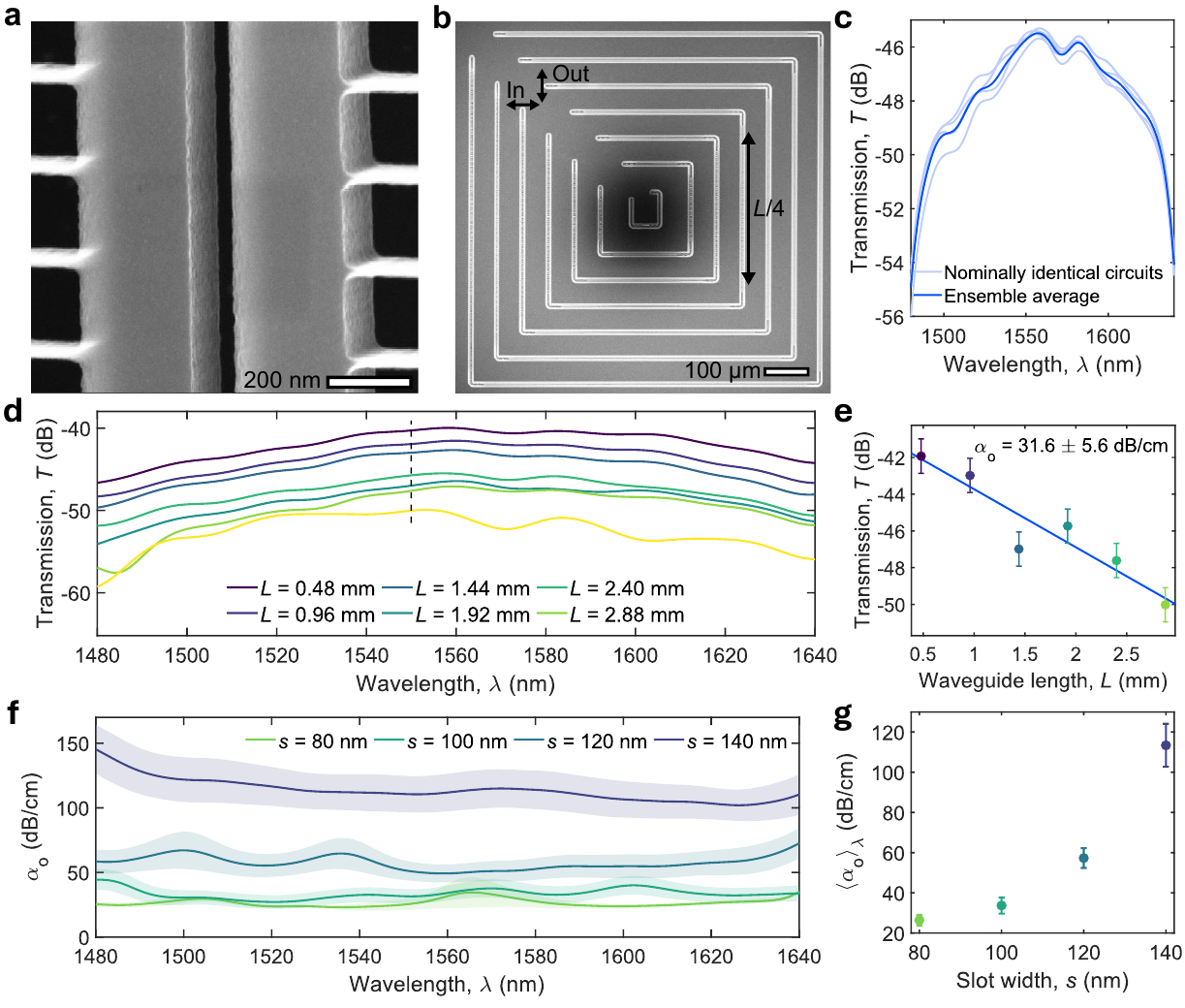}
    \caption{\textbf{Cutback measurements of propagation losses in subwavelength grating slot waveguides (SWGSW).} \textbf{a}, Tilted high-magnification SEM image of the central region of a SWGSW. \textbf{b}, Low-magnification SEM image of one set of varying-length devices for extracting the propagation losses. Each of these circuits has $S$ = 4 nominally identical copies on the chip. \textbf{c}, Transmission spectra of the $S$ = 4 circuits for one particular length (light blue lines) and their average (dark blue line). \textbf{d}, Ensemble-averaged transmission spectra as a function of length for SWGSW of slot $s$ = 100 nm. The dashed line represents the wavelength ($\lambda$ = 1550 nm) for which the linear fit in \textbf{e} is shown. \textbf{f}, Extracted propagation losses, $\alpha_\text{o}$, and their standard error as a function of wavelength and slot width $s$. \textbf{g}, Wavelength-averaged propagation losses as a function of $s$.}
    \label{fig:SWGSW_Loss} 
\end{figure*}

The slot-confined nature of the optical mode sustained by the SWGSW is a key feature of the optomechanical interactions underpinning the phase shifters we explore. At the same time, the strongly confined optical fields and the large electric field at the air-silicon interfaces can lead to strong scattering losses in the presence of fabrication imperfections~\cite{arregui_cavity_2023}. Despite employing a high-resolution nanofabrication process (see Section~\ref{sec:fabrication} for details), a finite amount of line-edge roughness and sidewall roughness respectively introduced during the electron-beam lithography and pattern transfer steps are unavoidable. Figure~\ref{fig:SWGSW_Loss}\textbf{a} shows a high-magnification SEM of the SWGSW, where a small yet visible amount of roughness is apparent both on the rectangular beam sidewalls and on the narrow tethers. We evaluate the extrinsic propagation losses of the SWGSW via the measurement of $S$ = 4 nominally-identical sets of devices that include photonic circuits with 6 different SWGSW lengths, ranging from $L$ = 0.48 mm to $L$ = 2.88 mm. To prevent the large spread in circuit transmittances that may result from waveguides that cross stitched electron-beam write fields~\cite{bolten_at_2014}, we include all circuits in a single 1x1 mm$^2$ writefield in 'O'-shaped circuits, as shown in Fig.~\ref{fig:SWGSW_Loss}\textbf{b}. The total length $L$ of the SWGSW is distributed in four equal parts which are interfaced to rectangular waveguides --used for the bends-- using the two devices described in Section~\ref{subsec:additional_devices}. As for all other circuits in this work, we use cross-polarized confocal transmittance spectroscopy. The experiment is carried out for SWGSWs of four different slot widths, $s$ = 80, 100, 120, and 140 nm, in an attempt to capture the relevant slot width range spanned by the phase shifters as they deform.

Figures~\ref{fig:SWGSW_Loss}\textbf{c}--\textbf{g} present the propagation loss measurements. As discussed in subsection~\ref{subsec:smoothing}, all displayed wavelength-dependent transmission traces are smoothed. The first step to find the losses is to take the ensemble average of each set of $S$ = 4 nominally identical circuits, a characteristic example of which is shown in Fig.~\ref{fig:SWGSW_Loss}\textbf{c}. The ensemble-averages as a function of $L$ for $s$ = 100 nm are displayed in Fig.~\ref{fig:SWGSW_Loss}\textbf{d} and exhibit a monotonic decrease in transmission with length. We then fit, at each probed wavelength, a first order polynomial to the transmission data as a function of length following Eq.(~\ref{eq:cutback}). A characteristic fit is shown in Fig.~\ref{fig:SWGSW_Loss}\textbf{e} and the extracted propagation loss, $\alpha_\text{o}$, indicated in text, including the standard error of the fit. The loss $\alpha_\text{o}$ as a function of wavelength and slot width are shown in Fig.~\ref{fig:SWGSW_Loss}\textbf{f} and representative values summarized in Table~\ref{tab:propagationlosses}. We observe the wavelength dependence to be considerably weaker than the slot width dependence. In Fig.~\ref{fig:SWGSW_Loss}\textbf{g} we report the mean propagation losses in the range $\lambda \in [1450,1640]$ nm, $\langle\alpha_\text{o}\rangle_\lambda$ for the different slot widths $s$. The observed trend is counterintuitive: one would expect narrower slots to produce higher losses, because the optical field is more concentrated at the slot sidewalls and the narrow geometry may also incur more roughness during fabrication. Instead, we observe the opposite, with losses increasing with slot width. We attribute this to the tighter optical confinement provided by narrower slots pushing the modal field away from the tethers: the dominant scattering mechanism appears to be sidewall roughness on the narrow tethers rather than on the slot facets, and the tighter confinement in narrow slots reduces the field overlap with those outer surfaces. As a consequence, SWGSWs with a larger slot, which confines the field less tightly and therefore increases its overlap with the tether sidewalls, are more lossy. We do not, however, have conclusive experimental data to unambiguously support this hypothesis.

\begin{table}[t!]
    \begingroup
    \setlength{\tabcolsep}{8pt}
    \setlength{\extrarowheight}{3pt}
    \begin{tabular}{|c|c|c|c|c|}
        \hline
        Slot width [nm] & Mean loss [dB/cm] & $\lambda$ = 1500 nm & $\lambda$ = 1550 nm & $\lambda$ = 1600 nm\\
        \hline
        80 & 26.4 $\pm$ 2.9 & 29.1 $\pm$ 1.9 & 25.7 $\pm$ 3.0 & 24.1 $\pm$ 1.3 \\
        100 & 33.7 $\pm$ 4.0 & 31.8 $\pm$ 4.1 & 32.3 $\pm$ 4.0 & 39.2 $\pm$ 5.0\\
        120 & 57.3 $\pm$ 4.9 & 61.8 $\pm$ 10.6 & 49.2 $\pm$ 5.6 & 53.4 $\pm$ 6.5\\
        140 & 113.4 $\pm$ 8.2 & 105.8 $\pm$ 15.3 & 99.2 $\pm$ 11.9 & 99.7 $\pm$ 10.7\\
        \hline
    \end{tabular}
    \endgroup
    \caption{\textbf{Propagation losses}}
    \label{tab:propagationlosses}
\end{table}

\subsection{S4.2. Insertion loss of slot waveguide to subwavelength grating slot waveguide couplers}
\setlabel{S4.2}{subsec:insertion_loss_SW_to_SWGSW}

The constant term $\overline{\log(T_o(\lambda))}$ in Eq.(~\ref{eq:cutback}) lumps together all power-independent losses in the measurement chain: the free-space optical setup, the grating coupler in- and out-coupling, and the insertion losses of all photonic circuit components used to interface with the element under study. On its own this quantity is of little interest, but comparing it against the transmission of an otherwise identical circuit from which a known element has been removed isolates the loss of that element. Here, since the same circuits used to extract $\alpha_\text{o}$ each contain $n = 8$ slot-waveguide to SWGSW couplers (Section~\ref{subsec:additional_devices}), the intercept $\overline{\log(T_o(\lambda))}$ can be used to extract the single-coupler transmittance, $T_\text{SW-SWGSW}$, by comparison against reference circuits containing no couplers or SWGSW sections. These reference circuits, which we term \emph{shunts}, correspond to the innermost 'O'-shaped circuit in Fig.~\ref{fig:SWGSW_Loss}\textbf{b}.

\begin{figure*}[t!]
    \includegraphics[width=\textwidth]{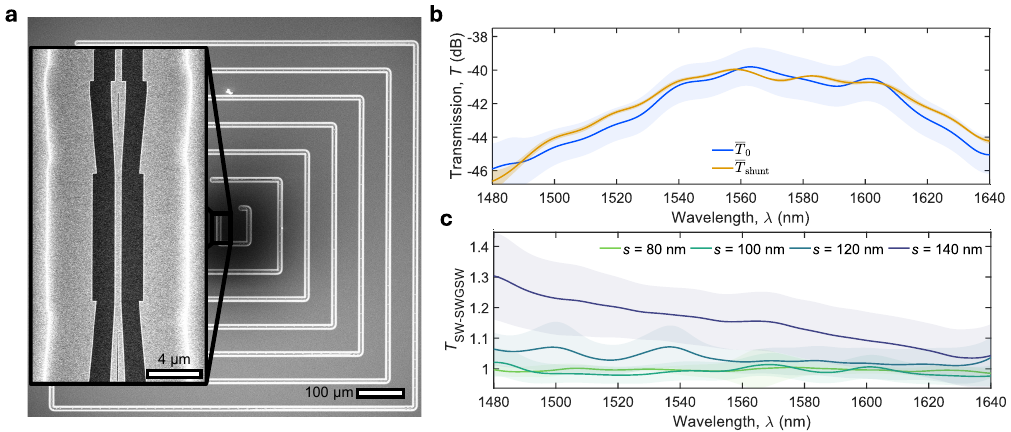}
    \caption{\textbf{Transmission measurement of the slot-waveguide to SWGSW coupler. a,} SEM image of the cutback circuits, including an inset showing a collapsed slot waveguide in the shortest \emph{shunt} circuit. \textbf{b,} Comparison of the transmission of the \emph{shunt} circuit and the transmission inferred for SWGSWs of $L$ = 0. \textbf{c,} Inferred transmittance of the slot-waveguide to SWGSW, $T_\text{SW-SWGSW}$, from the traces in \textbf{b}, shown for varying slot width. The fact that $T_\text{SW-SWGSW} > 1$ indicates that the \emph{shunt} circuits may have been compromised during fabrication (see inset in \textbf{a}).}
    \label{fig:SW_SWGSW_losses} 
\end{figure*}

Denoting the average transmission of the $S = 4$ shunts as $\overline{T}_\text{shunt}$, the single-coupler transmittance is
\begin{equation}
    T_\text{SW-SWGSW} = \sqrt[n]{\frac{10^{\overline{\log(T_o(\lambda))}/10}}{\overline{T}_\text{shunt}}},
    \label{eq:coupler_trans}
\end{equation}
with $n = 8$. A comparison of $\overline{T_0} \equiv 10^{\overline{\log(T_o(\lambda))}/10}$ and $\overline{T}_\text{shunt}$ shows that the two spectra differ only slightly (Fig.~\ref{fig:SW_SWGSW_losses}\textbf{b}), and over certain wavelength ranges $\overline{T_0}$ actually exceeds $\overline{T}_\text{shunt}$. This can be attributed to several causes: resonances arising from finite reflections between photonic components; differences in pattern-transfer fidelity due to short-range proximity-effect dose variations at the center of the 'O'-shaped circuits; and, most likely, modified fabrication-induced deformations (from stress release or surface forces) depending on whether SWGSWs are present. SEM inspection of the shunts revealed that some had collapsed slot waveguide sections (inset of Fig.~\ref{fig:SW_SWGSW_losses}\textbf{a}), which further compromises the comparison. As a result, the extracted transmittance values go even above one, as illustrated in Fig.~\ref{fig:SW_SWGSW_losses}\textbf{c}. That the loss falls below the detection limit of this method is itself evidence of the high optical quality of the SW-SWGSW interfaces. In the following and in the main text, we therefore consider the simulated transmittance (Fig.~\ref{fig:SWtoSWGSW}) as if it were achieved.

\subsection{S4.3. Insertion loss of V-groove rectangular waveguide to slot waveguide couplers}
\setlabel{S4.3}{subsec:insertion_loss_RectW_to_SW}

\begin{figure*}[t!]
    \includegraphics[width=\textwidth]{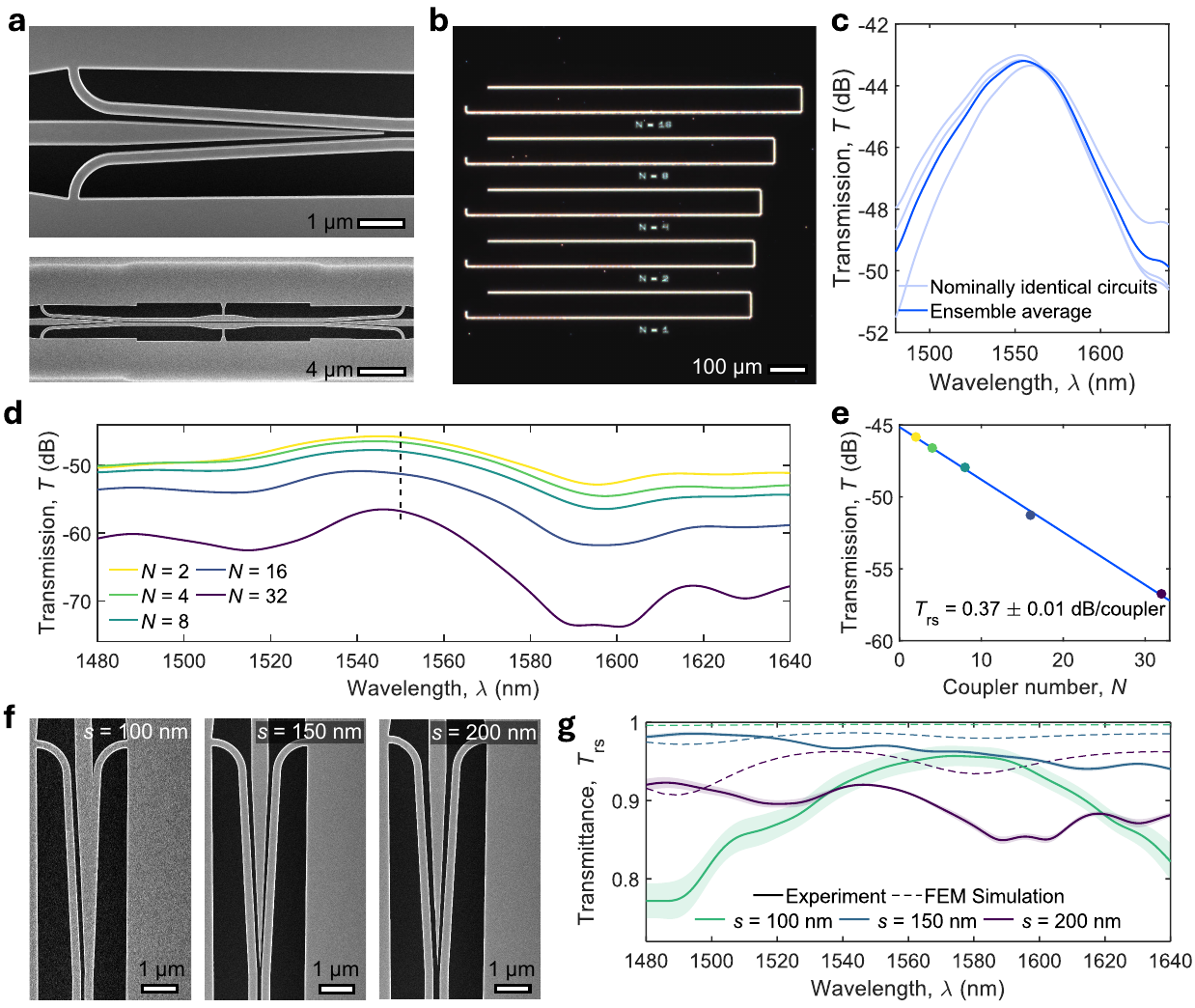}
    \caption{\textbf{Insertion loss of the V-groove rectangular-to-slot waveguide coupler.} \textbf{a,} SEM images of the V-groove coupler. \textbf{b,} Optical microscope image of the concatenated coupler circuits ($N = 2$ to $32$) used for the cutback measurements. \textbf{c,} Transmission spectra of $S = 4$ nominally identical circuits for $N = 2$ and $s = 200$~nm (light lines) and their ensemble average (dark line). \textbf{d,} Ensemble-averaged transmission spectra for $N = 2, 4, 8, 16, 32$ at $s = 200$~nm. The dashed line marks $\lambda = 1550$~nm. \textbf{e,} Transmission at $\lambda = 1550$~nm as a function of the number of couplers in series. A linear fit to the data gives $T_\text{rs} = 0.37 \pm 0.01$~dB/coupler. \textbf{f,} SEM images of couplers fabricated with slot widths $s = 100$, $150$, and $200$~nm. \textbf{g,} Extracted single-coupler transmittance as a function of wavelength for the three slot widths (solid lines), compared to FEM simulations (dashed lines). The shaded region represents the standard error.}
    \label{fig:Vtaperlosses}
\end{figure*}

The insertion loss of the V-groove rectangular-to-slot waveguide coupler (Fig.~\ref{fig:Vtaperlosses}\textbf{a}) is extracted via the cutback method, using circuits with $N = 2, 4, 8, 16$, and $32$ couplers in series (Fig.~\ref{fig:Vtaperlosses}\textbf{b}). For each value of $N$, the transmission spectra of $S = 4$ nominally identical circuits are measured and ensemble-averaged to suppress extrinsic variability (Fig.~\ref{fig:Vtaperlosses}\textbf{c}). The resulting averaged spectra, shown in Fig.~\ref{fig:Vtaperlosses}\textbf{d} for $s = 200$~nm, decrease systematically with increasing $N$ across the full measured wavelength range. Applying the cutback fit of Eq.(~\ref{eq:cutback}) at each wavelength, and extracting the per-coupler loss from the slope, yields $T_\text{rs} = 0.37 \pm 0.01$~dB/coupler at $\lambda = 1550$~nm for $s = 200$~nm (Fig.~\ref{fig:Vtaperlosses}\textbf{e}).

The measurement is repeated for three slot widths, $s = 100$, $150$, and $200$~nm, whose geometries are shown in Fig.~\ref{fig:Vtaperlosses}\textbf{f}. The extracted transmittances are plotted in Fig.~\ref{fig:Vtaperlosses}\textbf{g} alongside FEM predictions. Across all three slot widths, the experimental transmittance falls systematically below the FEM values, indicating that fabrication imperfections introduce additional losses not captured by the simulations. Among the three cases, $s = 150$~nm yields the highest transmittance, remaining above 95\% across the full measured bandwidth. The $s = 200$~nm coupler performs somewhat worse, as expected from the simulation trends in Section~\ref{subsec:additional_devices}. The $s = 100$~nm coupler shows markedly lower and more variable transmittance than the other two, with substantially larger uncertainty. The SEM images in Fig.~\ref{fig:Vtaperlosses}\textbf{f} offer a direct explanation: while the $s = 150$ and $200$~nm couplers retain their slot geometry intact, the $s = 100$~nm coupler shows clear signs of in-plane collapse of the narrow silicon beams defining the slot, which would drastically reduce the transmission of individual couplers within the cutback series and simultaneously inflate the extracted losses and increase the device-to-device variability.

All SWGSWs characterized in this work include $s = 200$~nm V-groove input and output couplers. This choice was made conservatively in a first-generation design to guarantee collapse-free operation and to keep the initial slot width of the coupler ($s_\text{t} = s = 200$~nm) well separated from the SWGSW slot widths, so that the coupler contribution to the total optical force profile does not strongly perturb the modelling. Switching to $s = 150$~nm couplers, for which no collapses are observed, would reduce the per-coupler insertion loss by approximately $0.27$~dB, yielding a total gain of $\approx 0.5$~dB per phase shifter (two couplers) without any change to the SWGSW itself.

\subsection{S4.4. Total insertion loss of the phase shifter}
\setlabel{S4.4}{subsec:total_insertion_loss}

\begin{figure*}[t!]
    \includegraphics[width=\textwidth]{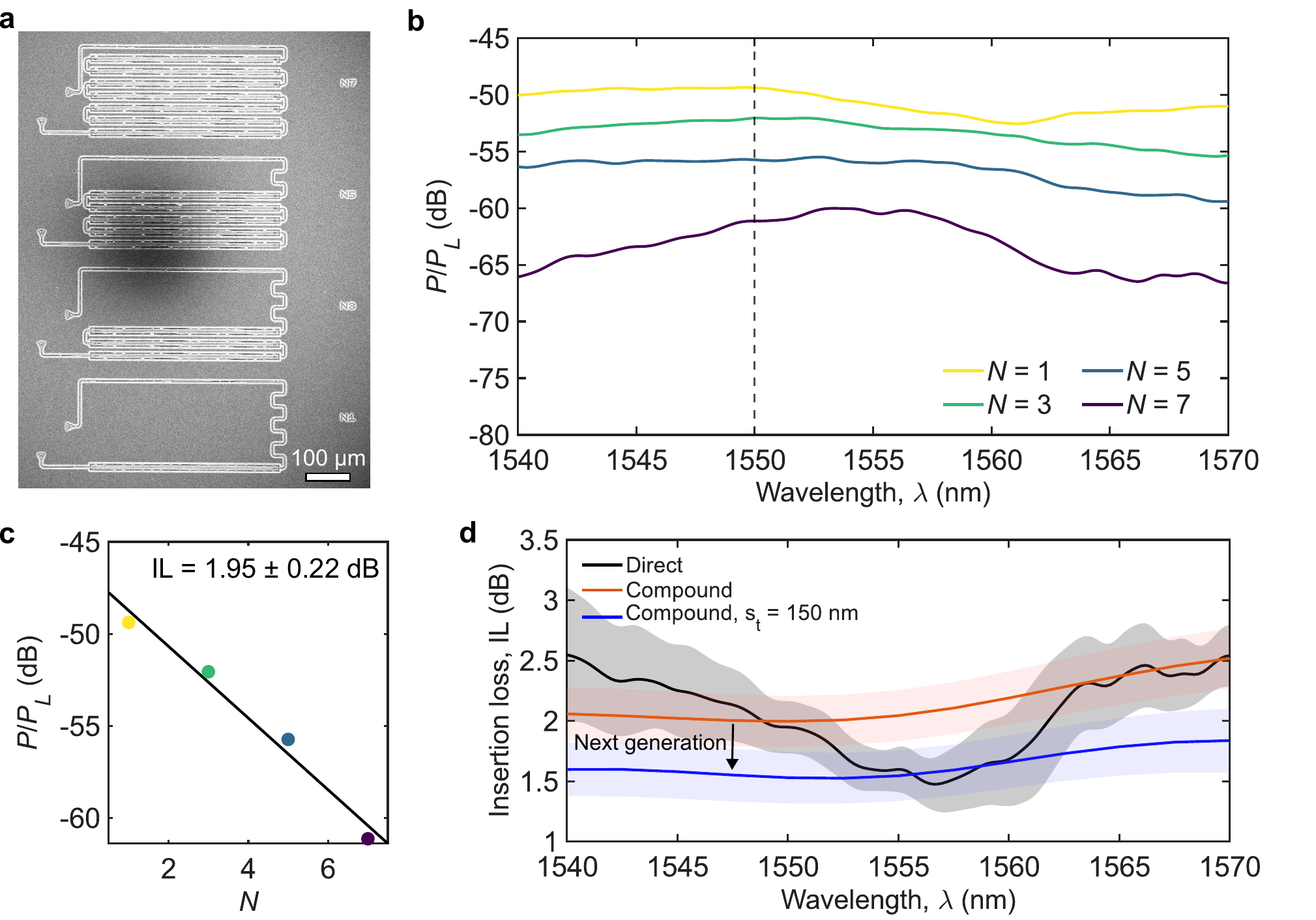}
    \caption{\textbf{Total insertion loss of the optomechanical phase shifter.} \textbf{a,} SEM image of the cutback measurement circuits, showing the $N = 1$, $3$, $5$, and $7$ phase shifters connected in series (scale bar: 100\,\textmu m). \textbf{b,} Transmission spectra $P/P_\text{L}$ (in dB) for $N = 1$, $3$, $5$, and $7$ nominally identical phase shifters with $\{s, L, k\} = \{100\,\text{nm},\, 400\,\mu\text{m},\, 0.564\,\text{N/m}\}$, plotted over the wavelength range where all circuits are above the power-meter noise floor. The dashed line marks $\lambda = 1550$\,nm. \textbf{c,} Transmission at $\lambda = 1550$\,nm as a function of $N$, with a linear cutback fit yielding an insertion loss of $\text{IL}_\text{PS} = 1.95 \pm 0.22$\,dB per phase shifter. \textbf{d,} Comparison of the directly extracted insertion loss $T_\text{PS}$ (black line, grey uncertainty band) and the compound estimate $T_\text{compound}$ (orange) assembled from the individual loss contributions characterized in Sections~\ref{subsec:propagation_losses}--\ref{subsec:insertion_loss_RectW_to_SW}.}
    \label{fig:phaseshifterloss}
\end{figure*}

To verify the consistency of the individual loss characterizations and to quantify the total insertion loss of the full device, we perform a cutback experiment on complete phase shifters with $\{s, L, k\} = \{100\,\text{nm},\, 400\,\mu\text{m},\, 0.564\,\text{N/m}\}$, using $N = 1$, $3$, $5$, and $7$ devices in series. The circuits are shown in the SEM of Fig.~\ref{fig:phaseshifterloss}\textbf{a}. The transmission spectra (Fig.~\ref{fig:phaseshifterloss}\textbf{b}) are plotted only over the wavelength range where all four circuits lie above the power-meter noise floor; outside this window, the $N = 5$ and $N = 7$ spectra fall below the noise level. This reduced bandwidth arises from a grating coupler etching defect on this chip that narrows the usable coupling window. The defect does not affect the insertion loss inferred at $\lambda = 1550$~nm, which lies well within the usable range for all four values of $N$.

Restricting the cutback analysis to $\lambda = 1550$\,nm, where all four values of $N$ yield measurable signals, yields $\text{IL}_\text{PS} = 1.95 \pm 0.22$\,dB per phase shifter (Fig.~\ref{fig:phaseshifterloss}\textbf{c}). This integrated figure of merit captures all loss contributions of a complete device: two V-groove rectangular-to-slot couplers, two SW-to-SWGSW couplers, and the SWGSW propagation losses over the $L = 400\,\mu\text{m}$ length. Figure~\ref{fig:phaseshifterloss}\textbf{d} compares the directly extracted insertion loss against the compound estimate assembled by combining the per-element losses extracted in Sections~\ref{subsec:propagation_losses}--\ref{subsec:insertion_loss_RectW_to_SW}. The two estimates are broadly consistent, although the agreement is imperfect. This is not unexpected as the individual loss characterizations entering the compound loss were performed on the main chip of this manuscript, fabricated more than one year earlier, while the direct measurement is on a separate chip with the grating coupler defect noted above. The two chips may therefore have different per-element loss budgets. Taken together, the broad consistency between the two estimates validates the individual loss characterizations carried out in Sections~\ref{subsec:propagation_losses}--\ref{subsec:insertion_loss_RectW_to_SW}, and confirms that the complete phase shifter achieves a total insertion loss of approximately 2\,dB over a 30\,nm bandwidth. As discussed in Section~\ref{subsec:insertion_loss_RectW_to_SW}, replacing the V-groove couplers with ones designed for an initial gap $s_t = 150$\,nm would reduce the per-coupler insertion loss by $\approx 0.27$\,dB, yielding an overall improvement of $\approx 0.5$\,dB per device. The corresponding compound estimate, assembled using the experimental $s_t = 150$\,nm coupler data of Fig.~\ref{fig:Vtaperlosses}\textbf{g}, is shown as the blue solid line in Fig.~\ref{fig:phaseshifterloss}\textbf{d}.

\subsection{S4.5. Propagation losses in circuits with reduced tether width and slot width}
\setlabel{S4.5}{subsec:propagation_losses_extended}
\begin{figure}[t!]
    \includegraphics[width=0.9\columnwidth]{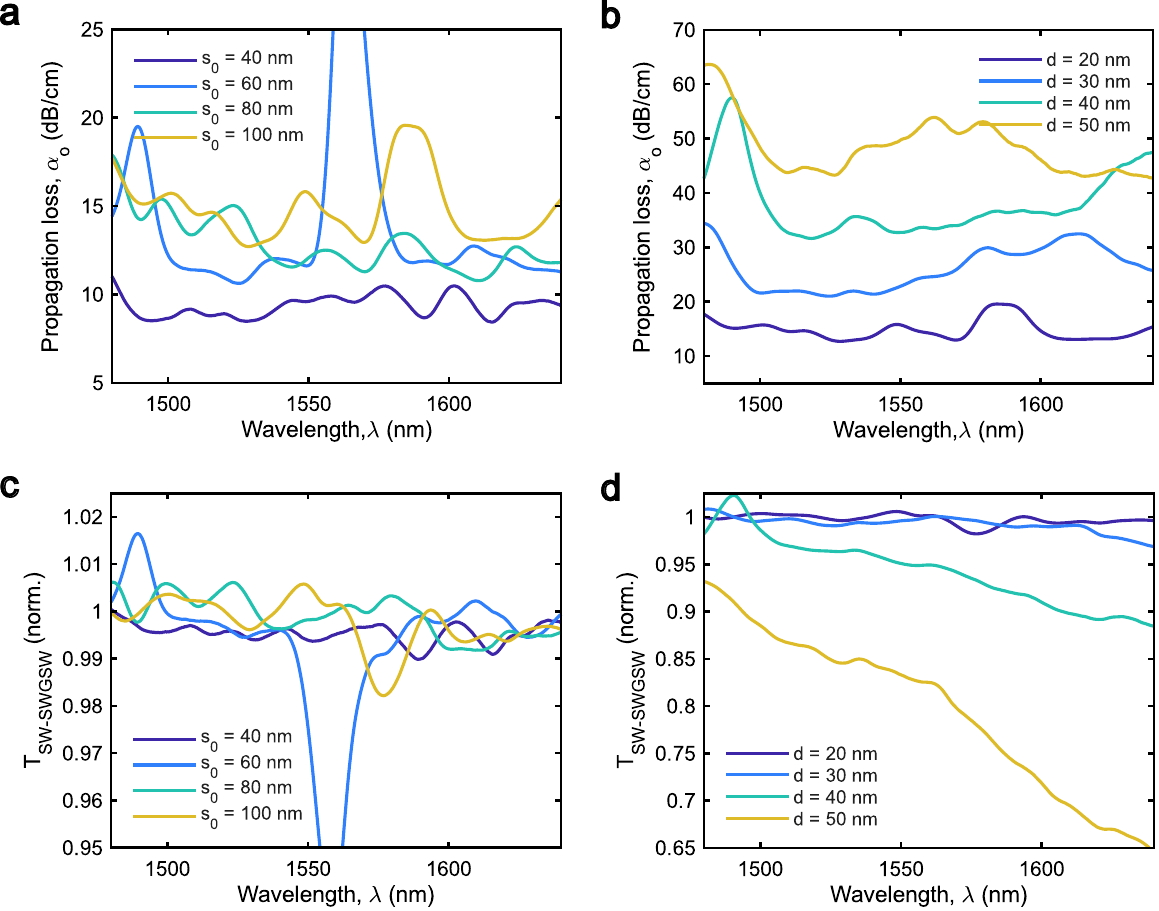}
    \caption{\textbf{Propagation losses in suspended circuits with reduced tether and slot widths.} \textbf{a}, Propagation loss as a function of wavelength for tether width $d = 20$~nm and slot widths $s_0 \in \{40, 60, 80, 100\}$~nm. \textbf{b}, Propagation loss for $s_0 = 100$~nm and tether widths $d \in \{20, 30, 40, 50\}$~nm. \textbf{c} Slot-waveguide to SWGSW inferred transmittance for geometries in \textbf{a}. \textbf{d} Equivalent to \textbf{c} with geometries in \textbf{b}.}
    \label{fig:prop_losses_extended}
\end{figure}
To assess the potential for propagation loss reduction beyond the geometries characterized above and used for the phase shifters, we report measurements performed on suspended-but-unsprung circuits of the same design as the cutback structures in Section~\ref{subsec:propagation_losses}, fabricated on a separate chip processed soon after the chip employed for all the results shown in this manuscript. The SWGSW tether width, $d$, and slot width, $s_0$, are varied independently over a wider range: $d \in \{20, 30, 40, 50\}$~nm and $s_0 \in \{40, 60, 80, 100\}$~nm. Figure~\ref{fig:prop_losses_extended}\textbf{a} shows the propagation losses for fixed $d = 20$~nm and varying $s_0$. The loss decreases monotonically with $s_0$, except for an anomalous peak at 1560 nm for $s_0$ = 60 nm. At $s_0 = 40$~nm the losses are $\sim$10~dB/cm across much of the C-band, approximately three times lower than the value measured at $s_0 = 80$~nm in Fig.~\ref{fig:SWGSW_Loss}\textbf{f,g}. The behaviour with $s_0$ is consistent with the trend observed on the main chip, albeit with lower propagation losses for equivalent values of $s_0$ due to the use of $d$ = 20 nm for the tethers. This is evidenced by Fig.~\ref{fig:prop_losses_extended}\textbf{b} that shows the complementary dependence at fixed $s_0 = 100$~nm and varying $d$: reducing $d$ from 30~nm to 20~nm lowers the propagation loss by between 5 and 10~dB/cm, further supporting the conclusion that sidewall roughness on the narrow tethers is the dominant scattering mechanism and that the loss reduction with decreasing slot is consistent with a picture in which tighter optical confinement in narrow slots reduces the field overlap with the tether sidewalls. Together, these measurements suggest that combining $s_0 \lesssim 60$~nm with $d = 20$~nm tethers could bring the propagation loss contribution to $\lesssim 0.2$~dB for a $L = 150~\mu$m device, and a total insertion loss below 0.5-0.7~dB when used with the improved couplers of Section~\ref{subsec:insertion_loss_RectW_to_SW}. In addition, we also provide in Fig.~\ref{fig:prop_losses_extended}\textbf{c,d} the extracted loss for the slot waveguide to SWGSW adiabatic taper, following the same protocol described in Section~\ref{subsec:insertion_loss_SW_to_SWGSW}. The dependence with $d$ (Fig.~\ref{fig:prop_losses_extended}\textbf{d}) evidences that the device transmission drops fast with $d$ and, unlike in Fig.~\ref{fig:SW_SWGSW_losses}\textbf{c}, we are able to reliably pinpoint considerable losses for $d > 40$~nm. This behaviour is also observed in simulations (Fig.~\ref{fig:SWtoSWGSW}\textbf{c}). In contrast, the dependence with slot width (Fig.~\ref{fig:prop_losses_extended}\textbf{c}) does not exhibit a clear trend and the extracted transmittance exceeds unity over some wavelength ranges, consistent with the detection-limit argument of Section~\ref{subsec:insertion_loss_SW_to_SWGSW}.

\end{document}